\documentclass[
  aps,
  prb,
  twocolumn,
  reprint,
  superscriptaddress,
  floatfix,
  citeautoscript,
  longbibliography,
]{revtex4-2}

\usepackage[T1]{fontenc}
\usepackage[utf8]{inputenc}

\usepackage{newtxtext}
\usepackage{newtxmath}

\usepackage{array}
\newcolumntype{C}[1]{>{\centering}m{#1}}

\usepackage{chemformula}
\usepackage{siunitx}
\DeclareSIUnit\atom{atom}
\DeclareSIUnit\gauss{G} 
\DeclareSIUnit\torr{Torr}
\DeclareSIUnit{\wtpercent}{wt.~\%}
\DeclareSIUnit{\ions}{ions}
\DeclareSIUnit\ppm{\text{ppm}}
\DeclareSIUnit\emu{\text{emu}}
\DeclareSIUnit\barn{b} 
\DeclareSIUnit\atomicmassunit{u} 

\usepackage{graphicx}
\graphicspath{{./figures}}

\usepackage[
   unicode,
   colorlinks = true,
   allcolors = blue,
]{hyperref}
\usepackage{booktabs,multirow}

\newcommand*{\figref}[2][]{%
  \hyperref[{fig:#2}]{%
    Figure~\ref*{fig:#2}%
    \ifx\\#1\\%
    \else
      \,#1%
    \fi
  }%
}

\usepackage{cleveref}

\usepackage{microtype}

\usepackage{comment}
\usepackage[caption=false]{subfig} 
\usepackage[normalem]{ulem}

\usepackage{tikz}
\PassOptionsToPackage{usenames,dvipsnames}{xcolor}

\usepackage{glossaries}
\glsdisablehyper
\newacronym{1d}{1D}{one-dimensional}
\newacronym{2d}{2D}{two-dimensional}
\newacronym{3d}{3D}{three-dimensional}

\newacronym{ac}{AC}{alternating current}
\newacronym{afm}{AFM}{atomic force microscopy}
\newacronym{alc}{ALC}{avoided level crossing}
\newacronym{ald}{ALD}{atomic layer deposition}
\newacronym{api}{API}{application programming interface}
\newacronym{ariel}{ARIEL}{Advanced Rare Isotope Laboratory}
\newacronym{arpes}{ARPES}{angle-resolved photoemission spectroscopy}
\newacronym{atp}{ATP}{adenosine triphosphate}
\newacronym{accg}{\ensuremath{E_{\mathrm{acc}}}}{accelerating gradient}
\newacronym{alcr}{ALCR}{avoided level-crossing resonance}
\newacronym{H0}{\ensuremath{H_{0}}}{applied field}

\newacronym[sort={b-NMR}]{bnmr}{\ensuremath{\beta}-NMR}{\ensuremath{\beta}-detected nuclear magnetic resonance}
\newacronym[sort={b-NQR}]{bnqr}{\ensuremath{\beta}-NQR}{\ensuremath{\beta}-detected nuclear quadrupole resonance}
\newacronym{bca}{BCA}{binary collision approximation}
\newacronym{bcc}{BCC}{body-centred cubic}
\newacronym{bcp}{BCP}{buffered chemical polishing}
\newacronym{bcs}{BCS}{Bardeen-Cooper-Schrieffer}
\newacronym{bl}{BL}{Bean-Livingston}
\newacronym{bpp}{BPP}{Bloembergen-Purcell-Pound}
\newacronym{bsc}{BSC}{\ch{Bi2Se3:Ca}}
\newacronym{btm}{BTM}{\ch{Bi2Te3:Mn}}
\newacronym{bts}{BTS}{\ch{Bi2Te2Se}}

\newacronym{camp}{CAMP}{control and monitor program}
\newacronym{ccd}{CCD}{charge-coupled device}
\newacronym{cdw}{CDW}{charge density wave}
\newacronym{cgs}{CGS}{centimetre-gram-second system of units}
\newacronym{cmms}{CMMS}{Centre for Molecular and Materials Science}
\newacronym{codata}{CODATA}{Committee on Data for Science and Technology}
\newacronym{cpu}{CPU}{central processing unit}
\newacronym{create}{CREATE}{Collaborative Research and Training Experience Program}
\newacronym{cw}{CW}{continuous wave}

\newacronym{daq}{DAQ}{data acquisition}
\newacronym{dc}{DC}{direct current}
\newacronym{dft}{DFT}{density functional theory}
\newacronym{dos}{DOS}{density of states}
\newacronym{dqt}{DQT}{double-quantum transition}
\newacronym{dfts}{DFTS}{dispersive Fourier transform spectrometer}

\newacronym{efg}{EFG}{electric field gradient}
\newacronym{emim-ac}{EMIM-Ac}{1-ethyl-3-methylimidazolium acetate}
\newacronym{emim-dca}{EMIM-DCA}{1-ethyl-3-methylimidazolium dicyanamide}
\newacronym{epr}{EPR}{electron paramagnetic resonance}
\newacronym{esr}{EPR}{electron spin resonance}
\newacronym{endor}{ENDOR}{electron nuclear double resonance}
\newacronym{epics}{EPICS}{Experimental Physics and Industrial Control System}
\newacronym{ep}{EP}{electropolishing}
\newacronym{edx}{EDX}{energy dispersive X-ray spectroscopy}

\newacronym{fcc}{FCC}{face-centred cubic}
\newacronym{fll}{FLL}{flux-line lattice}
\newacronym{fft}{FFT}{fast Fourier transform}
\newacronym{fom}{FoM}{figure of merit}
\newacronym{fwhm}{FWHM}{full width at half maximum}
\newacronym{ffvp}{\ensuremath{B_{\mathrm{vp}}}}{field of first vortex penetration}

\newacronym{gga}{GGA}{generalized gradient approximation}
\newacronym{gl}{GL}{Ginzburg-Landau}
\newacronym{gixrd}{GIXRD}{Grazing incidence x-ray diffraction}

\newacronym{hb}{HB}{hole-burning}
\newacronym{hfqs}{HFQS}{high-field \ensuremath{Q} slope}
\newacronym{hv}{HV}{high-voltage}
\newacronym{hwhm}{HWHM}{half width at half maximum}
\newacronym{hpcvd}{HPCVD}{hybrid physical chemical vapour deposition}
\newacronym{hs}{HS}{Hebel-Slichter}

\newacronym{iaea}{IAEA}{International Atomic Energy Agency}
\newacronym{il}{IL}{ionic liquid}
\newacronym{is}{IS}{impedance spectroscopy}
\newacronym{isac}{ISAC}{isotope separator and accelerator}
\newacronym{isol}{ISOL}{isotope separation online}
\newacronym{isosim}{IsoSiM}{Isotopes for Science and Medicine}

\newacronym{jlab}{JLab}{Thomas Jefferson National Accelerator Facility}

\newacronym{lcao}{LCAO}{linear combination of atomic orbitals}
\newacronym{lda}{LDA}{local density approximation}
\newacronym{led}{LED}{light-emitting diode}
\newacronym{leis}{LEIS}{low-energy ion scattering}
\newacronym{lib}{LIB}{lithium-ion battery}
\newacronym{lsat}{LSAT}{\ch{(La,Sr)(Al,Ta)O3}}
\newacronym{ltb}{LTB}{low-temperature baking}

\newacronym{mfm}{MFM}{magnetic force microscopy}
\newacronym{mas}{MAS}{magic angle spinning}
\newacronym{mpms}{MPMS}{magnetic property measurement system}
\newacronym{mbe}{MBE}{molecular beam epitaxy}
\newacronym{md}{MD}{molecular dynamics}
\newacronym{midas}{MIDAS}{Maximum Integrated Data Acquisition System}
\newacronym{mit}{MIT}{metal-insulator transition}
\newacronym{mnr}{MNR}{Meyer-Neldel rule}
\newacronym{mqt}{mqt}{multi-quantum transition}
\newacronym{mud}{MUD}{muon data}
\newacronym{ms}{MS}{mass spectrometry}
\newacronym{bmax}{\ensuremath{B_\mathrm{max}}}{maximum field in superconducting
heterostructures that can be sustained while remaining in the
Meissner state}

\newacronym{nbm}{NBM}{neutral beam monitor}
\newacronym{neb}{NEB}{nudged elastic band}
\newacronym{nim}{NIM}{nuclear instrumentation module}
\newacronym{nmr}{NMR}{nuclear magnetic resonance}
\newacronym{no}{NO}{nuclear orientation}
\newacronym{nqr}{NQR}{nuclear quadrupole resonance}
\newacronym{nrc}{NRC}{National Research Council of Canada}
\newacronym{nserc}{NSERC}{Natural Sciences and Engineering Research Council of Canada}

\newacronym{oa}{OA}{optical absorption}

\newacronym{pac}{PAC}{perturbed angular correlation}
\newacronym{pad}{PAD}{perturbed angular distribution}
\newacronym{pas}{PAS}{principle axis system}
\newacronym{pchip}{PCHIP}{piecewise cubic Hermite interpolating polynomial}
\newacronym{pdf}{PDF}{probability density function}
\newacronym{pld}{PLD}{pulsed laser deposition}
\newacronym{ppms}{PPMS}{physical property measurement system}
\newacronym{psi}{PSI}{Paul Scherrer Institute}
\newacronym{pct}{PCT}{point contact tunneling}

\newacronym{qens}{QENS}{quasielastic neutron scattering}
\newacronym{ql}{QL}{quintuple layer}
\newacronym{qo}{QO}{quantum oscillations}
\newacronym{qf}{\ensuremath{Q}}{quality factor}

\newacronym{rbs}{RBS}{Rutherford backscattering}
\newacronym{rf}{RF}{radio frequency}
\newacronym{rheed}{RHEED}{reflection high-energy electron diffraction}
\newacronym{rib}{RIB}{radioactive ion beam}
\newacronym{rkky}{RKKY}{Ruderman–Kittel–Kasuya–Yosida}
\newacronym{rrr}{RRR}{residual-resistivity ratio}
\newacronym{rtil}{RTIL}{room temperature ionic liquid}

\newacronym{sae}{SAE}{spin-alignment echo}
\newacronym{sans}{SANS}{small angle neutron scattering}
\newacronym{si}{SI}{International System of Units}
\newacronym{sis}{SIS}{superconductor-insulator-superconductor}
\newacronym{sims}{SIMS}{secondary ion mass spectrometry}
\newacronym{slr}{SLR}{spin-lattice relaxation}
\newacronym[sort={S/N}]{snr}{\textit{S}/\textit{N}}{signal-to-noise ratio}
\newacronym{squid}{SQUID}{superconducting quantum interference device}
\newacronym{srf}{SRF}{superconducting radio frequency}
\newacronym{srim}{SRIM}{Stopping and Range of Ions in Matter}
\newacronym{ss}{SS}{superconductor-superconductor}
\newacronym{ssid}{SSID}{solid-state ionic device}
\newacronym{ssr}{SSR}{spin-spin relaxation}
\newacronym{stm}{STM}{scanning tunnelling microscopy}
\newacronym{sts}{STS}{scanning tunnelling spectroscopy}
\newacronym{sh}{\ensuremath{H_\mathrm{sh}}}{superheating field}
\newacronym{sem}{SEM}{scanning electron microscope}

\newacronym{ti}{TI}{topological insulator}
\newacronym{tem}{TEM}{transmission electron microscopy}
\newacronym{trim}{TRIM}{Transport and Range of Ions in Matter}
\newacronym{tss}{TSS}{topological surface state}
\newacronym{tmd}{TMD}{transition metal dichalcogenide}
\newacronym{tofsims}{TOF-SIMS}{time-of-flight secondary ion mass spectrometry}
\newacronym{tds}{TDS}{time domain spectroscopy}

\newacronym{uhv}{UHV}{ultra-high vacuum}

\newacronym{vdw}{vdW}{van der Waals}
\newacronym{vft}{VFT}{Vogel-Fulcher-Tammann}

\newacronym{whh}{WHH}{Werthamer-Helfand-Hohenberg}

\newacronym{xrd}{XRD}{x-ray diffraction}
\newacronym{xrr}{XRR}{x-ray reflection}
\newacronym{xps}{XPS}{x-ray photoelectron spectroscopy}

\newacronym{ybco}{YBCO}{\ch{YBa2Cu3O_{6+x}}}
\newacronym{ysz}{YSZ}{yttria-stabilized zirconia}
\newacronym{vsm}{VSM}{vibrating sample magnetometer}

\newacronym[sort={muSR}]{musr}{\ensuremath{\mu}SR}{muon spin rotation}
\newacronym{alc-musr}{ALC-\ensuremath{\mu}SR}{avoided level crossing muon spin rotation}
\newacronym{le-musr}{LE-\ensuremath{\mu}SR}{low-energy muon spin rotation}
\newacronym{lf-musr}{LF-\ensuremath{\mu}SR}{longitudinal field muon spin rotation}
\newacronym{rf-musr}{RF-\ensuremath{\mu}SR}{radio frequency muon spin rotation}
\newacronym{tf-musr}{TF-\ensuremath{\mu}SR}{transverse field muon spin rotation}
\newacronym{zf-musr}{ZF-\ensuremath{\mu}SR}{zero field muon spin rotation}

\begin{document}

\title{
	Superconducting properties of thin film \ch{Nb_{1-x}Ti_xN} studied via the NMR of implanted \ch{^8Li}
}

\author{Md~Asaduzzaman}
\email[E-mail: ]{asadm@uvic.ca}
\affiliation{Department of Physics and Astronomy, University of Victoria, 3800 Finnerty Road, Victoria, BC V8P~5C2, Canada}
\affiliation{TRIUMF, 4004 Wesbrook Mall, Vancouver, BC V6T~2A3, Canada}

\author{Ryan~M.~L.~McFadden}
\affiliation{Department of Physics and Astronomy, University of Victoria, 3800 Finnerty Road, Victoria, BC V8P~5C2, Canada}
\affiliation{TRIUMF, 4004 Wesbrook Mall, Vancouver, BC V6T~2A3, Canada}

\author{Edward~Thoeng}
\affiliation{Department of Physics and Astronomy, University of British Columbia, 6224 Agricultural Road, Vancouver, British Columbia V6T 1Z1, Canada}
\affiliation{TRIUMF, 4004 Wesbrook Mall, Vancouver, BC V6T~2A3, Canada}

\author{Yasmine~Kalboussi}
\affiliation{Institut des lois fondamentales de l’univers, Commissariat de l’énergie atomique-centre de saclay, Paris-Saclay University, 91191 Gif-sur-Yvette, France}

\author{Ivana~Curci}
\affiliation{Institut des lois fondamentales de l’univers, Commissariat de l’énergie atomique-centre de saclay, Paris-Saclay University, 91191 Gif-sur-Yvette, France}

\author{Thomas~Proslier}
\affiliation{Institut des lois fondamentales de l’univers, Commissariat de l’énergie atomique-centre de saclay, Paris-Saclay University, 91191 Gif-sur-Yvette, France}

\author{Sarah~R.~Dunsiger}
\affiliation{TRIUMF, 4004 Wesbrook Mall, Vancouver, BC V6T~2A3, Canada}
\affiliation{Department of Physics, Simon Fraser University, 8888 University Drive, Burnaby, BC V5A~1S6, Canada}

\author{W.~Andrew~MacFarlane}
\affiliation{TRIUMF, 4004 Wesbrook Mall, Vancouver, BC V6T~2A3, Canada}
\affiliation{Department of Chemistry, University of British Columbia, 2036 Main Mall, Vancouver, BC V6T~1Z1, Canada}
\affiliation{Stewart Blusson Quantum Matter Institute, University of British Columbia, Vancouver, BC V6T~1Z4, Canada}

\author{Gerald~D.~Morris}
\affiliation{TRIUMF, 4004 Wesbrook Mall, Vancouver, BC V6T~2A3, Canada}

\author{Ruohong~Li}
\affiliation{TRIUMF, 4004 Wesbrook Mall, Vancouver, BC V6T~2A3, Canada}

\author{John~O.~Ticknor}
\affiliation{Department of Chemistry, University of British Columbia, 2036 Main Mall, Vancouver, BC V6T 1Z1, Canada}
\affiliation{Stewart Blusson Quantum Matter Institute, University of British Columbia, Vancouver, BC V6T~1Z4, Canada}

\author{Robert~E.~Laxdal}
\affiliation{Department of Physics and Astronomy, University of Victoria, 3800 Finnerty Road, Victoria, BC V8P~5C2, Canada}
\affiliation{TRIUMF, 4004 Wesbrook Mall, Vancouver, BC V6T~2A3, Canada}
	
\author{Tobias~Junginger}
\email[E-mail: ]{junginger@uvic.ca}
\affiliation{Department of Physics and Astronomy, University of Victoria, 3800 Finnerty Road, Victoria, BC V8P~5C2, Canada}
\affiliation{TRIUMF, 4004 Wesbrook Mall, Vancouver, BC V6T~2A3, Canada}

\date{\today}

\begin{abstract}
	We report measurements of the normal-state and superconducting
	properties of thin-film \ch{Nb_{1-x}Ti_{x}N} using \ch{^{8}Li} \gls{bnmr}.
	In these experiments, radioactive \ch{^{8}Li^{+}} probes were implanted \qty{\sim 21}{\nm}
	below the surface of a \ch{Nb_{0.75}Ti_{0.25}N}(\qty{91}{\nm}) film in \ch{Nb_{0.75}Ti_{0.25}N}(\qty{91}{\nm})/\ch{AlN}(\qty{4}{\nm})/\ch{Nb}
	and
	its \acrshort{nmr} response recorded (via \ch{^{8}Li}'s $\beta$-emissions)
	between \qty{4.6}{\kelvin} and \qty{270}{\kelvin} in
	a \qty{4.1}{\tesla} field applied normal to its surface.
	Resonance measurements reveal wide, symmetric lineshapes 
	at all temperatures,
	with significant additional broadening below the film's superconducting transition temperature $T_\mathrm{c}(\qty{0}{\tesla}) = \qty{15.4 \pm 0.7}{\kelvin}$
	due to vortex lattice formation.
	Fits to a broadening model find a magnetic penetration depth $\lambda(\qty{0}{\kelvin}) = \qty{180.57 \pm 0.30}{\nm}$
	and
	upper critical field $B_\mathrm{c2}(\qty{0}{\kelvin}) = \qty{18 \pm 4}{\tesla}$,
	consistent with literature estimates.
	\Gls{slr} measurements find a Korringa response at low temperatures,
	with dynamic (i.e., thermally activated) contributions dominating above \qty{\sim 100}{\kelvin}.
	Below $T_\mathrm{c}$, we observe a small Hebel-Slichter coherence peak
	characterized by a superconducting energy gap $\Delta(\qty{0}{\kelvin}) = \qty{2.60 \pm 0.12}{\milli\electronvolt}$
	and
	modest Dynes-like broadening.
	Our measurements suggest a
	gap ratio $2\Delta(\qty{0}{\kelvin})/k_\mathrm{B}T_\mathrm{c}(\qty{0}{\tesla}) = \num{3.92 \pm 0.25}$,
	consistent with strong-coupling behavior.
	Sources for the dynamic high-$T$ relaxation are suggested. 
\end{abstract}

\maketitle
\glsresetall

\section{
  Introduction
  \label{sec:introduction}
 }

\begin{figure}
	\centering
	\includegraphics*[width=\columnwidth]{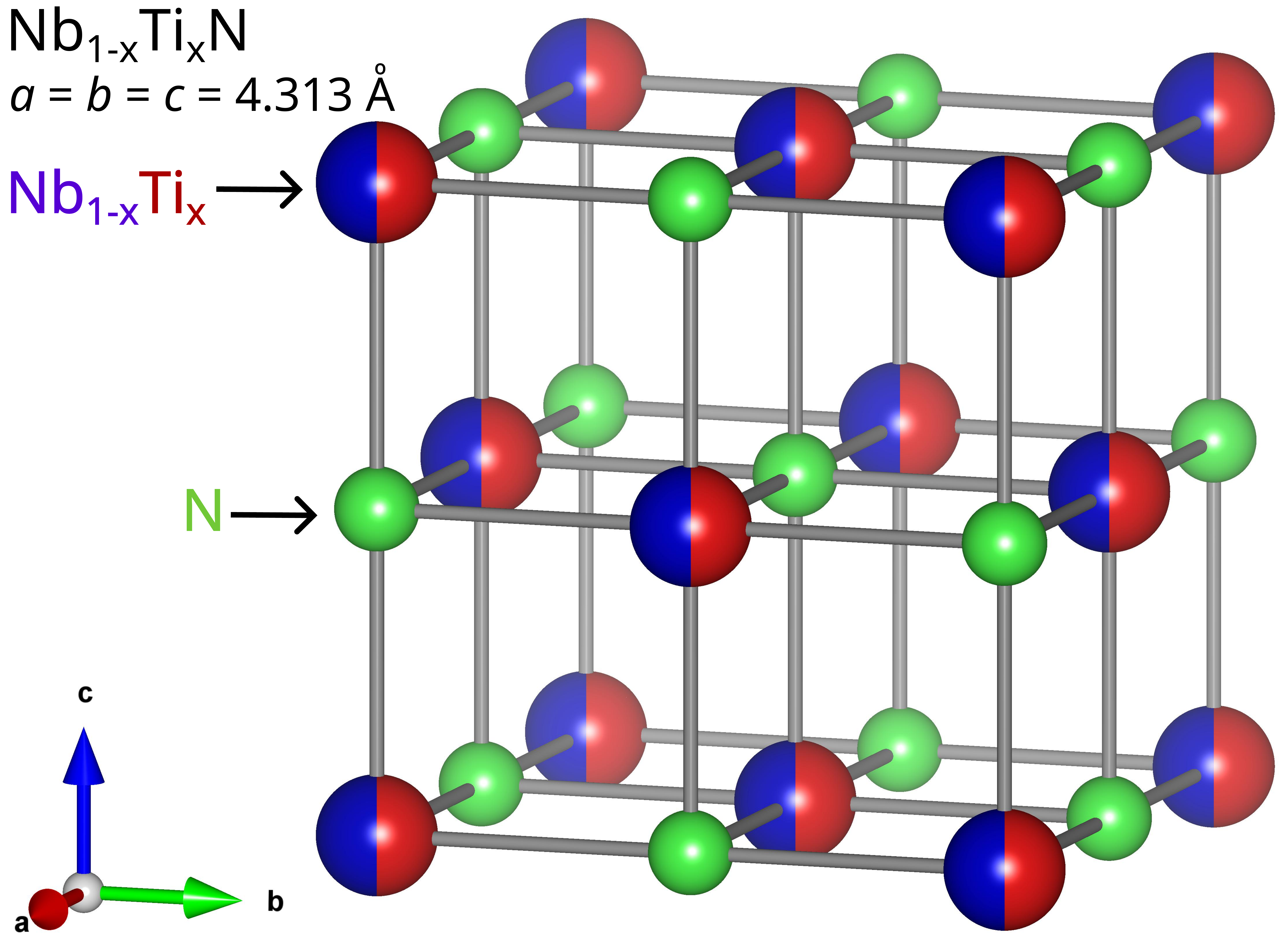}
	\caption{
	\label{fig:NbTiN-crystal-structure}
	Crystal structure of cubic $B1$ (rocksalt) \ch{Nb_{1-x}Ti_xN}
	(space group $Fm\bar{3}m$, number 225).
	The metal atoms \ch{Nb} and \ch{Ti} atoms (blue and red spheres) randomly occupy
	the Wyckoff $4a$ site,
	which forms an FCC sublattice.
	Similarly,
	the \ch{N} atoms (green spheres) occupy the Wyckoff $4b$ site,
	can be viewed as filling the octahedral ``intersticies'' of
	the FCC metal sublattice.
	The gray bonds highlight the octahedral 
	coordination environment of each equilibrium position.
	Lattice constants $a, b, c$ for the $\mathrm{x} = 0.25$ stoichiometry are indicated in the inset.
	The structures were drawn using VESTA~\cite{2011-Momma-JAC-44-1272}.
	}
\end{figure}

\ch{Nb_{1-x}Ti_xN} is a ternary alloy with a cubic $B1$ (rocksalt) crystal structure
(see \Cref{fig:NbTiN-crystal-structure}),
derived from Group IV and Group V transition metal nitrides~\cite{1969-Gavaler-JAP-15-329,1997-Benvenuti-NIMPRSB-124-106}.
It forms a fully miscible,
quasi-binary solid solution with its end members \ch{TiN} and \ch{NbN} over the entire $0 \leq \mathrm{x} \leq 1$ composition
range~\cite{2015-Zhang-APL-107-122603,2023-Pratiksha-SST-36-085017},
with structural details that closely follow Vegard's law~\cite{2009-Matenoglou-SCT-204-911,2012-Vasu-JMS-47-3522}.
Similar to its end members, the alloy is a type-II superconductor with a relatively high critical temperature $T_\mathrm{c}$ (up to~\qty{17}{\kelvin}~\cite{2012-Toth-nitrie-book,1996-Kikkawa-CTMCN-9-175}).
Thanks to the alloy's facile synthesis in the form of thin films~\cite{1969-Zbasnik-JAP-40-2147,1969-Gavaler-JAP-15-329,2011-Makise-IEEETAS-21-139,2018-Hazra-PRB-97-144518}, 
\ch{Nb_{1-x}Ti_xN} finds use in numerous technical applications
that employ superconducting coatings
(e.g., tunnel junctions~\cite{2019-Cyberey-IEEETAS-29-1,2024-Cyberey-IEETAS-34-1},
\unit{\tera\hertz} receivers~\cite{2024-Zhukova-IEETAS-34-1,2015-Uzawa-IEEETAS-25-1}
and mixers~\cite{2013-Westig-JAP-114-124504,2023-Fedor-IEETTST-13-627},
etc.).
In particular,
the alloy has emerged as a promising candidate~\cite{2016-Anne-Marie-SST-29-113002}
for coating conventional \ch{Nb} \gls{srf} cavities
--- common components of modern particle accelerators~\cite{2023-Padamsee-SRTA} ---
which we consider in detail below.

\Gls{srf} cavities accelerate charged particle beams via the electric fields created under
resonant \gls{rf} excitation.
Essential for maintaining a high quality factor $Q_{0}$ is low electrical resistivity of the cavity material,
making superconductors like \ch{Nb} ideal~\cite{2023-Padamsee-SRTA}.
The performance of 
\ch{Nb} cavities is ultimately limited by the
element's 
so-called superheating field
$B_{\mathrm{sh}} \approx \qty{240}{\milli\tesla}$~\cite{2015-Posen-PRL-115-047001,2017-Junginger-SST-30-125012},
beyond which vortex penetration occurs
and resistive losses cause $Q_{0}$ to plummet.
To address this limitation,
coatings with superconductors having higher $T_\mathrm{c}$
and
superheating field $B_\mathrm{sh}$ 
than \ch{Nb}
(e.g., \ch{Nb3Sn}, \ch{Nb_{1-x}Ti_{x}N})
have been proposed~\cite{2006-Gurevich-APL-88-012511,2014-Kubo-APL-104-032603,2017-Kubo-SST-30-023001}.
Such coatings
(with or without an insulating ``buffer'' layer)
are predicted~\cite{2006-Gurevich-APL-88-012511,2014-Kubo-APL-104-032603,2017-Kubo-SST-30-023001,2019-Kubo-JJAP-58-088001,2021-Kubo-SST-34-045006}
to allow such devices to operate in field regimes beyond the capabilities of ``bare'' \ch{Nb} cavities,
a claim supported by recent
experiments~\cite{2017-Junginger-SST-30-125012,2016-Tan-SR-6-35879,2019-Antoine-SST-32-085005,2024-Asaduzzaman-SST-37-025002,2024-Asaduzzaman-SST-37-085006,2023-Roach-IEETAS-23-8600203}.
Central to modelling the \emph{macroscopic} behavior of these heterostructures,
however,
is knowledge of the coating film's \emph{microscopic} superconducting properties,
which are not well-established for \ch{Nb_{1-x}Ti_{x}N}.

Part of the uncertainty in \ch{Nb_{1-x}Ti_{x}N}'s intrinsic properties stems from
their compositional ``tunability'', 
with a transition temperature $T_\mathrm{c}$ that can be tuned with stoichiometry. 
For instance,
a $T_\mathrm{c} \approx \qty{17}{\kelvin}$ has been reported for $\mathrm{x} = 0.34$~\cite{1968-Cody-PR-173-481,1968-Bell-JAP-39-2797},
while a slightly lower value of \qty{\sim 15}{\kelvin} is observed for $\mathrm{x} \lesssim 0.5$~\cite{1990-DiLeo-JLTP-78-41,1969-Gavaler-JAP-15-329,1997-Benvenuti-NIMPRSB-124-106,2016-Burton-JVSTA-34-021518}.
As $\mathrm{x} \to 1$, $T_\mathrm{c}$ decreases further to approximately \qty{4}{\kelvin}~\cite{1990-DiLeo-JLTP-78-41,1967-Yen-JAP-38-2268}.
This trend may be attributed to the properties of the end-member compositions, where \ch{NbN} exhibits $T_\mathrm{c} \gtrsim \qty{16}{\kelvin}$~\cite{1983-Pan-Cryo-23-258,1981-Isagawa-JAP-52-921}, while \ch{TiN} has a significantly lower $T_\mathrm{c} \approx \qty{4}{\kelvin}$~\cite{2018-Torgovkin-SST-31-055017,2013-Michael-TSF-548-485}.
Less well characterized is how stoichiometry affects its other superconducting properties.
For example,
measurements indicate a magnetic penetration depth $\lambda \gtrsim \qty{150}{\nm}$~\cite{2023-Fedor-IEETTST-13-627,2024-Asaduzzaman-SST-37-025002,1990-DiLeo-JLTP-78-41,2005-LeiYu-IEETAS-15-44,2022-Khan-IEEETAS-32-1,2018-Junginger-IPAC-3921,2013-Hong-JAP-114-243905} for $\mathrm{x} \leq 0.46$,
a \gls{gl} coherence length $\xi_\mathrm{GL} \approx \qty{4}{\nm}$~\cite{2021-Sidorova-PRB-104-184514,2002-Lei-IEEETAS-12-1795,2018-Hazra-PRB-97-144518,2023-Pratiksha-SST-36-085017},
a lower critical field $B_\mathrm{c1} \approx \qty{30}{\milli\tesla}$~\cite{2016-Burton-JVSTA-34-021518,2023-Gonzalez-JAP-134-035301,2015-Valente-Feliciano-SRF-TUBA08},
and
an upper critical field $B_\mathrm{c2} \gtrsim \qty{15}{\tesla}$~\cite{1967-Yen-JAP-38-2268,1969-Gavaler-JAP-15-329,2021-Sidorova-PRB-104-184514,1969-Hechler-JLTP-1-29,1969-Zbasnik-JAP-40-2147,2023-Rezinovsky-PC-607-1354241,2023-Pratiksha-SST-36-085017}.
Similarly,
the alloy's superconducting energy gap $\Delta$ is also known to vary with $\mathrm{x}$,
with gap ratios $2\Delta(\qty{0}{\kelvin})/k_\mathrm{B}T_\mathrm{c}(\qty{0}{\tesla})$ 
ranging from \numrange{3.53}{5.0}~\cite{2024-Cyberey-IEETAS-34-1,2019-Cyberey-IEEETAS-29-1,2023-Fedor-IEETTST-13-627,2022-Khan-IEEETAS-32-1,2013-Hong-JAP-114-243905,2013-Westig-JAP-114-124504,2024-Zhukova-IEETAS-34-1,2014-Groll-APL-104-092602,2015-Uzawa-IEEETAS-25-1,2021-Lap-APL-119-152601,2010-Barends-APL-97-033507}.
This complexity can be further compounded by their variability
with other factors such as
film thickness,
substrate material,
and post-deposition annealing~\cite{2012-Driessen-PRL-109-107003,2011-Makise-IEEETAS-21-139,2018-Hazra-PRB-97-144518,2015-Zhang-APL-107-122603,2023-Kalboussi-thesis,2011-Proslier-ET-41-237,2021-Kalboussi-SRF-2021}.

As \ch{Nb_{1-x}Ti_{x}N} films are most important for technical applications
---
particularly in form of \gls{sis} structures
(e.g., \ch{Nb_{1-x}Ti_xN}/\ch{AlN}/\ch{Nb}), which are widely used in tunnel junctions, THz mixers, and \gls{srf} cavities
---
we focus our attention on this material class.
One means of elucidating its properties is through the
study of the superconductor's internal magnetic field distribution $p(B)$
in the vortex state~\cite{1957-Abrikosov-SPJ-5-1174,1995-Brandt-RPP-58-1465}.
Techniques such as
\gls{nmr}~\cite{1976-MacLaughlin-SSP-31-1,2008-Walstedt-NMRPHTM-2-13}
and
\gls{musr}~\cite{2000-Sonier-RMP-72-769,2024-Amato-musr-book-new}
are effective approaches for bulk superconductors,
but are less suited for thin films or layered heterostructures.
On the other hand,
closely related techniques based on low-energy implanted spin-probes like
\gls{le-musr}~\cite{2004-Morenzoni-JPCM-16-S4583,2021-Blundell-Book-OUP}
and
\gls{bnmr}~\cite{2015-MacFarlane-SSNMR-68-1,2022-MacFarlane-ZPC-236-757}
are well-suited to such a situation,
enabling spatially resolved measurements 
at subsurface depths $\lesssim \qty{150}{\nm}$.

In this work,
we investigate the superconducting
and normal-state properties of thin film \ch{Nb_{1-x}Ti_{x}N}
using \ch{^{8}Li} \gls{bnmr} spectroscopy~\cite{2015-MacFarlane-SSNMR-68-1,2022-MacFarlane-ZPC-236-757}.
Specifically, we report measurements
in the normal and vortex states of a
\ch{Nb_{0.75}Ti_{0.25}N}(\qty{91}{\nm})/\ch{AlN}(\qty{4}{\nm})/\ch{Nb} heterostructure
under a \qty{4.1}{\tesla} field applied normal to its surface.
Resonance measurements find wide \ch{^{8}Li} lineshapes that (symmetrically) broaden
below the film's $T_\mathrm{c}$ due to the formation of vortex field lines.
\Gls{slr} data reveal distinct $T$-dependent behavior,
with a Korringa response~\cite{1950-Korringa-Physica-16-601} below \qty{\sim 100}{\kelvin} that
displays a Hebel-Slichter coherence peak~\cite{1957-Hebel-PR-107-901,1959-Hebel-PR-113-1504} below $T_\mathrm{c}$,
and
relaxation that is dominated by thermally activated fluctuations at higher-$T$.
From an analysis of these features,
we quantify the parameters governing the film's superconducting properties
and
compare them against literature values.
Our findings provide key insight into the superconducting behavior of \ch{Nb_{1-x}Ti_{x}N} thin films
in an \gls{sis} arrangement.

\section{
	Experiment
	\label{sec:experiment}
}

\begin{figure}
	\centering
	\includegraphics*[width=\columnwidth]{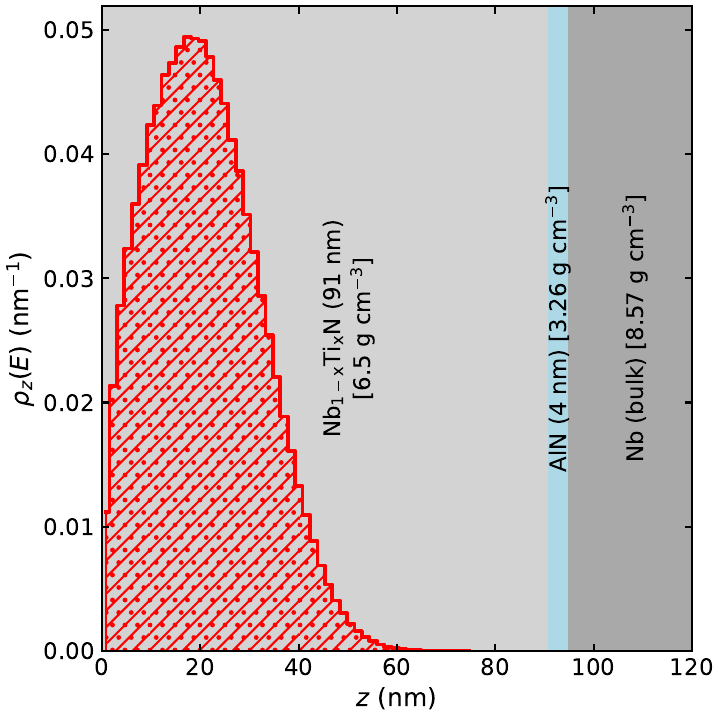}
	\caption{
	\label{fig:stopping-profile} 
	Simulated stopping profile for \num{e6} \ch{^8Li+} implanted in
	\ch{Nb_{0.75}Ti_{0.25}N}(\qty{91}{\nm})/\ch{AlN}(\qty{4}{\nm})/\ch{Nb} at energy $E = \qty{4.85}{\kilo\electronvolt}$,
	obtained using the \gls{srim} Monte Carlo code~\cite{2008-Ziegler-book-srim}.
	The target's layer thicknesses and material densities are indicated in the inset.
	The stopping profile $\rho_z(E)$
	(i.e., the distribution of implantation depths $z$)
	is represented as a histogram,
	whose mean implantation depth $\langle z \rangle \approx \qty{21}{\nano\meter}$
	and
	straggle
	(i.e., standard deviation)
	$\sigma_z \approx \qty{11}{\nano\meter}$. 
	}
\end{figure}

\Gls{bnmr}~\cite{2015-MacFarlane-SSNMR-68-1,2022-MacFarlane-ZPC-236-757}
experiments were conducted at TRIUMF's \gls{isac} facility in Vancouver, BC, Canada.
The local spin probe \ch{^8Li}
(nuclear spin $I = 2$, radioactive lifetime $\tau = \qty{1.21}{\second}$ (half-life ~\qty{848}{\milli\second}), gyromagnetic ratio $\gamma_{\ch{^8Li}}/(2\pi) = \qty{6.30198 \pm 0.00008}{\mega\hertz\per\tesla}$, nuclear electric quadrupole moment $Q = \qty[retain-explicit-plus=true]{+32.6}{\milli\barn}$, and mass $m_{\ch{^8Li}} = \qty{8.023}{\atomicmassunit}$)
was introduced into the sample by ion-implantation using a beam of \ch{^{8}Li^{+}}.
The incident ion beam had a typical flux of \qty{\sim e6}{\ions\per\second} over a beam spot \qty{\sim 3}{\milli\meter} in diameter,
with a beam implantation energy $E = \qty{4.85}{\kilo\electronvolt}$,
corresponding to a mean stopping depth $\langle z \rangle \sim \qty{21}{\nm}$,
as calculated by the \gls{srim}~\cite{2008-Ziegler-book-srim} Monte Carlo code
(see \Cref{fig:stopping-profile}).
Prior to implantation,
the probe was spin-polarized in-flight by collinear optical pumping with circularly polarized laser light~\cite{2013-Levy-HI-225-165},
achieving a polarization $p_{z} \approx \qty{70}{\percent}$~\cite{2003-Levy-NIMPRS-204-689}.
All \gls{bnmr} measurements were performed in an applied field
$B_{0} = \qty{4.1}{\tesla}$ perpendicular to the sample's surface.

During the measurements,
\ch{^{8}Li}'s $p_{z}$, defined by:
\begin{equation}
	p_z = \frac{1}{I}\mathrm{Tr}(\rho I_z)
\end{equation}
where $I_z$ is the operator for the $z$-component of the nuclear spin and $\rho$ is the corresponding density matrix, 
was monitored through its $\beta$-decay anisotropy.
In this process,
an electron is preferentially emitted opposite to the direction of the nuclear polarization at the time of decay. 
This was done
using a pair of fast plastic scintillation counters positioned \qty{180}{\degree} apart with respect to each other along the beam axis~\cite{2015-MacFarlane-SSNMR-68-1,2014-Morris-HI-225-173}. 
During the data acquisition,
the handedness of the laser polarization used during optical pumping was periodically alternated to control the \ch{^8Li^+} beam's helicity
(i.e., positive and negative)~\cite{2015-MacFarlane-SSNMR-68-1},
with data recorded separately for each polarization sense $\pm$.
The four-counter method was used to form the $\beta$-decay asymmetry $\mathscr{A}$,
which is proportional to $p_{z}$~\footnote{Forming the asymmetry in this manner has the advantage of implicitly removing select detection systematics (e.g., different detector efficiencies).}:
\begin{equation}
	\mathscr{A} \equiv \dfrac{r-1}{r+1} = A_{0} p_{z},
	\label{eq:asym-bnmr}
\end{equation}
where
\begin{equation*}
	r \equiv \sqrt{\dfrac{ \left ( N_F^{+}  / N_B^{+} \right ) }{ \left ( N_F^{-} / N_B^{-} \right ) }},
\end{equation*}
$N_B^{\pm}$ and $N_F^{\pm}$
are the beta decay electron events recorded 
in the forward ($F$) and backward ($B$) detectors for the $\pm$ polarization senses,
and
$A_{0} \approx 0.1$ is a proportionality factor that depends on the experimental setup
(e.g., detection geometry, probe $\beta$-decay properties, etc.).

In this work,
two types of experiments were conducted
(i) resonance measurements, where the steady-state
(i.e., time-integral) spin-polarization was monitored as a function 
of the frequency of the small \gls{rf} magnetic field $B_{1}$,
which is used to map the (static)
local field distribution;
and (ii) \gls{slr} measurements, where the temporal decay of $p_{z}$ caused by stochastic fluctuations in the probe's local field
is monitored both during and following implantation.
In the former,
a continuous
\ch{^8Li+} beam was used with the frequency of the small transverse \gls{rf} field stepped slowly near \ch{^8Li}'s Larmor frequency:
\begin{equation}
	\omega_{\mathrm{0}} = \gamma_{\ch{^8Li}} B_0.
	\label{eq:larmor-frequency}
\end{equation}

On resonance, the \ch{^8Li} spin precesses rapidly due to the \gls{rf} field, resulting in a loss of the time-averaged asymmetry.
Multi-frequency techniques were also used to search for small ``quadrupolar'' features
(see, e.g., Refs.~\cite{1993-Minamisono-HI-80-1315,2022-Adelman-PRB-106-035205}). 
For the~\gls{slr} measurement,
a pulsed \ch{^8Li+} beam was used with a typical duration $\Delta \sim \qty{4}{\second}$.
As the initial state of the probe nuclei are \emph{very} far from thermal equilibrium,
no \gls{rf} field is required to measure \gls{slr}, unlike conventional \gls{nmr}.
During the pulse,
the polarization approaches a dynamic equilibrium value,
while afterward it relaxes to \num{\sim 0}.
Notably,
data acquired in this manner has a characteristic bipartite form
with (statistical) error bars that are 
governed by Poisson statistics.
This uncertainty is minimized near the pulse's trailing edge,
but increases exponentially with the \ch{^{8}Li} lifetime afterward
(see, e.g.,~\cite{2022-MacFarlane-ZPC-236-757}).
In the present study,
the typical duration of either measurement was \qtyrange{\sim 15}{\sim 30}{\minute}.

\subsection{
	Sample Preparation
	\label{sec:experiment:samples}
}
First, flat \ch{Nb} substrates were prepared by cutting fine-grain \ch{Nb} stock sheets (Wah Chang Corporation) with a \gls{rrr} $> 150$ and machining them into flat plates approximately 12 mm by 8 mm by 0.5 mm. Following machining, the samples underwent \gls{bcp} (see, e.g.,~\cite{2011-Ciovati-JAE-41-721}) to remove the topmost \qty{\sim 100}{\micro\meter} of material from the surface. Subsequently, the samples were annealed at \qty{1400}{\celsius} for \qty{5}{\hour} to relieve any remaining mechanical stresses in the metal. After annealing, an additional round of \gls{bcp} was performed to remove the topmost \qty{\sim 10}{\micro\meter} of material from the surface, effectively eliminating any contaminants introduced during the annealing process.

The \ch{Nb_{1-x}Ti_{x}N}/\ch{AlN} bilayer was deposited on a \ch{Nb} substrate using thermal ~\gls{ald} in a custom-built reactor at CEA Saclay~\cite{2013-Miikkulainen-JAP-113-021301}. 
The \ch{AlN} layer was grown using a standard process with \ch{AlCl3} and \ch{NH3} precursors~\cite{2018-Rontu-JVSTA-36-021508}, while the \ch{Nb_{1-x}Ti_{x}N} film was deposited at \qty{450}{\celsius} by alternating \ch{NbN} and \ch{TiN} cycles. Its composition can be controlled by adjusting the number of \ch{TiN} and \ch{NbN} cycles~\cite{2011-Proslier-ET-41-237,2023-Kalboussi-thesis,2021-Kalboussi-SRF-2021}.
In this work, each \ch{Nb_{1-x}Ti_{x}N} supercycle consisted of
4 \ch{(TiCl4 + NH3)} cycles followed by 1 \ch{(NbCl5 + NH3)} cycle, as the subsequent \ch{NbCl5} pulse etches surface \ch{Ti} in the form of volatile \ch{TiCl4}. \ch{NH3} was pulsed for ~\qty{0.5}{\second}, \ch{TiCl4} for \qty{2.5}{\second}, and \ch{NbCl5} for ~\qty{1}{\second}, with ~\qty{10}{\second} purges after each step.
This yielded a \ch{Ti}/\ch{Nb} ratio of 0.25 ($\mathrm{x} = 0.25$), confirmed by~\gls{xps}~\cite{2023-Kalboussi-thesis}. 
Excess nitrogen incorporated during deposition was effectively removed by high-vacuum annealing at \qty{900}{\celsius}~\cite{2023-Kalboussi-thesis}.

The film's characteristic superconducting transition temperature $T_\mathrm{c}$ was determined to be \qty{\sim 15}{\kelvin} using a~\gls{vsm}. \Gls{pct}~\cite{2015-Groll-RSI-86-095111,2025-Kalboussi-PRA-23-044023} measurements on a similarly prepared sample show a spatial variation in the density of states near the Fermi level, which is encapsulated by an energy gap $\Delta = \qty{2.49 \pm 0.29}{\milli\electronvolt}$ and a Dynes-like~\cite{1978-Dynes-PRL-41-1509} broadening parameter $\Gamma_\mathrm{D} = \qty{0.10 \pm 0.06}{\milli\electronvolt}$.
Full characterization details, along with complementary measurements on similarly prepared samples, can be found in the Supplemental Material~\cite{supp}.

\section{
  Results and analysis
  \label{sec:results}
 }

\subsection{Resonance Spectra
	\label{sec:results:resonance}
}

\begin{figure}
	\centering
	\includegraphics*[width=\columnwidth]{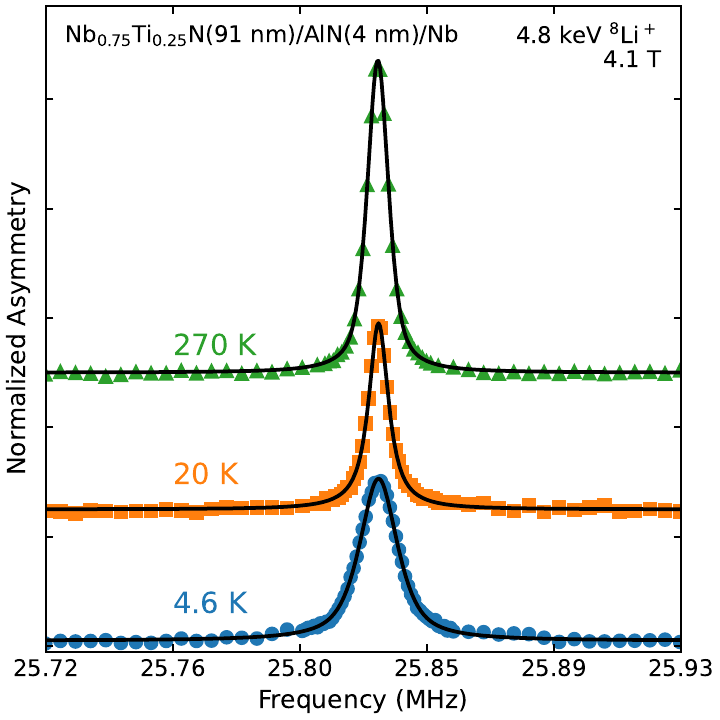}
	\caption{
	\label{fig:resonance-spectra}
	Typical \ch{^8Li} \gls{nmr} lineshapes in \ch{Nb_{0.75}Ti_{0.25}N}(\qty{91}{\nm})/\ch{AlN}(\qty{4}{\nm})/\ch{Nb},
	measured in an applied field $B_0 = \qty{4.1}{\tesla}$ perpendicular to the sample's surface,
	at select temperatures $T$
	(indicated in the figure)
	above and below the film's critical temperature $T_\mathrm{c} \approx \qty{15}{\kelvin}$.
	For $T > T_\mathrm{c}$,
	the resonance linewidth remains roughly $T$-independent,
	but broadens by up to a factor of \num{\sim 2} at lower temperatures.
	The solid black lines represent fits to the data using \Cref{eq:psedo-voigt}.
	}
\end{figure}

\begin{figure}
	\centering
	\includegraphics*[width=\columnwidth]{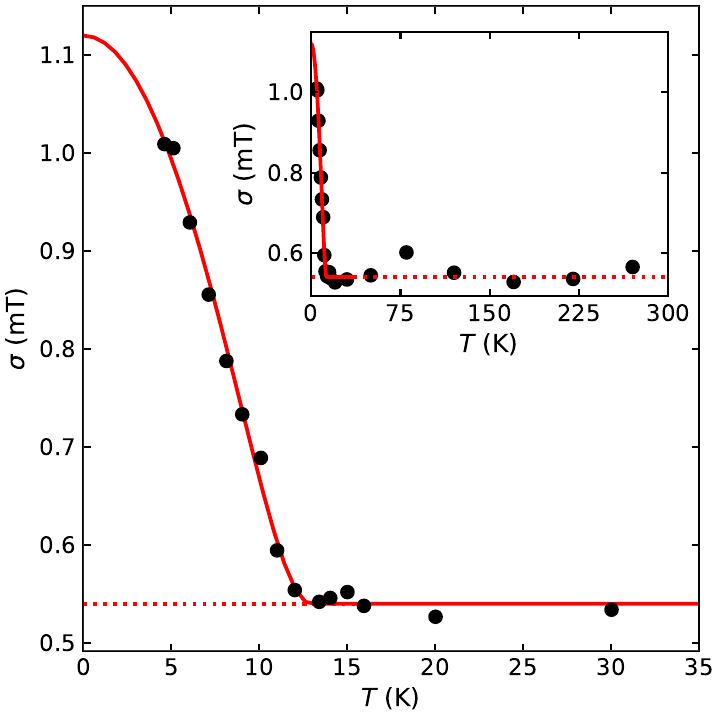}
	\caption{
	\label{fig:resonance-linewidth}
	Temperature $T$ dependence of the Gaussian component $\sigma$ of the resonance linewidth 
	\ch{Nb_{0.75}Ti_{0.25}N}(\qty{91}{\nm})/\ch{AlN}(\qty{4}{\nm})/\ch{Nb}
	in an applied field of $B_0 = \qty{4.1}{\tesla}$ perpendicular to its surface.
	Note that the parameter $\sigma$ is expressed as a magnetic field in units of \unit{\milli\tesla}.
	The solid red line represents a fit to the data for $T < \qty{35}{\kelvin}$
	using the model given by \Cref{eq:total-width,eq:brandt-eq-11,eq:pearl-lambda,eq:Tc-dependence-H0,eq:lambda-T-dependence},
	capturing the line broadening below the superconducting transition temperature $T_\mathrm{c}$,
	with the $T$-independent (normal state) contribution highlighted as a dotted line.
	Measured values at temperatures up to \qty{270}{\kelvin}
	are shown in the inset,
	where (apart from some small scatter)
	$\sigma$'s $T$-independence is evident.
	}
\end{figure}

Typical resonance spectra in \ch{Nb_{0.75}Ti_{0.25}N} are presented in \Cref{fig:resonance-spectra}.
At all temperatures,
the lineshape consists of a single, broad resonance whose amplitude decreases with decreasing temperature.
The simplicity of this spectral shape suggests
that all implanted \ch{^{8}Li^{+}} stop in high-symmetry lattice positions
where the local \gls{efg} is vanishing~\footnote{To confirm the absence of any (static) non-zero \glspl{efg}, we also performed frequency comb measurements (see, e.g., Refs.~\cite{1993-Minamisono-HI-80-1315,2022-Adelman-PRB-106-035205}), which greatly amplifies the sensitivity to such features. No evidence for any finite \glspl{efg} was found at all measured temperatures.}.
These sites are likely the tetrahedral interstitial positions (e.g., (1/4, 1/4, 1/4) in \Cref{fig:NbTiN-crystal-structure}), which are vacant in the ideal rocksalt structure. 
Such a site assignment aligns with the cubic symmetry of \ch{Nb_{1-x}Ti_{x}N} (see \Cref{fig:NbTiN-crystal-structure}), as well as with observations in isostructural compounds like \ch{MgO}~\cite{2014-MacFarlane-JPCS-551-012033}.
In the film's normal state,
the resonance's width is approximately temperature-independent,
with a \gls{fwhm} of ~\qty{8.126 \pm 0.018}{\kilo\hertz}.
This large width is likely due to the high natural abundance of spin-active isotopes
that make up the film's elemental composition
(see \Cref{tab:host-spins}), and is consistent with values predicted from dipolar broadening by host nuclear spins for the likely interstitial or substitutional sites in an ideal, unperturbed host lattice. 
Below $T_\mathrm{c}$,
the resonance broadens substantially,
with the linewidth gradually increasing by a factor \num{\sim 2}.
Such a broadening is typical of a superconductor
upon formation of the vortex state's \gls{fll}~\cite{2003-Brandt-PRB-68-054506}.

\begin{table}
    \centering
	\caption{
		\label{tab:host-spins}
		Stable spin-active nuclei present in \ch{Nb_{1-x}Ti_{x}N}.
		Here,
		$n_{A}$ is the natural isotopic abundance,
		$I$ is the nuclear spin,
		$\gamma$ is the gyromagnetic ratio,
		and
		$Q$ is the electric quadrupole moment.
		For comparison,
		properties of our \gls{bnmr} probe \ch{^{8}Li} are also listed.
		Inapplicable properties are marked by an asterisk (*).    
	}
	\begin{ruledtabular}
    \begin{tabular}{l S l S S[retain-explicit-plus=true]}
        {Isotope} & {$n_{A}$ (\unit{\percent})} & {$I$} & {$\gamma / (2 \pi)$ (\unit{\mega\hertz\per\tesla})} & {$Q$ (\unit{mb})} \\
		\hline
        \ch{^{14}N}  &  99.58 & 1   &   3.07771 &   +20.44 \\
		\ch{^{15}N}  &   0.42 & 1/2 &  -4.31727 &    {*} \\
        \ch{^{47}Ti} &   7.44 & 5/2 &  -2.40410 &  +302 \\
        \ch{^{49}Ti} &   5.41 & 7/2 &  -2.40475 &  +247 \\
        \ch{^{93}Nb} & 100    & 9/2 &  10.43956 &  -320 \\
		\hline
		\ch{^{8}Li}  &  {*} & 2   &   6.30198 &   +32.6 \\
    \end{tabular}
	\end{ruledtabular}
\end{table}

To quantify these features,
the resonances were fitted using a phenomenological pseudo-Voigt ($pV$) function:
\begin{equation} 
	\label{eq:psedo-voigt} 
	pV(\nu) = \mathcal{A} \left[\alpha G(\nu) + (1-\alpha) L(\nu) \right], 
\end{equation}
which represents a linear combination of Gaussian ($G$) and Lorentzian ($L$) components:
\begin{align*}
	G(\nu) &\equiv \exp\left(-\dfrac{(\nu-\nu_0)^2}{2 \sigma^2} \right) , \\
	L(\nu) &\equiv \frac{\Gamma^2}{(\nu-\nu_0)^2+\Gamma^2} .
\end{align*}
and it mimics features of the true Voigt function which is a convolution of the two.
Here,
$\mathcal{A}$ is the resonance amplitude,
$\alpha \in [0,1]$ is a mixing term that defining the lineshape's Gaussian fraction, 
$\nu_{0}$ is the resonance frequency,
$\sigma = \gamma / ( 2\sqrt{2 \ln 2} )$ is the Gaussian width parameter,
$\Gamma = \gamma/2$ defines the Lorentzian width,
and
$\gamma$ denotes the line's (common) \gls{fwhm}.
Consistent with the qualitative description above,
the $T$-dependence of the line's \gls{fwhm} remains its most salient feature.
Assuming the Lorentzian contribution to the line is systematic
(i.e., ``power broadening'' from the single-tone \gls{cw} technique described in \Cref{sec:experiment}),
we encapsulate this broadening by $\sigma$'s $T$-dependence,
which is shown in~\Cref{fig:resonance-linewidth}.
Although some scatter is evident,
$\sigma$ remains approximately constant above $T_\mathrm{c}$,
but increases monotonically below the superconducting transition.

To understand this broadening,
we consider the following model.
In the normal conducting (nc) state,
the Gaussian width $\sigma$ can be approximated by a $T$-independent constant,
whereas in the vortex state the measured value
consists of a convolution of the superconducting (sc) and nc contributions:
\begin{equation}
	\sigma^2 =
	\begin{cases}
		\sigma_\mathrm{sc}^2 + \sigma_\mathrm{nc}^2, & T < T_\mathrm{c},    \\
		\sigma_\mathrm{nc}^2,                        & T \geq T_\mathrm{c} . \\
	\end{cases}
	\label{eq:total-width}
\end{equation}
In an isotropic type-II superconductor with ~\gls{gl} parameter $\kappa \gg 1$,
$\sigma_\mathrm{sc}$
is related to the material's effective magnetic penetration depth $\lambda_{\mathrm{eff}}$ by the
expression~\cite{1988-Brandt-PRB-37-2349,2003-Brandt-PRB-68-054506}:
\begin{equation}
	\sigma_\mathrm{sc}^2 (T) = 0.00371 \dfrac{\Phi_{0}^2}{\lambda_\mathrm{eff}^4 (T)}, 
	\label{eq:brandt-eq-11}
\end{equation}
where $\Phi_{0} = \qty{2.068e-15}{\weber}$ is the magnetic flux quantum.
In thin-film superconductors subjected to a magnetic field normal to their surface,
when the film thickness $d$ is smaller than the material's (bulk) penetration depth $\lambda$,
$\lambda_{\mathrm{eff}}$ is given by the Pearl length~\cite{1964-Pearl-APL-5-65}:
\begin{equation}
	\lambda_\mathrm{eff}(T) = \frac{\lambda^{2}(T)}{d},
	\label{eq:pearl-lambda}
\end{equation}
where we account for $\lambda$'s $T$-dependence
using the analytic approximation~\footnote{The form of \Cref{eq:lambda-T-dependence} closely approximates the $T$-dependence predicted by \gls{bcs} theory (see, e.g.,~\cite{2024-Amato-musr-book-new}).}:
\begin{equation}
	\lambda (T) \approx \dfrac{\lambda (\qty{0}{\kelvin})}{\sqrt{1- \left [ T / T_\mathrm{c}(B_{0}) \right ]^2}},
	\label{eq:lambda-T-dependence}
\end{equation}
where $\lambda(\qty{0}{\kelvin})$ is the penetration depth at \qty{0}{\kelvin}
and
$T_\mathrm{c}(B_{0})$ accounts for the suppression of the transition temperature in an applied
field~\footnote{Note that the exponent in the denominator of \Cref{eq:lambda-T-dependence} is 2, unlike its more common value of 4 found in the two-fluid model. This choice is intentional, as it better describes $\lambda(T)$ in \ch{Nb_{1-x}Ti_xN}~\cite{2013-Hong-JAP-114-243905} and \ch{NbTi}~\cite{2024-Yeonkyu-PS-99-065963}.}.
We treat this suppression empirically by inverting
an analytic approximation
for the simplest solution to 
the \gls{whh}~\cite{1996-Werthamer-PR-147-295} expression for
$B_{\mathrm{c2}}$~\cite{2014-Baumgartner-SST-27-015005}:
\begin{equation}
	h^{\ast}( t ) \approx {0.693} \frac{B_{0}}{B_{\mathrm{c2}}(\qty{0}{\kelvin})},
	\label{eq:Tc-dependence-H0}
\end{equation}
where
$B_\mathrm{c2}(\qty{0}{\kelvin})$ is the upper critical field at $\qty{0}{\kelvin}$,
\begin{equation*}
	h^{\ast} (t) \equiv 1-t-0.153(1-t)^2-0.152(1-t)^4,
\end{equation*}
and
$t \equiv T_\mathrm{c}(B) /T_\mathrm{c} (\qty{0}{\tesla})$.
Further details are given in \Cref{sec:app:Tc-vs-B}.
We note that \gls{whh} theory has been used describe
the $T$-dependence of $B_{\mathrm{c2}}$ in
\ch{Nb_{1-x}Ti_{x}N}~\cite{2023-Rezinovsky-PC-607-1354241,1969-Gavaler-JAP-15-329}
and
related materials
(e.g., \ch{NbTi}~\cite{2024-Yeonkyu-PS-99-065963} and \ch{Nb3Sn}~\cite{2014-Baumgartner-SST-27-015005}),
implying the correctness of this choice here, where fitting our data to this expression yields $B_{\mathrm{c2}}(\qty{0}{\kelvin}) = \qty{18 \pm 4}{\tesla}$.

A fit of \Cref{eq:total-width,eq:brandt-eq-11,eq:pearl-lambda,eq:lambda-T-dependence,eq:Tc-dependence-H0}
to the $\sigma (T)$ data is shown in \Cref{fig:resonance-linewidth}.
In the fit,
both the film thickness $d = \qty{91}{\nm}$
and
the applied field $B_0 = \qty{4.1}{\tesla}$ were fixed to their known values
(see \Cref{sec:experiment}),
with all other parameters left free.
The fit is in good agreement with the data,
with the optimal values for the free parameters summarized in \Cref{tab:results-frequency}.
At this juncture,
we note the good agreement of the extracted $T_\mathrm{c}$ 
with the value identified by magnetometry measurements (see \Cref{sec:experiment:samples}),
as well as the $\lambda(\qty{0}{\kelvin})$
measured in other films~\cite{2024-Asaduzzaman-SST-37-025002}.
We shall consider these results further in \Cref{sec:discussion:metallic}.

\begin{table}
	\centering
	\caption{
		\label{tab:results-frequency}
		Fit parameters describing the temperature dependence of resonance linewidth's
		Gaussian component $\sigma$ 
		(shown in \Cref{fig:resonance-linewidth})
		using \Cref{eq:total-width,eq:brandt-eq-11,eq:pearl-lambda,eq:Tc-dependence-H0,eq:lambda-T-dependence}.
		Here, $d_{\ch{Nb_{0.75}Ti_{0.25}N}}$ is the thickness of the \ch{Nb_{0.75}Ti_{0.25}N} thin film,
		$T_\mathrm{c} (\qty{0}{\tesla})$ is the critical temperature at \qty{0}{\tesla},
		$\lambda (\qty{0}{\kelvin})$ is the penetration depth of
		at \qty{0}{\kelvin},
		$B_{0}$ is the applied magnetic field,
		and
		$B_\mathrm{c2} (\qty{0}{\kelvin})$ represents the upper critical field at \qty{0}{\kelvin}.
		Both $d_{\ch{Nb_{0.75}Ti_{0.25}N}}$ and $B_{0}$ were determined independently
		and fixed during fitting.
	}
	\begin{ruledtabular}
	\begin{tabular}{l S l l}
		Parameter       & {Value}        & Unit & Comment  \\
		\hline
		$d_{\ch{Nb_{0.75}Ti_{0.25}N}}$ & 91   & \unit{\nm} &  fixed (from \Cref{sec:experiment:samples})  \\
		$T_\mathrm{c} (\qty{0}{\tesla})$       & 15.4 \pm 0.7           & \unit{\kelvin} & \\
		$\lambda (\qty{0}{\kelvin})$            & 180.57 \pm 0.30        & \unit{\nm} &    \\
		$\sigma_\mathrm{n}$      & 540.1 \pm 1.9       & \unit{\micro\tesla} & \\
		$B_{0}$                  & 4.1   & \unit{\tesla} & fixed (from \Cref{sec:experiment}) \\
		$B_\mathrm{c2} (\qty{0}{\kelvin})$      & 18 \pm 4           & \unit{\tesla} & \\
	\end{tabular}
	\end{ruledtabular}
\end{table}

Besides the changes to the resonance lineshape induced by the superconducting transition,
we also quantified \ch{^{8}Li}'s \gls{nmr} shift $K^{c}$ in the film.
Using the resonance position in single crystal \ch{MgO}
with $B_{0} \parallel (100)$ at ~\qty{295}{\kelvin} as a reference~\cite{2014-MacFarlane-JPCS-551-012033},
$ K^\mathrm{c}$ (in ppm) is obtained using the expression (see, e.g.,~\cite{2022-McFadden-ACIE-61-e202207137}):
\begin{equation}
	K^\mathrm{c} = 10^6 \left( \dfrac{\nu_0/\zeta_0 - \nu_{\ch{MgO}}/\zeta_{\ch{MgO}}}{\nu_{\ch{MgO}}/\zeta_{\ch{MgO}}} \right),
	\label{eq:corrected-shift}
\end{equation}
where the factor
\begin{equation}
	\zeta_i = 1 + \left( \frac{1}{3} - N_i \right) \chi_i 
	\label{eq:shift-factor}
\end{equation}
corrects for contributions from demagnetization~\footnote{Note that the form of \Cref{eq:corrected-shift} correctly accounts for the general case when both the sample and reference have different magnetic susceptibilities and demagnetization factors.},
with $N_{i}$ denoting the demagnetization factor 
and $\chi_{i}$ denoting the material's volume susceptibility.
While details of the full calculation can be found in \Cref{sec:app:N-chi},
we summarize the main results below.
We find that $K^{c}$
varies between \qtyrange[retain-explicit-plus=true]{+15}{+35}{\ppm} over the measured
$T$-ranges.
This magnitude is typical of \ch{^{8}Li} in many materials~\cite{2015-MacFarlane-SSNMR-68-1,2022-MacFarlane-ZPC-236-757},
but as is common with small \gls{nmr} shifts,
corrections for demagnetization are a dominant contribution
(\qty[retain-explicit-plus=true]{+25.4}{\ppm} here)~\footnote{The ``raw'' (i.e., uncorrected) \ch{^{8}Li} \gls{nmr} shifts were measured to range from \qtyrange[retain-explicit-plus=true]{-10}{+10}{\ppm}.}.
This range of $K^{c}$ is small compared to other metals
(see, e.g.,~\cite{2019-Parolin-PRB-100-209904}),
suggesting weak hybridization with the ternary alloy's conduction band.
We shall consider this quantity further in \Cref{sec:discussion:metallic}.

\subsection{\GLS{slr} Spectra
	\label{sec:results:slr}
}

\begin{figure}
	\centering
	\includegraphics*[width=\columnwidth]{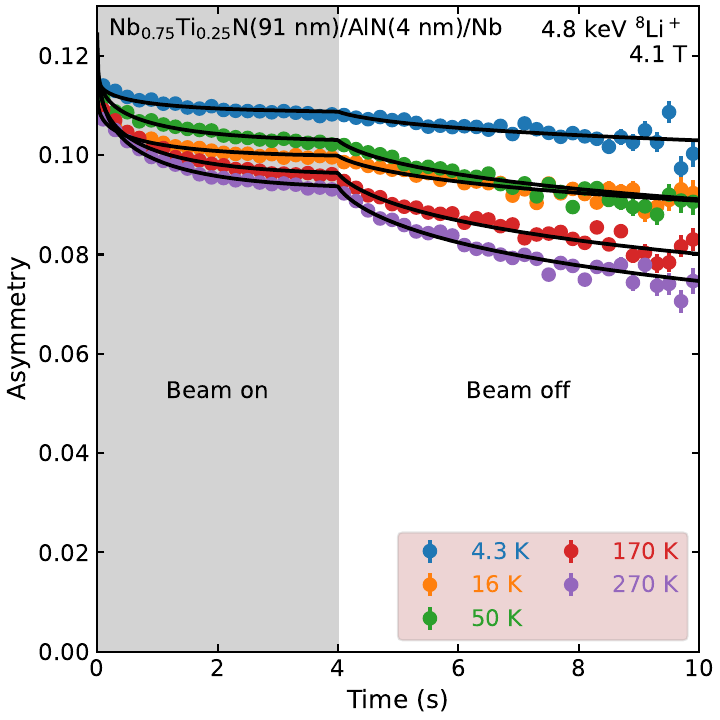}
	\caption{
	\label{fig:slr-asymmetry}
	\ch{^8Li} \gls{slr} data at various temperatures $T$ in \ch{Nb_{0.75}Ti_{0.25}N}(\qty{91}{\nm})/\ch{AlN}(\qty{4}{\nm})/\ch{Nb},
	measured under a perpendicular applied field of $B_0 = \qty{4.1}{\tesla}$.
	The shaded region indicates the duration of the \ch{^8Li+} beam pulse (\qty{4}{\second}).
	The data exhibit $T$-dependent relaxation
	that is non-monotonic with temperature, with a significant non-relaxing or very slow-relaxing component.
	The solid black lines represent fits to a stretched exponential [\Cref{eq:stretched-exp}]
	convoluted with the \ch{^8Li+} beam pulse
	using a common stretching exponent $\beta$
	(described in \Cref{sec:results:slr}).
	The displayed data have been binned by a factor of 20 for clarity.
	}
\end{figure}

Representative time-differential \gls{slr} data at various temperatures
are shown in \Cref{fig:slr-asymmetry}.
At low temperatures the relaxation is very slow,
approaching the limit of what is measurable by \ch{^{8}Li} due to its
radioactive lifetime~\cite{2015-MacFarlane-SSNMR-68-1}.
As the temperature is raised,
so too does the relaxation,
though it remains slow compared to other cubic metals
(e.g., \ch{Ag}~\cite{2004-Morris-PRL-93-157601} and \ch{Nb}~\cite{2009-Parolin-PRB-80-174109}),
but comparable to other metallic compounds where the probe's hybridization
with the conduction band is weak~\cite{2006-Wang-PBC-374-239,2019-McFadden-PRB-99-125201,2020-McFadden-PRB-102-235206}.
Interestingly,
this increase in relaxation is non-monotonic with temperature,
with a local maximum observed near \qty{\sim 140}{\kelvin}.
At all temperatures,
the \gls{slr} is non-exponential,
as might be expected for a host with an abundance of
spin-active nuclei
(see \Cref{tab:host-spins})
and
\gls{3d} disorder
(see, e.g.,~\cite{1984-Stockmann-JNCS-66-501}).

\begin{figure}
	\centering
	\includegraphics*[width=\columnwidth]{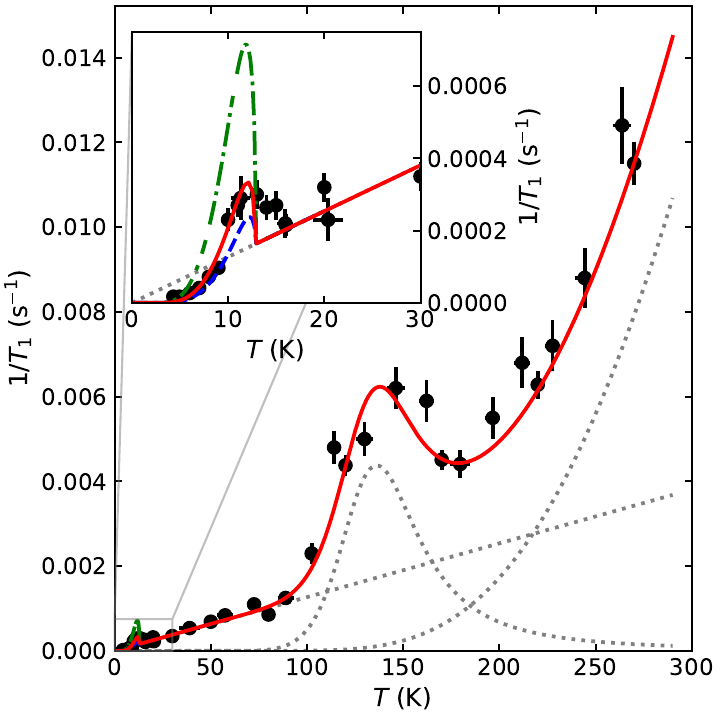}
	\caption{
	\label{fig:slr-rate}
	Temperature $T$-dependence of the \ch{^{8}Li} \gls{slr} rate $1/T_1$ in \ch{Nb_{0.75}Ti_{0.25}N}(\qty{91}{\nm})/\ch{AlN}(\qty{4}{\nm})/\ch{Nb} at $B_0 = \qty{4.1}{\tesla}$.
	$1/T_1$ varies nonmonotonically with temperature,
	with $T$-linear behavior below \qty{\sim 100}{\kelvin}
	that is modified by a Hebel-Slichter coherence peak below the film's $T_\mathrm{c}$
	(see inset for a detailed view).
	Near \qty{\sim 140}{\kelvin},
	a BPP peak is observed,
	while at higher temperatures $1/T_{1}$ increases exponentially with increasing $T$.
	The solid red line shows the fit to \Cref{eq:Tc-dependence-H0,eq:total-slr,eq:hs-slr-rob-kiefl,eq:sc-dos,eq:anamoly-dos,eq:gap-T-dependence,eq:spectral-density-3d,eq:slr-dyn,eq:arrhenius-relation,eq:slr-trapping}
	(described in \Cref{sec:results:slr}),
	with the individual contributions from the linear slope, BPP peak, and exponential rise shown as dotted lines.
	The green dash-dotted and blue dashed curves show the expected Hebel-Slichter coherence peak  for fixed $\Gamma_\mathrm{D} = 0$ and \qty{0.1}{\milli\electronvolt}, respectively, using the same fit parameters.
	}
\end{figure}

To quantify these observations,
we adopt a phenomenological approach used to analyze other
disordered metal-like compounds~\cite{2020-McFadden-PRB-102-235206,2019-McFadden-PRB-99-125201}
and
fit the \gls{slr} data using a stretched exponential model.
Explicitly,
for an \ch{^8Li} ion implanted at time $t^{\prime}$,
the spin polarization at a later time $t > t^{\prime}$ is given by:
\begin{equation}
	R(t , t^{\prime}) = \exp \left( -\left[\frac{(t - t^{\prime})}{T_1}\right]^\beta  \right),
	\label{eq:stretched-exp}
\end{equation}
where $1/T_1$ is the \gls{slr} rate
(i.e., the reciprocal of the signal decays to $1/e$ of its initial value)
and
$0 < \beta \leq 1$ is the stretching exponent.
This model provides a simple yet effective fit to the data,
while minimizing the number of free parameters.
To further avoid overparameterization,
all \gls{slr} data were fit simultaneously
using \Cref{eq:stretched-exp} convoluted with the \qty{4}{\second} beam pulse
and a shared $\beta = \num{0.216 \pm 0.006}$~\footnote{This suggests a significant fraction of \ch{^8Li} relax much more slowly than the fitted $1/T_1$, likely due to weak coupling to the electronic system in low-density or poorly metallic regions, beyond simple site-to-site Korringa variation.} 
While this approach yields an excellent fit
(reduced $\chi^2 \approx 1.02$),
as pointed out by others~\cite{2017-Sugiyama-PRB-96-094402,2016-Cortie-PRL-116-106103},
this choice has the caveat of imparting some temperature dependence to $A_0$.
We assert that this choice does not impede the quantitation of
$1/T_{1}$~\footnote{For example, sharing $A_0$ instead of $\beta$ yields a very similar fit, with a virtually identical $T$-dependence to $1/T_1$.}
and 
that this model provides a simple,
accurate fit across all measured
conditions~\footnote{We note that the small $\beta$ value suggests that a biexponental relaxation model would also work; however, we find that it's ``extra'' degrees-of-freedom lead to overparameterization when applied to the present data.}.
The resulting $1/T_1$ values in both normal and superconducting states
are shown in \Cref{fig:slr-rate}.

Consistent with the above observations,
the $T$-dependence of $1/T_{1}$ exhibits a rich assortment of behavior.
At temperatures below \qty{\sim 100}{\kelvin},
$1/T_{1}$ varies linearly with $T$,
typical of metallic systems~\cite{1950-Korringa-Physica-16-601}.
Near the film's $T_\mathrm{c}$,
this linear proportionality is modified,
revealing a small Hebel-Slichter coherence peak~\cite{1957-Hebel-PR-107-901,1959-Hebel-PR-113-1504}
that decays exponentially to zero for $T \ll T_\mathrm{c}$.
Such a feature is expected for an $s$-wave \gls{bcs} superconductor,
but a rare observation by \ch{^{8}Li} \gls{bnmr}
(see, e.g.,~\cite{2009-Hossain-PRB-79-144518}).
Above \qty{\sim 100}{\kelvin},
additional relaxation contributions appear superimposed
on 
the $T$-linear contribution.
At \qty{\sim 140}{\kelvin},
$1/T_{1}$ goes through a local maximum,
reminiscent of a \gls{bpp} peak~\cite{1948-Bloembergen-PR-73-679}.
At higher temperature,
the peak vanishes
and
$1/T_{1}$ increases exponentially,
suggesting another distinct contribution
to the \gls{slr}.

With these qualitative features in mind,
we now consider a quantitative model to describe them.
We postulate that the distinct $T$-dependencies of $1/T_{1}$
correspond to the presence of distinct relaxation mechanisms,
with the measured \gls{slr} rate corresponding to their sum.
We make the \emph{ansatz} that the rate contributions can be added linearly:
\begin{equation}
	\dfrac{1}{T_1} = \left(\dfrac{1}{T_1}\right)_\mathrm{e} + \left(\dfrac{1}{T_1}\right)_\mathrm{BPP} + \left(\dfrac{1}{T_1}\right)_\mathrm{exp},
	\label{eq:total-slr}
\end{equation}
where each $( 1/ T_1 )_{i}$ term represents a specific contribution $i$.
Here,
we assign $i = \mathrm{e}$ to relaxation due to conduction electrons, 
$i = \mathrm{BPP}$ to the \gls{bpp} peak,
and,
$i = \mathrm{exp}$ to the exponential takeoff at high-$T$.
We now consider the detailed form of each mechanism.

The \Gls{slr} in metallic systems is governed by spin-flip scattering
``collisions'' between conduction electrons and the probe nuclear spins.
The magnitude of this contribution is proportional to the \gls{dos}
at the Fermi level $E_{\mathrm{F}}$,
as well as the probe's hybridization with the host's conduction band,
yielding a relaxation rate that varies linearly with $T$~\cite{1950-Korringa-Physica-16-601}.
In \gls{bcs} superconductors below $T_{\mathrm{c}}$,
the condensation of Cooper pairs along with the
opening of a gap in the \gls{dos} at $E_{\mathrm{F}}$ modifies this linearity,
producing a coherence peak~\cite{1957-Hebel-PR-107-901,1959-Hebel-PR-113-1504}
just below $T_{\mathrm{c}}$
and exponential decay of $(1/T_{1})_{\mathrm{e}}$ to zero as $T \rightarrow \qty{0}{\kelvin}$
(i.e., due to the freezing-out of all cross-gap thermal excitations).
Quantitatively,
this can be described by
(see, e.g.,~\cite{1993-Kiefl-PRL-70-3987,1976-MacLaughlin-SSP-31-1,2001-Kotegawa-PRL-87-127001}):
\begin{widetext}
	\begin{equation}
		\left(\dfrac{1}{T_1}\right)_\mathrm{e}  = \begin{cases}
			mT,                                                                                  & T>T_\mathrm{c} \\
			mT \times \dfrac{2}{k_\mathrm{B}T} \displaystyle\int_{0}^{\infty} f(E)[1-f(E')]
			[N_\mathrm{s}(E)N_\mathrm{s}(E') + M_\mathrm{s}(E)M_\mathrm{s}(E')] dE  , & T \leq T_\mathrm{c} 
		\end{cases}
		\label{eq:hs-slr-rob-kiefl}
	\end{equation}
\end{widetext}
where
$m$ is the so-called Korringa slope~\cite{1950-Korringa-Physica-16-601},
$E$ and $E' = E + \hbar \omega_0$ are the superconducting quasiparticle energies
in their initial and final scattering states
(the latter defined by the \gls{nmr} probe's Larmor frequency),
$f(E)$ is Fermi function:
\begin{equation*}
	f(E) = \dfrac{1}{ \exp \left [ E / (k_\mathrm{B} T ) \right ] + 1 } ,
	\label{eq:fermi-function}
\end{equation*}
$k_\mathrm{B}$ is the Boltzmann constant,
$N_\mathrm{s}(E)$ denotes the superconducting \gls{dos},
and
$M_\mathrm{s}(E)$ refers to the anomalous quasiparticle density arising from the
coherence factor~\cite{2005-Curro-Nature-434-622,1976-MacLaughlin-SSP-31-1}.
These latter two quantities can be expressed as:
\begin{align}
	\label{eq:sc-dos}
	N_\mathrm{s}(E) &= \mathrm{Re}\left\{(E - i\Gamma_\mathrm{D})/[(E-i\Gamma_\mathrm{D})^2-\Delta^2]^{1/2}\right\}, \\
	\label{eq:anamoly-dos}
	M_\mathrm{s}(E) &= \mathrm{Re}\left\{\Delta/[(E-i\Gamma_\mathrm{D})^2-\Delta^2]^{1/2}\right\},
\end{align}
where $\Delta$ is the superconducting energy gap~\cite{1966-Sheahen-PR-149-368}:
\begin{equation}
	\Delta (T) \approx \Delta (\qty{0}{\kelvin}) \sqrt{  \cos \left [ \frac{\pi}{2} \left( \frac{T}{T_\mathrm{c}}\right)^2 \right ]  },
	\label{eq:gap-T-dependence}
\end{equation}
$\Delta(\qty{0}{\kelvin})$ is the gap's value at \qty{0}{\kelvin},
and
$\Gamma_\mathrm{D}$
is an empirical broadening parameter
(e.g., accounting for finite quasiparticle lifetimes)~\cite{1978-Dynes-PRL-41-1509}.
Additionally, the dependence of $T_\mathrm{c}$ under an applied magnetic field follows the same relationship as described in \Cref{eq:Tc-dependence-H0}, and also quantifies the $B_\mathrm{c2}$ independently.

We now consider the \gls{slr} contribution from the \gls{bpp} peak.
Quite generally,
a local maximum in the \gls{slr} rate manifests when the correlation rate
$\tau_\mathrm{c}^{-1}$
of
the fluctuating interaction causing the relaxation matches the probe's Larmor frequency
[see \Cref{eq:larmor-frequency}].
Far away from this ``resonance,''
the contribution to the relaxation  is negligible,
leaving a ``peak'' that is superimposed atop the other \gls{slr} contributions.
Explicitly,
this contribution can be described by~\cite{1948-Bloembergen-PR-73-679,1988-Beckmann-PR-171-85}:
\begin{equation}
	\left(\dfrac{1}{T_1}\right)_\mathrm{BPP} = c(J_1 + 4J_2)
	\label{eq:slr-dyn}
\end{equation}
where $c$ is a coupling constant proportional to the mean-squared transverse fluctuating field,
and $J_n$ is the $n$-quantum \gls{nmr} spectral density function~\cite{1988-Beckmann-PR-171-85}:
\begin{equation}
	J_n = \dfrac{\tau_\mathrm{c}}{1 + (n\omega_0 \tau_\mathrm{c})^{2}} .
	\label{eq:spectral-density-3d}
\end{equation}
As $\omega_{0}$ is fixed in the measurements,
the temperature dependence of \Cref{eq:slr-dyn} arises from $\tau_\mathrm{c}^{-1}$,
which we assume follows an Arrhenius form:
\begin{equation}
	\tau_\mathrm{c}^{-1} = \tau_0^{-1} \exp \left[-E_\mathrm{A}/(k_\mathrm{B} T)\right],
	\label{eq:arrhenius-relation}
\end{equation}
where $\tau_0^{-1}$ is the attempt frequency,
and
$E_\mathrm{A}$ is the activation energy.
Note that these expressions are agnostic the source
of the (thermally activated) dynamics causing the relaxation ``peak.''

Finally,
we consider a model for the growth of the \gls{slr} observed
for $T \gtrsim \qty{200}{\kelvin}$.
Recalling the exponential-like take-off noted above,
we describe this behavior empirically using:
\begin{equation}
	\left(\dfrac{1}{T_1}\right)_\mathrm{exp} \approx g  \exp \left[-E_\mathrm{exp}/(k_\mathrm{B} T)\right] ,
	\label{eq:slr-trapping}
\end{equation}
where $g$ is a prefactor,
$E_\mathrm{exp}$ is the activation energy.
Note that the form of \Cref{eq:slr-trapping} is identical
to \Cref{eq:slr-dyn,eq:spectral-density-3d,eq:arrhenius-relation} in the
``slow fluctuating'' limit
(i.e., when $\tau_\mathrm{c}^{-1} \ll \omega_{0}$).

Combining the above models
[\Cref{eq:total-slr,eq:hs-slr-rob-kiefl,eq:sc-dos,eq:anamoly-dos,eq:gap-T-dependence,eq:slr-dyn,eq:spectral-density-3d,eq:arrhenius-relation,eq:slr-trapping,eq:Tc-dependence-H0}],
we fit the $1/T_{1}$ vs.\ $T$ data shown in \Cref{fig:slr-rate}.
For consistency with the lineshape analysis in \Cref{sec:results:resonance},
we fixed both $T_\mathrm{c}(\qty{0}{\tesla})$
and
$B_{\mathrm{c}2}(\qty{0}{\kelvin})$ at the values listed in
\Cref{tab:results-frequency}.
Similarly,
we fixed $\Gamma_\mathrm{D}$ at \qty{0.02}{\milli\electronvolt}
to mitigate its strong correlation with $\Delta (\qty{0}{\kelvin})$
(i.e., the present data lacks to the precision to simultaneously identify both quantities). We note that this value small compared to what is expected from \gls{pct} (see~\Cref{sec:experiment:samples}); however, it is consistent with the range reported by others~\cite{2010-Barends-APL-97-033507,2009-Barends-IEEETAS-19-936,2008-Barends-APL-92-223502}.
This (empirical) restriction notwithstanding, our approach enables a robust description of $1/T_1$'s $T$-dependence
across both \ch{Nb_{1-x}Ti_{x}N}'s normal and superconducting states,
with the resulting fit in good agreement with the data
(see \Cref{fig:slr-rate}).
Values of the extracted fit parameters are summarized in \Cref{tab:results-slr}.
We will return to considering these results in \Cref{sec:discussion:dynamics}.

\begin{table*}
	\centering
	\caption{
		\label{tab:results-slr}	
		Fit parameters describing the temperature dependence of the \gls{slr} rate $1/T_1$ (shown in \Cref{fig:slr-rate})
		using
		\Cref{eq:Tc-dependence-H0,eq:total-slr,eq:hs-slr-rob-kiefl,eq:sc-dos,eq:anamoly-dos,eq:gap-T-dependence,eq:spectral-density-3d,eq:slr-dyn,eq:arrhenius-relation,eq:slr-trapping}.
		Here,
		$\Delta (\qty{0}{\kelvin})$ represents the superconducting gap at \qty{0}{\kelvin},
		$T_\mathrm{c}(\qty{0}{\tesla})$ is the critical temperature at \qty{0}{\tesla},
		$\hbar\omega_0$ corresponds to the final scattering state associated with the \ch{^{8}Li} \gls{nmr} frequency,
		$\Gamma_\mathrm{D}$ is the broadening parameter,
		$m$ is the Korringa slope,
		$B_{0}$ is the applied magnetic field,
		$B_\mathrm{c2} (\qty{0}{\kelvin})$ is the upper critical field at \qty{0}{\kelvin},
		$\tau_0^{-1}$ and $E_\mathrm{A}$ are the Arrhenius prefactor and activation energy from \Cref{eq:arrhenius-relation},
		$c$ is a coupling constant,
		while
		$g$ and $E_\mathrm{exp}$ are the prefactor and activation energy from \Cref{eq:slr-trapping}.
		The values of $\hbar\omega_0$, $B_{0}$, $T_\mathrm{c}(\qty{0}{\tesla})$ and $B_\mathrm{c2}(\qty{0}{\kelvin})$ are fixed,
		as the first two quantities are known independently,
		while the latter two are determined from resonance linewidth analysis
		(\Cref{tab:results-frequency}).
	}
	\begin{ruledtabular}
	\begin{tabular}{l S l l}
		Parameter  & {Value}           & Unit & Comment              \\
		\hline
		$\Delta (\qty{0}{\kelvin})$        & 2.60 \pm 0.12              & \unit{\milli\electronvolt} &   \\
		$T_\mathrm{c}(\qty{0}{\tesla})$   & 15.4   & \unit{\kelvin}          &  fixed (from \Cref{tab:results-frequency})     \\
		$\hbar\omega_{0}$       & 1.1e-4    & \unit{\milli\electronvolt}  &  fixed (from \Cref{sec:experiment}) \\
		$\Gamma_\mathrm{D}$ & 0.02      & \unit{\milli\electronvolt}   &  fixed (from \cite{2010-Barends-APL-97-033507,2009-Barends-IEEETAS-19-936,2008-Barends-APL-92-223502}) \\
		$m$                 & 12.7 \pm 0.6e-6            & \unit{\per\second\per\kelvin} & \\
		$B_{0}$             & 4.1      & \unit{\tesla}                & fixed (from \Cref{sec:experiment}) \\
		$B_\mathrm{c2}(\qty{0}{\kelvin})$  & 18      & \unit{\tesla}      &  fixed (from \Cref{tab:results-frequency})       \\
		$c$                 & 0.50 \pm 0.05e6            & \unit{\per\square\second}   &  \\
		$\tau_0^{-1}$       & 1e12  & \unit{\per\second}          &  fixed (see~\Cref{sec:discussion:dynamics}) \\
		$E_\mathrm{A}$      & 0.0968 \pm 0.0014          & \unit{\electronvolt}        &  \\
		$g$                 & 0.6       & \unit{\per\second}          &  fixed (see~\Cref{sec:discussion:dynamics}) \\
		$E_\mathrm{exp}$      & 0.1011 \pm 0.0015          & \unit{\electronvolt}      &    \\
	\end{tabular}%
	\end{ruledtabular}
\end{table*}

\section{
	Discussion
	\label{sec:discussion}
}

The \ch{^{8}Li} \gls{bnmr} data presented in \Cref{sec:results}
display a rich range of behavior.
At temperatures below \qty{\sim 100}{\kelvin},
the \gls{nmr} response is metallic,
with the most salient features occurring at or below the superconducting transition,
where significant lineshape broadening coincides with a coherence peak in the \gls{slr}.
At higher temperatures,
the lineshape remains $T$-insensitive,
while the \gls{slr} is dominated by additional sources of relaxation,
deviating significantly from the $T$-linear Korringa response.
Following this dichotomy,
we divide our discussion into two parts.
First, we consider the low-$T$ metallic behavior in \Cref{sec:discussion:metallic},
followed by the dynamics observed at higher temperatures in \Cref{sec:discussion:dynamics}.

\subsection{
	Metallic \& Superconducting Response
	\label{sec:discussion:metallic}
}

To start this section,
we proceed with a discussion of the resonance lineshapes,
whose salient feature is a broadening upon transition to the vortex state
below $T_\mathrm{c}$.
From the analysis
in \Cref{sec:results:resonance},
we find that $\lambda (\qty{0}{\kelvin}) = \qty{180.57 \pm 0.30}{\nm}$ for our film,
in excellent agreement with a direct measurement using \gls{le-musr}~\cite{2024-Asaduzzaman-SST-37-025002}
and
comparable to a crude estimate made for different \ch{Nb_{1-x}Ti_{x}N} stoichiometries~\cite{1990-DiLeo-JLTP-78-41}.
Other studies have reported values where
$\lambda \gtrsim \qty{200}{\nano\meter}$~\cite{2023-Fedor-IEETTST-13-627,2005-LeiYu-IEETAS-15-44,2022-Khan-IEEETAS-32-1,2013-Hong-JAP-114-243905},
suggesting that our measurement
is likely closer to the alloy's \emph{intrinsic} value
(i.e., its London penetration depth $\lambda_{\mathrm{L}} \sim \qty{150}{\nano\meter}$~\cite{2016-Anne-Marie-SST-29-113002}).
While $\lambda_\mathrm{L}$ value is not well-defined,
we may use it to estimate the \gls{gl} parameter $\kappa$ for our film
via~\cite{1996-Tinkham-Book-2-McGraw,2012-Tobi-thesis}:
\begin{equation} 
	\label{eq:ginzberg_landau}
	\kappa = \frac{2 \sqrt{3}}{\pi}\frac{\lambda^2 (\qty{0}{\kelvin})}{\xi_0 \lambda_\mathrm{L}} ,
\end{equation}
which makes use of the relationship between $\Phi_0$ and the coherence lengths within \gls{bcs} ($\xi_0$) and \gls{gl} theory.
Using the values for $\lambda (\qty{0}{\kelvin})$ and $\lambda_\mathrm{L}$ noted above,
along with $\xi_0 = \qty{2.4 \pm 0.3}{\nm}$~\cite{2002-Lei-IEEETAS-12-1795},
\Cref{eq:ginzberg_landau} yields
$\kappa = \num{100 \pm 12}$,
placing our film in the extreme type-II limit.
While a value on this order is typical of \ch{Nb_{1-x}Ti_{x}N},
we note that it justifies our use of \Cref{eq:brandt-eq-11},
which requires $\kappa \gtrsim 70$.

Similar to $\lambda$,
the extracted critical temperature $T_\mathrm{c}(0) =\qty{15.4 \pm 0.7}{\kelvin}$
from the lineshape analysis
agrees with the magnetometry measurements (see~\Cref{sec:experiment:samples} and Sec. II of the Supplemental Material~\cite{supp}) and range
expected for our film's
stoichiometry~\cite{1990-DiLeo-JLTP-78-41,1969-Gavaler-JAP-15-329,1997-Benvenuti-NIMPRSB-124-106,2016-Burton-JVSTA-34-021518}.
Given \ch{Nb_{1-x}Ti_{x}N}'s variability in properties,
particularly with synthesis and form
(see \Cref{sec:introduction}),
this level of agreement is encouraging.
Unique to our analysis approach though,
is using $T_\mathrm{c}$'s suppression by $B_{0}$ to make
an estimate of our film's upper
critical field $B_\mathrm{c2}(\qty{0}{\kelvin})$.
This value turned out to be \qty{18 \pm 4}{\tesla},
which has a high relative uncertainty (\qty{\sim 22}{\percent}),
largely due to $B_\mathrm{c2}$'s strong correlation with $T_\mathrm{c}(\qty{0}{\tesla})$.
Nonetheless,
its value is consistent with the range reported by
others~\cite{2021-Sidorova-PRB-104-184514,1969-Hechler-JLTP-1-29,2023-Pratiksha-SST-36-085017,2023-Rezinovsky-PC-607-1354241}. To further validate this estimate, we calculate the $\xi_\mathrm{GL}$ using~\cite{1996-Tinkham-Book-2-McGraw}:
\begin{equation}
	\xi_\mathrm{GL}(\qty{0}{\kelvin}) = \sqrt{\frac{\Phi_0}{2 \pi B_\mathrm{c2}(\qty{0}{\kelvin})}}
	\label{eq:gl-xi}
\end{equation}
which yields $\xi_\mathrm{GL}(\qty{0}{\kelvin}) = ~\qty{4.3\pm 0.5}{\nm}$, in excellent agreement with prior reports~\cite{2021-Sidorova-PRB-104-184514,2002-Lei-IEEETAS-12-1795,2023-Pratiksha-SST-36-085017},
further suggesting the reliability of our analysis approach.

We now turn our attention to the Knight shifts $K^{c}$ extracted from the resonance data.
As noted in \Cref{sec:results:resonance},
$K^{c}$ is small and positive,
ranging from \qtyrange[retain-explicit-plus=true]{+15}{+35}{\ppm} over the measured temperature range.
The observed shift arises from a (presumably isotropic) contact hyperfine interaction,
stemming from the hybridization of \ch{^{8}Li^{+}}'s vacant $2s$ orbital with
\ch{Nb_{1-x}Ti_{x}N}'s conduction band.
This weak coupling is a general feature of \ch{^8Li} \gls{bnmr},
and similar observations have been noted for other cubic metals
(see, e.g.,~\cite{2019-Parolin-PRB-100-209904}).
While this static (i.e., time-average) contribution of the contact interaction
gives rise to the resonance shift,
its dynamic (i.e., time-dependent) component
dominates \ch{^{8}Li}'s low-$T$ \gls{slr},
which we consider below.

The $T$-linear relaxation observed below \qty{\sim 100}{\kelvin}
is one of the hallmarks of the metallic state probed by \gls{nmr}~\cite{1950-Korringa-Physica-16-601,1990-Slichter-Book-3-Springer}.
Over this temperature range,
a linear fit with zero intercept yields a slope $m = \qty{12.7 \pm 0.6e-6}{\per\second\per\kelvin}$.
This value is surprisingly small
and at least an order of magnitude less than in other
cubic metals
(cf.~\cite{2019-Parolin-PRB-100-209904}).
From relaxation in metals,
this slope may be written as~\cite{1990-Slichter-Book-3-Springer,1950-Korringa-Physica-16-601}:
\begin{equation}
	\label{eq:korringa-slope}
	m \equiv \frac{1}{T_1T} = \frac{2 \pi k_B}{\hbar} A^2 \rho_e(E)^2  ,
\end{equation}
where
$k_{\mathrm{B}}$ is the Boltzmann constant,
$\hbar$ is the reduced Planck constant,
$A$ is the hyperfine coupling (in units of energy),
and
$\rho_e(E )$ is the electronic \gls{dos}.
It follows from \Cref{eq:korringa-slope} that two sources can contribute to a small $m$:
a small $A$
or
a small $\rho_e(E)$
(or both).
Given the metallic nature of \ch{Nb_{1-x}Ti_{x}N}~\cite{2023-Gonzalez-JAP-134-035301,2023-Pratiksha-SST-36-085017},
The implication of this means that a small hyperfine coupling is the most probable source
for the small $m$,
but note that a small $A$ is also unexpected given \ch{Nb_{1-x}Ti_{x}N}'s structure
(see \Cref{fig:NbTiN-crystal-structure}).
That is,
in the rocksalt structure
there are a limited number of sites for implanted \ch{^{8}Li^{+}} that could facilitate this
(cf.\ the van~der~Waals gap in layered chalcogenides~\cite{2006-Wang-PBC-374-239,2019-McFadden-PRB-99-125201,2020-McFadden-PRB-102-235206}).
Noting the absence of quadrupolar splittings to our resonance data
(see \Cref{fig:resonance-spectra}),
symmetry constrains the possible candidates.
We suggest the most plausible interstitial position to be the Wyckoff $8c$ site i.e.,  the tetrahedral interstitial sites in the ~\gls{fcc} lattice,
situated in the center of the sub-cubes within the alloy's unit cell
(see \Cref{fig:NbTiN-crystal-structure}),
but note that
\ch{^{8}Li^{+}} may simply be substitutional for \ch{Nb}/\ch{Ti}
in Wyckoff site $4a$. 
The small $m$ is even more surprising when one considers the (intrinsic) disorder
present in the film,
which generally \emph{increases} its
value~\cite{1983-Gotze-ZPB-54-49,1994-Shastry-PRL-72-1933}.

Further insight into the film's metallic behavior can be inferred from considering
$K^\mathrm{c}$ and $m$ in unison.
Following a so-called Korringa analysis,
we compute the
(dimensionless)
Korringa ratio~\cite{1990-Slichter-Book-3-Springer,2009-Parolin-PRB-80-174109}:
\begin{equation}
	\mathcal{K} \equiv \dfrac{(K^{\mathrm{c}})^2 T_1 T}{\mathcal{S}} ,
	\label{eq:korringa-ratio}
\end{equation}
where $\mathcal{S}$ is a constant specific to the \gls{nmr} probe nucleus
($\mathcal{S} \approx \qty{1.20e-5}{\second\kelvin}$ for \ch{^{8}Li}). 
Using our measured value for $m$,
along with $K^\mathrm{c}(\qty{270}{\kelvin}) = \qty{14.0 \pm 1.3}{\ppm}$, 
we obtain $\mathcal{K} = \num{1.29 \pm 0.25}$ ~\footnote{If different values were chosen, such as $K^\mathrm{c} (\qty{20}{\kelvin}) = \qty{22.5 \pm 1.4}{\ppm} $ or  $K^\mathrm{c} (\qty{120}{\kelvin}) = \qty{20.3 \pm 1.3}{\ppm}$, the corresponding Korringa ratios would be $\mathcal{K} = \num{3.4 \pm 0.5}$ and $\mathcal{K} = \num{2.7 \pm 0.4}$, respectively.}.
In the limit that the conduction electrons are non-interacting,
one expects $\mathcal{K} = 1$;
however,
deviations from this can indicate (anti)ferromagnetic electronic correlations
and
are often encountered for weakly interacting systems~\cite{1990-Slichter-Book-3-Springer}
(see, e.g.,~\cite{2019-Parolin-PRB-100-209904}).
Alternatively,
a $\mathcal{K} > 1$ may be understood in terms of an enhancement factor $\eta$
resulting from disorder~\cite{1983-Gotze-ZPB-54-49,1994-Shastry-PRL-72-1933}:
\begin{equation*}
	\mathcal{K}_{\mathrm{measured}} = \mathcal{K} \eta ,
\end{equation*}
caused by, for example, an increased residence time of the conduction electrons at the probe's site.
Assuming electronic correlations in our film are negligible,
an $\eta \approx 1.3$ at \qty{270}{\kelvin} is implied.
While this is quite modest on the scale of (disordered) conductors,
it agrees with
predictions based on our film's conductivity inferred from 
measured resistivity (see Sec. I of the Supplemental
Material~\cite{supp}) 
[see, e.g., the ``Warren plot'' in Ref.~\cite{1983-Gotze-ZPB-54-49}].
Despite the film's intrinsic disorder,
the above suggests that its electronic behavior is consistent with nearly-free electrons,
but close to the crossover to diffusive transport.

Having discussed the metallic \gls{slr} in the film's normal state,
we now consider its modification below the superconducting transition,
whose main feature is the Hebel-Slichter coherence peak~\cite{1957-Hebel-PR-107-901,1959-Hebel-PR-113-1504}
below $T_\mathrm{c}$
(see the inset in \Cref{fig:slr-rate}).
While such an observation is rare for \gls{bnmr}
(see, e.g., \ch{^{8}Li} \gls{slr} in \ch{NbSe2}~\cite{2009-Hossain-PRB-79-144518}),
its presence is not unexpected.
For example,
such a feature is also observed in the \ch{^{93}Nb} \gls{nmr} of the alloy's end member
\ch{NbN}~\cite{2004-Nishihara-JAC-383-308,2009-Lascialfari-PRB-80-104505}.
One might anticipate that \ch{Nb_{1-x}Ti_{x}N}'s intrinsic disorder may suppress such a feature,
but given the alloy's modest enhancement factor $\eta$
(see above),
any suppression is likely minimal~\cite{1992-Bahlouli-PLA-164-206,1993-Devereaux-ZPBCM-90-65}.
While our data is in good qualitative agreement with the theoretical description
outlined in \Cref{sec:results:slr},
it does not provide a perfect match.
Notably,
a rise in $1/T_{1}$ is observed above the onset of film's
(average)
$T_\mathrm{c}$.
Indeed,
transport measurements of a similarly prepared sample on an \ch{Al2O3} substrate show a sharp transition at approximately \qty{15}{\kelvin},
but with a small deviation beginning earlier around $\qty{\sim 16}{\kelvin}$~\cite{2023-Kalboussi-thesis},
suggesting a range of $T_\mathrm{c}$ values.
While the relaxation model captures the most prominent $T_\mathrm{c}$,
the \gls{slr} data appear much more sensitive to this distribution of transition temperatures 
than the resonance measurements
(see \Cref{sec:results:resonance}).
The precise reason for this is unclear;
however,
we remark that these deviations are small
and that
the \gls{slr} model does an excellent job of describing the data below
the film's ``characteristic'' $T_\mathrm{c}$.
As will be clear below,
this aberration does affect our ability to quantify
parameters governing the coherence peak.

Following \Cref{eq:hs-slr-rob-kiefl,eq:sc-dos,eq:anamoly-dos,eq:gap-T-dependence},
the magnitude of the Hebel-Slichter coherence peak is dictated primarily by two quantities:
a zero-temperature gap $\Delta(\qty{0}{\kelvin}) = \qty{2.60 \pm 0.12}{\milli\electronvolt}$~\footnote{This value reflects a spatial average over many vortex unit cells within the $\qty{\sim 3}{\milli\meter}$ \gls{bnmr} beam spot. Given the short~\gls{gl} coherence length $\xi_\mathrm{GL} = \qty{4.3 \pm 0.5}{\nm}$, the vortex cores occupy a negligible volume fraction, so the spatial variation of $\Delta$ in the mixed state can be neglected.},
and
a Dynes~\cite{1978-Dynes-PRL-41-1509} broadening parameter $\Gamma_\mathrm{D} = \qty{0.02}{\milli\electronvolt}$.
The extracted gap agrees with \gls{pct} measurements on similarly prepared samples (see~\Cref{sec:experiment:samples}). 
In contrast, the smaller $\Gamma_\mathrm{D}$ value inferred from \gls{bnmr} differs from the \gls{pct} value of $\qty{0.1 \pm 0.06}{\milli\electronvolt}$, which is too large to account for the magnitude of the coherence peak (see~\Cref{fig:slr-rate}).
This discrepancy 
could reflect differences in each technique's spatial sensitivity
[\gls{pct} is only sensitive to depths up to \qty{\sim 10}{\nm}, but has micrometer lateral resolution; \gls{bnmr} has sensitivity to depths up to \qty{\sim 60}{\nm} (see~\Cref{fig:stopping-profile}), but averages over lateral distances on the order of millimeters (i.e., the \ch{^{8}Li^{+}} beam's spot size)].
The \gls{bnmr} value of $\Gamma_\mathrm{D}$ agrees with the \qtyrange{0.015}{0.02}{\milli\electronvolt} range reported for \ch{Nb_{1-x}Ti_xN} thin-film coplanar waveguide resonators, where a Dynes broadening parameter $\Gamma_\mathrm{D}$ was required in Mattis-Bardeen~\cite{1958-Mattis-PR-111-412} fits to microwave transmission measurements at \qty{100}{\milli\kelvin}~\cite{2010-Barends-APL-97-033507,2009-Barends-IEEETAS-19-936,2008-Barends-APL-92-223502}.
We note that without the broadening term (i.e., $\Gamma_\mathrm{D} = 0$) the coherence peak is much too sharp relative to the data (see~\Cref{fig:slr-rate}).

Using the values of $\Gamma_\mathrm{D}$ and $\Delta (\qty{0}{\kelvin})$, the dimensionless ratio $\Gamma_\mathrm{D} / \Delta(\qty{0}{\kelvin}) = \num{7.68 \pm 0.34 e-3}$ was obtained, serving as a sensitive metric for quantifying disorder within the superconducting film. 
This value 
is consistent with tunneling studies on disordered films of \ch{Nb_{1-x}Ti_{x}N}'s end members
\ch{NbN}~\cite{2009-Chockalingam-PRB-79-094509} and \ch{TiN}~\cite{2008-Sacepe-PRL-101-157006},
and
implies small disorder in the crystal lattice. 
A corresponding value extracted from \gls{pct} measurements, $\Gamma_\mathrm{D} / \Delta(\qty{0}{\kelvin}) = \num{0.040 \pm 0.024}$, appears larger due to its higher absolute $\Gamma_\mathrm{D}$, 
but still overlaps with the \gls{bnmr} result.
These results suggest that $\Gamma_\mathrm{D}$ at the surface is higher compared to that in the bulk of the film.
The modest peak suppression and the overall agreement with the relaxation model
[\Cref{eq:anamoly-dos,eq:hs-slr-rob-kiefl,eq:sc-dos,eq:gap-T-dependence}]
is suggestive of $s$-wave symmetry of the Cooper pairs,
consistent with other reports~\cite{2012-Driessen-PRL-109-107003}.
Returning to the zero-temperature gap,
upon combining $\Delta(\qty{0}{\kelvin})$ with the $T_\mathrm{c}(\qty{0}{\tesla})$ determined from the resonance lineshape analysis
we obtain a gap ratio $2\Delta(\qty{0}{\kelvin})/k_\mathrm{B}T_\mathrm{c}(\qty{0}{\tesla}) = \num{3.92 \pm 0.25}$.
This value aligns well with the \numrange{3.53}{5} range reported in the literature~\cite{2024-Cyberey-IEETAS-34-1,2019-Cyberey-IEEETAS-29-1,2023-Fedor-IEETTST-13-627,2022-Khan-IEEETAS-32-1,2013-Hong-JAP-114-243905,2013-Westig-JAP-114-124504,2024-Zhukova-IEETAS-34-1,2014-Groll-APL-104-092602,2015-Uzawa-IEEETAS-25-1,2021-Lap-APL-119-152601,2010-Barends-APL-97-033507},
and
matches with values observed for films of similar stoichiometry~\cite{2015-Uzawa-IEEETAS-25-1}.
This ratio is also consistent with the \gls{bcs} strong-coupling limit,
as well as 
other reports on \ch{Nb_{1-x}Ti_{x}N}~\cite{2021-Lap-APL-119-152601,2022-Khan-IEEETAS-32-1}.

With the alloy's superconducting properties established,
it would be interesting in the future to study how they translate into
preventing magnetic-flux-nucleation in \gls{sis} heterostructures
~\footnote{Particularly in ellipsoidal samples, which prevent field penetration from both sides of the superconducting layers and ensure a uniform magnetic response, making them a relevant proxy for~\gls{srf} cavities.}.
Similar \ch{^{8}Li} \gls{bnmr} measurements have been done on ``bare'' and ``baked'' \ch{Nb}~\cite{2024-Thoeng-SR-14-21487}
using a purpose-built \gls{bnmr} spectrometer~\cite{2023-Thoeng-RSI-94-023305}.
Such measurements would complement the findings of Ref.~\cite{2024-Asaduzzaman-SST-37-025002},
but under conditions closer to those of state-of-the-art
accelerator cavities~\cite{2023-Padamsee-SRTA}.

\subsection{
	High-$T$ Dynamics
	\label{sec:discussion:dynamics}
}

Similar to \Cref{sec:discussion:metallic},
we first discuss the \ch{^{8}Li} resonance at high temperatures.
Above $T_\mathrm{c}$,
the lineshape is approximately $T$-independent,
characterized by a field distribution width $\sigma_\mathrm{n} = \qty{540.1 \pm 1.9}{\micro\tesla}$
(see \Cref{fig:resonance-spectra,fig:resonance-linewidth}).
The absence of any pronounced linewidth reduction at elevated temperatures
(i.e., ``motional narrowing''~\cite{1990-Slichter-Book-3-Springer})
immediately rules out long-range translational dynamics as the source of the dominant \gls{slr}
contributions above \qty{\sim 100}{\kelvin}.
While it is difficult to be conclusive about their origin,
we consider some possibilities below.

To begin,
we consider the \gls{bpp} at \qty{\sim 140}{\kelvin}
(see \Cref{fig:slr-rate}).
The fit of this peak to the model given by
\Cref{eq:slr-dyn,eq:spectral-density-3d,eq:arrhenius-relation}
is characterized by three main terms:
a coupling constant $k = \qty{0.50 \pm 0.05e6}{\per\square\second}$,
an activation barrier $E_\mathrm{A} = \qty{0.0968 \pm 0.0014}{\electronvolt}$,
and an assumed attempt frequency $\tau_0^{-1} = \qty{e12}{\per\second}$
(see \Cref{tab:results-slr}),
the latter being comparable to optical phonon frequencies.
Consistent with the absence of ``motional narrowing,''
the small $E_\mathrm{A}$ is incompatible with \ch{^{8}Li^{+}} diffusion,
which is typically characterized by a larger energy barrier
(in all but the most exceptional lithium-ion conductors).
This is reasonable based on \ch{Nb_{1-x}Ti_{x}N}'s structure
(see \Cref{fig:NbTiN-crystal-structure}),
and the (nominal) ionic valence of its atomic constituents.
Thus,
the kinetic process causing relaxation must be \emph{highly localized}.
We shall hold off on speculating its origin for the time being,
but note that 
the magnitude of the coupling term $c$
is also rather small.
This is surprising given \ch{Nb_{1-x}Ti_{x}N}'s dense spin concentration
(see \Cref{tab:host-spins});
however,
the dynamic component of the local field may have an origin distinct
from the static component
(cf.\ the \ch{^{8}Li} \gls{bnmr} in \ch{Bi}~\cite{2014-MacFarlane-PRB-90-214422}).

We now consider the monotonic increase in $1/T_1$ above \qty{\sim 200}{\kelvin},
which is empirically described by an Arrhenius-like relation
[\Cref{eq:slr-trapping}].
As noted in \Cref{sec:results:slr},
\Cref{eq:slr-dyn,eq:spectral-density-3d,eq:arrhenius-relation}
reduce to this expression in the limit that $\tau_\mathrm{c}^{-1} \ll \omega_{0}$,
emphasizing its similarity with the \gls{bpp} peak. 
This \gls{slr} component is governed by an activation energy $E_\mathrm{exp} = \qty{0.1011 \pm 0.0015}{\electronvolt}$
and
a fixed prefactor $g = \qty{0.6}{\per\second}$ (assigned empirically). 
The similarity of $E_\mathrm{exp}$ to $E_\mathrm{A}$ is remarkable
and
possibly suggests a connection between the two.
Much like $E_\mathrm{A}$,
$E_\mathrm{exp}$'s magnitude is suggestive of a localized kinetic process.
A closer comparison with the former's kinetic details is,
however,
hampered by our ability to precisely identify $g$,
which is only correct to an order-of-magnitude.
This is often the case for fits in the ``slow fluctuating'' limit,
where the coupling term is inseparable from the (apparent) prefactor.
A crude calculation using $c$'s magnitude suggests the $g$ is equivalent to
a $\tau_{0}^{-1}$ on the order of \qty{\sim e6}{\per\second},
which is too low to be physical.
While the kinetic process' true prefactor remains uncertain,
we may infer that the coupling term for this likely exceeds
$c$ by several orders-of-magnitude.

Having discussed the high-$T$ relaxation details,
we now consider their possible origin.
One plausible source is the annealing of damage caused during \ch{^{8}Li^{+}} implantation.
During implantation, \ch{^8Li} ions lose energy via collisions with the host lattice,
displacing host atoms from their equilibrium positions
and
forming Frenkel pairs
(i.e., a vacancy and an interstitial defect).
After generating its final defect,
the probe ion often continues to a high-symmetry stopping site.
While this site is usually sufficiently distant from the damage to avoid affecting
the \gls{nmr} signal,
it may remain close enough to participate in a local re-arrangement of atoms.
For example,
in elemental \gls{fcc} metals,
\ch{^{8}Li^{+}} stops in a (metastable) interstitial site at low-$T$,
but undergoes a site-change-transition to a substitutional site
(i.e., via the vacancy provided by the Frenkel pair)
as the sample is ``annealed'' at higher temperature~\cite{2015-MacFarlane-SSNMR-68-1}.
While Knight shifts in excess of the resonance linewidth make this
process easily distinguishable in the ``simple'' metals,
that is not the case in \ch{Nb_{1-x}Ti_{x}N}
(see \Cref{sec:results:resonance}).

An alternative explanation involves the implantation-induced displacement of ``excess'' nitrogen atoms introduced during the growth process, but not purged during annealing (see~\Cref{sec:experiment:samples}). During implantation, the (energetic) \ch{^{8}Li^{+}} ions may dislodge nitrogen atoms from their lattice sites, creating closely spaced interstitials that may combine to form molecular \ch{N2}. Given the high bond strength,
these molecules are effectively inert; however, at elevated temperatures, thermally activated rotational or vibrational modes may become relevant. Such dynamics could couple weakly to the \ch{^8Li} spin system, contributing to the observed increase in relaxation rate, however 
their manifestation in disordered nitrides like \ch{Nb_{1-x}Ti_{x}N} remains largely unexplored.

In the present,
we propose a related, but different process is taking place.
If we suppose that all \ch{^{8}Li^{+}} stops in a substitutional
Wyckoff $4a$ site,
then the relaxation behavior observed may be ascribed to the annealing
of \ch{Ti/Nb} \emph{intersticies}.
In a purely ``ionic'' picture of the lattice,
the size and charge of atoms plays an important role in dictating their translational motion.
Assuming similar oxidations states,
both \ch{Ti} and \ch{Nb} have similar ionic radii~\cite{1976-Shannon-AC-A32-751},
and
one might postulate that the minimum energy path for either element back into a vacant $4a$ site
has a similar barrier.
This would be
consistent with our observed $E_\mathrm{A}$ and $E_\mathrm{exp}$.
As \ch{Ti} has both a smaller stoichiometry
and
a smaller spin-active fraction than \ch{Nb}
(see \Cref{tab:host-spins}),
we expect its ``migration'' to have a small coupling term $c$,
consistent with the observed \gls{bpp} peak at \qty{\sim 140}{\kelvin}
(see \Cref{fig:slr-rate}).
As noted above,
the higher-$T$ relaxation process likely has a much larger coupling term,
which would be consistent with the motion of \ch{^{93}Nb}.
In this picture,
the site-change of \ch{Ti} occurring at a lower-$T$ than \ch{Nb} is
difficult to rationalize,
but the similarity in activation barriers suggests it is related to the
vibrational modes of the two intersticial species.
For classical harmonic motion,
the vibrational mode along the ``reaction'' pathway scales as the inverse square root
of mass,
so one might naively expect the frequency $\tau_{0}^{-1}$ to be greater for \ch{Ti} than
\ch{Nb}, qualitatively in line with our interpretation.
Of course, this depiction is rather simplistic
and the purely ionic picture above breaks down in a metal;
however, the availability of conduction electrons to screen localized changes is consistent with the absence of any quadrupolar ``features'' in the resonance data.
That said, quadrupolar broadening could contribute to a minor extent,
which may explain the $T$-dependent scatter of the lineshape's \gls{fwhm} about its mean value.

While we have suggested an explanation for our observations,
further studies are necessary to test these ideas.
Additional \gls{slr} measurements at different applied fields
may help refine our parameterization of the underlying kinetic processes.
Central to this may be identifying \ch{^{8}Li}'s stopping site,
which, for a dense spin system like \ch{Nb_{1-x}Ti_{x}N},
may be accomplished by searching for \glspl{alcr} over a wide range of applied fields
(see, e.g.,~\cite{2012-Chow-PRB-85-092103}).
A recently upgraded \gls{bnmr} spectrometer at TRIUMF is
well-suited for this purpose~\cite{2023-Thoeng-RSI-94-023305}.

\section{
	Conclusion
	\label{sec:conclusion}
}

Using implanted-ion \ch{^{8}Li} \gls{bnmr},
we investigated the superconducting and normal-state properties of a
\ch{Nb_{0.75}Ti_{0.25}N}(\qty{91}{\nm})/\ch{AlN}(\qty{4}{\nm})/\ch{Nb} thin film
between $\qty{4.6}{\kelvin} \leq T \leq \qty{270}{\kelvin}$
in a \qty{4.1}{\tesla} field perpendicular to its surface.
Resonance measurements revealed a broad lineshape in the normal state
that (symmetrically) broadens below $T_\mathrm{c}$ due to \gls{fll} formation in the vortex state.
From a fit to a broadening model,
we find a superconducting transition temperature $T_\mathrm{c}(\qty{0}{\tesla}) = \qty{15.4 \pm 0.7}{\kelvin}$,
an upper critical field $B_\mathrm{c2}(\qty{0}{\kelvin}) = \qty{18 \pm 4}{\tesla}$,
and a magnetic penetration depth $\lambda(\qty{0}{\kelvin}) = \qty{180.57 \pm 0.30}{\nm}$,
in good agreement with independent measurements and
estimates from the literature.
\Gls{slr} measurements find a metallic response at low-$T$
with a Korringa slope $m = \qty{12.6 \pm 0.6 e-6}{\per\second\per\kelvin}$,
which is modified by below $T_\mathrm{c}$ by a Hebel-Slichter coherence peak
characterized by a superconducting gap $\Delta(\qty{0}{\kelvin}) = \qty{2.60 \pm 0.12}{\milli\electronvolt}$
and
a Dynes broadening parameter $\Gamma_\mathrm{D} = \qty{0.02}{\milli\electronvolt}$.
These findings yield a gap ratio $2\Delta(\qty{0}{\kelvin})/k_\mathrm{B}T_\mathrm{c}(\qty{0}{\tesla}) = \num{3.92 \pm 0.25}$,
consistent with strong-coupling behavior for the ternary alloy.
In the future,
it would be interesting to test explicitly how these properties translate into preventing
magnetic-flux penetration in \gls{sis} heterostructures
(e.g., for \gls{srf} cavity applications).

\begin{acknowledgments}
	Technical support during the \gls{bnmr} experiments from
	R.~Abasalti, D.~J.~Arseneau, B.~Hitti, and D.~Vyas
	(TRIUMF) is gratefully acknowledged.
	T.J.\ acknowledges financial support from NSERC.
\end{acknowledgments}

\section*{Author Declarations}

\subsection*{Conflict of Interest}
The authors have no conflicts to disclose.

\section*{Data Availability}
Raw data from the \gls{bnmr} experiments performed at TRIUMF
are publicly available for download from
\url{https://cmms.triumf.ca}
(experiment number M2265).

\appendix

\section{
	Magnetic Field Suppression of $T_\mathrm{c}$
	\label{sec:app:Tc-vs-B}
}

Key to our analysis of the \ch{^{8}Li} \gls{bnmr} data is accounting for the well-known
suppression of the superconducting transition temperature $T_\mathrm{c}$ in a
static magnetic field $B_{0}$.
As mentioned in \Cref{sec:results:resonance},
we treat this quantitatively by inverting an empirical expression for $B_\mathrm{c2}$
derived from \gls{whh} theory~\cite{1996-Werthamer-PR-147-295,2014-Baumgartner-SST-27-015005},
with explicit steps given below.

Starting from \Cref{eq:Tc-dependence-H0},
we re-arrange it into the form of a depressed quartic expression:
\begin{equation}
	u^4 + p u^2 - q u + r b = 0 ,
	\label{eq:depressed}
\end{equation}
where we have used the substitutions:
\begin{align*}
	u &\equiv 1 - t , \\
	b &\equiv \frac{B_0}{B_{\mathrm{c2}}(0)} , \\
	p &\equiv \frac{-0.153}{-0.152}, \\ 
	q &\equiv \frac{1}{-0.152}, \\
	r &\equiv \frac{-0.693}{-0.152} . \\
\end{align*}
To solve \Cref{eq:depressed},
we first write it 
as:
\begin{equation*}
	u^4 = -pu^2 + q u - rb ,
\end{equation*}
whereafter we introduce an auxiliary term $y$ that makes both sides of this expression perfect squares.
While the left-hand side may be rewritten using:  
\begin{equation*}
	(u^2 + y)^2 - (2y - p)u^2 - q u - (y^2 - r b) = 0 ,
\end{equation*}  
for the right-hand side to be a perfect square the condition:
\begin{equation*}
    y^2 - r b = \frac{q^2}{4(2y - p)}
    \label{eq:satisfy-condition}
\end{equation*}
must hold.
Upon substitution,
it can be shown that:
\begin{equation*}
	(u^2 + y)^2 - \left( u \sqrt{2y - p} + \frac{q}{2\sqrt{2y - p}} \right)^2  = 0.
\end{equation*}  
This leads to four possible solutions for $u$:
\begin{align*}
	u_{1\pm} &= \frac{1}{2}\left(-\sqrt{2y - p} \pm \sqrt{2y - p - 4y - \frac{2q}{\sqrt{2y - p}}} \right) \\
	u_{2\pm} &= \frac{1}{2} \left( \sqrt{2y - p} \pm \sqrt{2y - p - 4y + \frac{2q}{\sqrt{2y - p}}} \right) \\
\end{align*}
Among these,
only
\begin{equation}
	u_{2-} = \frac{1}{2} \left( \sqrt{2y - p} - \sqrt{2y - p - 4y + \frac{2q}{\sqrt{2y - p}}} \right)
	\label{eq:u-final}
\end{equation}
is physically meaningful,
based on the criterion that $u \in [0, 1]$
(i.e., ensuring a real and positive $T_\mathrm{c}$).

To determine $u_{2-}$,
we solve the corresponding cubic equation for $y$: 
\begin{equation*}
    2 y^3 - p y^2 - 2 r b y + \left( pr b - \frac{1}{4} q^2 \right) = 0 .
\end{equation*}
Its solution is:
\begin{equation}
    y = \frac{p}{6} + v - \frac{x}{3v},
    \label{eq:y-solution}
\end{equation}
where
\begin{align}
    v &= \sqrt[3]{-\frac{w}{2} + \sqrt{\frac{w^2}{4} + \frac{x^3}{27}}},
    \label{eq:v-formula} \\
    x &= - \frac{p^2}{12} - r b,
    \label{eq:x-formula} \\
    w &= -\frac{p^3}{108} + \frac{pr b}{3} - \frac{q^2}{8}.
    \label{eq:w-formula}
\end{align}
Substituting the known values of $m, n, p$, and $b$ into
\Cref{eq:y-solution,eq:v-formula,eq:x-formula,eq:w-formula}
provides the explicit solution for $y$,
which in turn determines $u_{2-}$
and
ultimately $T_\mathrm{c}$'s dependence on $B_0$ using \Cref{eq:u-final}.

\section{Knight Shift Calculation
	\label{sec:app:N-chi}
}

\begin{figure}
	\centering
	\includegraphics*[width=\columnwidth]{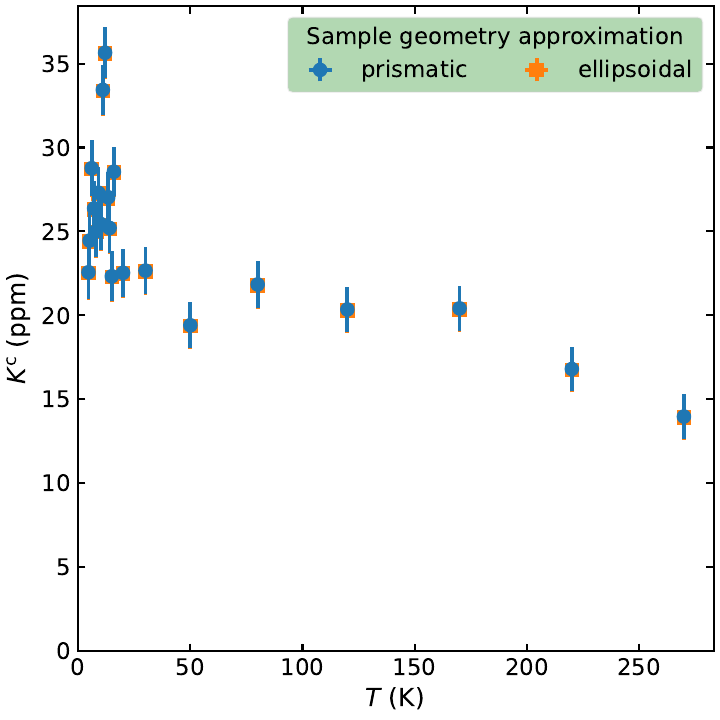}
	\caption{
	\label{fig:NMR-shift}
	Temperature $T$-dependence of the (corrected) \ch{^8Li} Knight shift $K^\mathrm{c}$
	in the \ch{Nb_{0.75}Ti_{0.25}N} thin film.
	The data are presented for two different sample geometry approximations:
	prismatic (blue circles) and ellipsoidal (orange squares).
	Both geometries yield nearly identical values across the measured temperature range. 
	}
\end{figure}

As described in \Cref{sec:results:resonance},
correcting for demagnetization contributions to the resonance position
are crucial for quantifying the Knight shift,
which can be calculated using \Cref{eq:corrected-shift,eq:shift-factor}
for the case when both the sample and reference
have different geometries
(see, e.g.,~\cite{2022-McFadden-ACIE-61-e202207137}).
Necessary for the computation are knowledge of each component's
demagnetization factor $N$ and (volume) magnetic susceptibility $\chi$.
Below,
we detail our estimates for these quantities.  

For thin films,
the demagnetization factor $ N = 1$~\cite{2008-Xu-JMR-191-47},
which applies to our \qty{91}{\nm} \ch{Nb_{0.75}Ti_{0.25}N} thin film sample.
For the reference compound \ch{MgO}, with dimensions \qtyproduct[product-units=power]{10 x 8 x 0.5}{\milli\metre},
the $N$ was calculated assuming both prismatic and ellipsoidal geometries~\cite{2018-Prozorov-PRA-10-014030}.
These calculations yielded consistent values of  $ N \approx 0.9$ ($ N = 0.922$ for a prism and $ N = 0.917$ for an ellipsoid),
confirming the minor geometric dependence.
The \gls{cgs} volume magnetic susceptibility of \ch{Nb_{0.75}Ti_{0.25}N} $\chi_0$  was calculated from the molar susceptibilities of its components,
\ch{NbN} ($\qty{31e-6}{\emu\per\centi\meter\cubed}$) and \ch{TiN} ($\qty{38e-6}{\emu\per\centi\meter\cubed}$)~\cite{2012-Toth-nitrie-book,1980-Rietschel-PRB-22-4284},
using their molar masses ($\qty{106.913}{\gram\per\mole}$) and ($\qty{61.874}{\gram\per\mole}$)~\cite{2016-CRC-handbook},
respectively.
Assuming our sample stoichiometry $x = 0.25$ and incorporating the measured density of the \ch{Nb_{0.75}Ti_{0.25}N} (\qty{6.5}{\gram\per\centi\meter\cubed}),
$ \chi_0$ was estimated to be $\qty{2.23e-6}{\emu\per\centi\meter\cubed}$.
For \ch{MgO}, the molar susceptibility $\chi_{\ch{MgO}}^{\mathrm{mol}} = \qty{-10.2e-6}{\emu\per\centi\meter\cubed}$~\cite{2016-CRC-handbook} is typical of weakly diamagnetic solids. This
was converted into a volume susceptibility $ \chi_{\ch{MgO}} = \qty{-9.11e-6}{\emu\centi\meter\cubed}$
using its density (\qty{3.6}{\gram\centi\meter\cubed}~\cite{2016-CRC-handbook}) and its molar mass (\qty{40.3044}{\gram\per\mol}~\cite{2016-CRC-handbook}).
Inserting these values into \Cref{eq:corrected-shift,eq:shift-factor}
along with the measured $\nu_{0}$s,
we obtain the corrected (Knight) shift $K^{c}$,
which varied between \qtyrange[retain-explicit-plus=true]{+15}{+35}{\ppm} over the measured
$T$-range
(see \Cref{fig:NMR-shift}),
with the correction for demagnetization
(\qty[retain-explicit-plus=true]{+25.4}{\ppm})
being the dominant contribution.

\bibliography{references.bib,new-refs.bib}

\begin{thebibliography}{143}%
\makeatletter
\providecommand \@ifxundefined [1]{%
 \@ifx{#1\undefined}
}%
\providecommand \@ifnum [1]{%
 \ifnum #1\expandafter \@firstoftwo
 \else \expandafter \@secondoftwo
 \fi
}%
\providecommand \@ifx [1]{%
 \ifx #1\expandafter \@firstoftwo
 \else \expandafter \@secondoftwo
 \fi
}%
\providecommand \natexlab [1]{#1}%
\providecommand \enquote  [1]{``#1''}%
\providecommand \bibnamefont  [1]{#1}%
\providecommand \bibfnamefont [1]{#1}%
\providecommand \citenamefont [1]{#1}%
\providecommand \href@noop [0]{\@secondoftwo}%
\providecommand \href [0]{\begingroup \@sanitize@url \@href}%
\providecommand \@href[1]{\@@startlink{#1}\@@href}%
\providecommand \@@href[1]{\endgroup#1\@@endlink}%
\providecommand \@sanitize@url [0]{\catcode `\\12\catcode `\$12\catcode `\&12\catcode `\#12\catcode `\^12\catcode `\_12\catcode `\%12\relax}%
\providecommand \@@startlink[1]{}%
\providecommand \@@endlink[0]{}%
\providecommand \url  [0]{\begingroup\@sanitize@url \@url }%
\providecommand \@url [1]{\endgroup\@href {#1}{\urlprefix }}%
\providecommand \urlprefix  [0]{URL }%
\providecommand \Eprint [0]{\href }%
\providecommand \doibase [0]{https://doi.org/}%
\providecommand \selectlanguage [0]{\@gobble}%
\providecommand \bibinfo  [0]{\@secondoftwo}%
\providecommand \bibfield  [0]{\@secondoftwo}%
\providecommand \translation [1]{[#1]}%
\providecommand \BibitemOpen [0]{}%
\providecommand \bibitemStop [0]{}%
\providecommand \bibitemNoStop [0]{.\EOS\space}%
\providecommand \EOS [0]{\spacefactor3000\relax}%
\providecommand \BibitemShut  [1]{\csname bibitem#1\endcsname}%
\let\auto@bib@innerbib\@empty
\bibitem [{\citenamefont {Gavaler}\ \emph {et~al.}(1969)\citenamefont {Gavaler}, \citenamefont {Deis}, \citenamefont {Hulm},\ and\ \citenamefont {Jones}}]{1969-Gavaler-JAP-15-329}%
  \BibitemOpen
  \bibfield  {author} {\bibinfo {author} {\bibfnamefont {J.~R.}\ \bibnamefont {Gavaler}}, \bibinfo {author} {\bibfnamefont {D.~W.}\ \bibnamefont {Deis}}, \bibinfo {author} {\bibfnamefont {J.~K.}\ \bibnamefont {Hulm}},\ and\ \bibinfo {author} {\bibfnamefont {C.~K.}\ \bibnamefont {Jones}},\ }\bibfield  {title} {\bibinfo {title} {{Superconducting Properties of Niobium-Titanium-Nitride Thin Films}},\ }\href {https://doi.org/10.1063/1.1652846} {\bibfield  {journal} {\bibinfo  {journal} {Appl. Phys. Lett.}\ }\textbf {\bibinfo {volume} {15}},\ \bibinfo {pages} {329} (\bibinfo {year} {1969})}\BibitemShut {NoStop}%
\bibitem [{\citenamefont {Benvenuti}\ \emph {et~al.}(1997)\citenamefont {Benvenuti}, \citenamefont {Chiggiato}, \citenamefont {Parrini},\ and\ \citenamefont {Russo}}]{1997-Benvenuti-NIMPRSB-124-106}%
  \BibitemOpen
  \bibfield  {author} {\bibinfo {author} {\bibfnamefont {C.}~\bibnamefont {Benvenuti}}, \bibinfo {author} {\bibfnamefont {P.}~\bibnamefont {Chiggiato}}, \bibinfo {author} {\bibfnamefont {L.}~\bibnamefont {Parrini}},\ and\ \bibinfo {author} {\bibfnamefont {R.}~\bibnamefont {Russo}},\ }\bibfield  {title} {\bibinfo {title} {{Production of niobium-titanium nitride coatings by reactive diffusion for superconducting cavity applications}},\ }\href {https://doi.org/10.1016/S0168-583X(97)00059-1} {\bibfield  {journal} {\bibinfo  {journal} {Nucl. Instrum. Methods Phys. Res., Sect. B}\ }\textbf {\bibinfo {volume} {124}},\ \bibinfo {pages} {106} (\bibinfo {year} {1997})}\BibitemShut {NoStop}%
\bibitem [{\citenamefont {Zhang}\ \emph {et~al.}(2015)\citenamefont {Zhang}, \citenamefont {Peng}, \citenamefont {You},\ and\ \citenamefont {Wang}}]{2015-Zhang-APL-107-122603}%
  \BibitemOpen
  \bibfield  {author} {\bibinfo {author} {\bibfnamefont {L.}~\bibnamefont {Zhang}}, \bibinfo {author} {\bibfnamefont {W.}~\bibnamefont {Peng}}, \bibinfo {author} {\bibfnamefont {L.~X.}\ \bibnamefont {You}},\ and\ \bibinfo {author} {\bibfnamefont {Z.}~\bibnamefont {Wang}},\ }\bibfield  {title} {\bibinfo {title} {{{Superconducting properties and chemical composition of NbTiN thin films with different thickness}}},\ }\href {https://doi.org/10.1063/1.4931943} {\bibfield  {journal} {\bibinfo  {journal} {Appl. Phys. Lett.}\ }\textbf {\bibinfo {volume} {107}},\ \bibinfo {pages} {122603} (\bibinfo {year} {2015})}\BibitemShut {NoStop}%
\bibitem [{\citenamefont {Pratap}\ \emph {et~al.}(2023)\citenamefont {Pratap}, \citenamefont {Nanda}, \citenamefont {Senapati}, \citenamefont {Aloysius},\ and\ \citenamefont {Achanta}}]{2023-Pratiksha-SST-36-085017}%
  \BibitemOpen
  \bibfield  {author} {\bibinfo {author} {\bibfnamefont {P.}~\bibnamefont {Pratap}}, \bibinfo {author} {\bibfnamefont {L.}~\bibnamefont {Nanda}}, \bibinfo {author} {\bibfnamefont {K.}~\bibnamefont {Senapati}}, \bibinfo {author} {\bibfnamefont {R.~P.}\ \bibnamefont {Aloysius}},\ and\ \bibinfo {author} {\bibfnamefont {V.}~\bibnamefont {Achanta}},\ }\bibfield  {title} {\bibinfo {title} {{Optimization of the superconducting properties of NbTiN thin films by variation of the N$_2$ partial pressure during sputter deposition}},\ }\href {https://doi.org/10.1088/1361-6668/ace3fa} {\bibfield  {journal} {\bibinfo  {journal} {Supercond. Sci. Technol.}\ }\textbf {\bibinfo {volume} {36}},\ \bibinfo {pages} {085017} (\bibinfo {year} {2023})}\BibitemShut {NoStop}%
\bibitem [{\citenamefont {Matenoglou}\ \emph {et~al.}(2009)\citenamefont {Matenoglou}, \citenamefont {Koutsokeras}, \citenamefont {Lekka}, \citenamefont {Abadias}, \citenamefont {Kosmidis}, \citenamefont {Evangelakis},\ and\ \citenamefont {Patsalas}}]{2009-Matenoglou-SCT-204-911}%
  \BibitemOpen
  \bibfield  {author} {\bibinfo {author} {\bibfnamefont {G.~M.}\ \bibnamefont {Matenoglou}}, \bibinfo {author} {\bibfnamefont {L.~E.}\ \bibnamefont {Koutsokeras}}, \bibinfo {author} {\bibfnamefont {C.~E.}\ \bibnamefont {Lekka}}, \bibinfo {author} {\bibfnamefont {G.}~\bibnamefont {Abadias}}, \bibinfo {author} {\bibfnamefont {C.}~\bibnamefont {Kosmidis}}, \bibinfo {author} {\bibfnamefont {G.~A.}\ \bibnamefont {Evangelakis}},\ and\ \bibinfo {author} {\bibfnamefont {P.}~\bibnamefont {Patsalas}},\ }\bibfield  {title} {\bibinfo {title} {Structure, stability and bonding of ternary transition metal nitrides},\ }\href {https://doi.org/10.1016/j.surfcoat.2009.06.032} {\bibfield  {journal} {\bibinfo  {journal} {Surf. Coat. Technol.}\ }\textbf {\bibinfo {volume} {204}},\ \bibinfo {pages} {911} (\bibinfo {year} {2009})}\BibitemShut {NoStop}%
\bibitem [{\citenamefont {Vasu}\ \emph {et~al.}(2012)\citenamefont {Vasu}, \citenamefont {Krishna},\ and\ \citenamefont {Padmanabhan}}]{2012-Vasu-JMS-47-3522}%
  \BibitemOpen
  \bibfield  {author} {\bibinfo {author} {\bibfnamefont {K.}~\bibnamefont {Vasu}}, \bibinfo {author} {\bibfnamefont {M.~G.}\ \bibnamefont {Krishna}},\ and\ \bibinfo {author} {\bibfnamefont {K.~A.}\ \bibnamefont {Padmanabhan}},\ }\bibfield  {title} {\bibinfo {title} {Effect of \ch{Nb} concentration on the structure{,} mechanical{,} optical{,} and electrical properties of nano-crystalline \ch{Ti_{1-x}Nb_{x}N} thin films},\ }\href {https://doi.org/10.1007/s10853-011-6197-x} {\bibfield  {journal} {\bibinfo  {journal} {J. Mater. Sci.}\ }\textbf {\bibinfo {volume} {47}},\ \bibinfo {pages} {3522} (\bibinfo {year} {2012})}\BibitemShut {NoStop}%
\bibitem [{\citenamefont {Toth}(2012)}]{2012-Toth-nitrie-book}%
  \BibitemOpen
  \bibfield  {author} {\bibinfo {author} {\bibfnamefont {L.}~\bibnamefont {Toth}},\ }\href {https://shop.elsevier.com/books/transition-metal-carbides-and-nitrides/toth/978-0-12-695950-5} {\emph {\bibinfo {title} {{Transition Metal Carbides and Nitrides}}}}\ (\bibinfo  {publisher} {Academic Press},\ \bibinfo {year} {2012})\BibitemShut {NoStop}%
\bibitem [{\citenamefont {Kikkawa}\ \emph {et~al.}(1996)\citenamefont {Kikkawa}, \citenamefont {Yamamoto}, \citenamefont {Ohta}, \citenamefont {Takahashi},\ and\ \citenamefont {Kanamaru}}]{1996-Kikkawa-CTMCN-9-175}%
  \BibitemOpen
  \bibfield  {author} {\bibinfo {author} {\bibfnamefont {S.}~\bibnamefont {Kikkawa}}, \bibinfo {author} {\bibfnamefont {T.}~\bibnamefont {Yamamoto}}, \bibinfo {author} {\bibfnamefont {K.}~\bibnamefont {Ohta}}, \bibinfo {author} {\bibfnamefont {M.}~\bibnamefont {Takahashi}},\ and\ \bibinfo {author} {\bibfnamefont {F.}~\bibnamefont {Kanamaru}},\ }\bibfield  {title} {\bibinfo {title} {Transition metal-based double nitrides},\ }in\ \href {https://doi.org/10.1007/978-94-009-1565-7_9} {\emph {\bibinfo {booktitle} {The Chemistry of Transition Metal Carbides and Nitrides}}},\ \bibinfo {editor} {edited by\ \bibinfo {editor} {\bibfnamefont {S.~T.}\ \bibnamefont {Oyama}}}\ (\bibinfo  {publisher} {Springer},\ \bibinfo {address} {Dordrecht},\ \bibinfo {year} {1996})\ Chap.~\bibinfo {chapter} {9}, pp.\ \bibinfo {pages} {175--190}\BibitemShut {NoStop}%
\bibitem [{\citenamefont {Zbasnik}\ \emph {et~al.}(1969)\citenamefont {Zbasnik}, \citenamefont {Toth}, \citenamefont {Shy},\ and\ \citenamefont {Maxwell}}]{1969-Zbasnik-JAP-40-2147}%
  \BibitemOpen
  \bibfield  {author} {\bibinfo {author} {\bibfnamefont {J.}~\bibnamefont {Zbasnik}}, \bibinfo {author} {\bibfnamefont {L.~E.}\ \bibnamefont {Toth}}, \bibinfo {author} {\bibfnamefont {Y.~M.}\ \bibnamefont {Shy}},\ and\ \bibinfo {author} {\bibfnamefont {E.}~\bibnamefont {Maxwell}},\ }\bibfield  {title} {\bibinfo {title} {{Superconducting Critical Fields and Currents of Nb–Ti–N Thin Films in Continuous Magnetic Fields to 175 kG}},\ }\href {https://doi.org/10.1063/1.1657940} {\bibfield  {journal} {\bibinfo  {journal} {J. Appl. Phys.}\ }\textbf {\bibinfo {volume} {40}},\ \bibinfo {pages} {2147} (\bibinfo {year} {1969})}\BibitemShut {NoStop}%
\bibitem [{\citenamefont {Makise}\ \emph {et~al.}(2011)\citenamefont {Makise}, \citenamefont {Terai}, \citenamefont {Takeda}, \citenamefont {Uzawa},\ and\ \citenamefont {Wang}}]{2011-Makise-IEEETAS-21-139}%
  \BibitemOpen
  \bibfield  {author} {\bibinfo {author} {\bibfnamefont {K.}~\bibnamefont {Makise}}, \bibinfo {author} {\bibfnamefont {H.}~\bibnamefont {Terai}}, \bibinfo {author} {\bibfnamefont {M.}~\bibnamefont {Takeda}}, \bibinfo {author} {\bibfnamefont {Y.}~\bibnamefont {Uzawa}},\ and\ \bibinfo {author} {\bibfnamefont {Z.}~\bibnamefont {Wang}},\ }\bibfield  {title} {\bibinfo {title} {{Characterization of NbTiN Thin Films Deposited on Various Substrates}},\ }\href {https://doi.org/10.1109/TASC.2010.2088350} {\bibfield  {journal} {\bibinfo  {journal} {IEEE Trans. Appl. Supercond.}\ }\textbf {\bibinfo {volume} {21}},\ \bibinfo {pages} {139} (\bibinfo {year} {2011})}\BibitemShut {NoStop}%
\bibitem [{\citenamefont {Hazra}\ \emph {et~al.}(2018)\citenamefont {Hazra}, \citenamefont {Tsavdaris}, \citenamefont {Mukhtarova}, \citenamefont {Jacquemin}, \citenamefont {Blanchet}, \citenamefont {Albert}, \citenamefont {Jebari}, \citenamefont {Grimm}, \citenamefont {Konar}, \citenamefont {Blanquet}, \citenamefont {Mercier}, \citenamefont {Chapelier},\ and\ \citenamefont {Hofheinz}}]{2018-Hazra-PRB-97-144518}%
  \BibitemOpen
  \bibfield  {author} {\bibinfo {author} {\bibfnamefont {D.}~\bibnamefont {Hazra}}, \bibinfo {author} {\bibfnamefont {N.}~\bibnamefont {Tsavdaris}}, \bibinfo {author} {\bibfnamefont {A.}~\bibnamefont {Mukhtarova}}, \bibinfo {author} {\bibfnamefont {M.}~\bibnamefont {Jacquemin}}, \bibinfo {author} {\bibfnamefont {F.}~\bibnamefont {Blanchet}}, \bibinfo {author} {\bibfnamefont {R.}~\bibnamefont {Albert}}, \bibinfo {author} {\bibfnamefont {S.}~\bibnamefont {Jebari}}, \bibinfo {author} {\bibfnamefont {A.}~\bibnamefont {Grimm}}, \bibinfo {author} {\bibfnamefont {A.}~\bibnamefont {Konar}}, \bibinfo {author} {\bibfnamefont {E.}~\bibnamefont {Blanquet}}, \bibinfo {author} {\bibfnamefont {F.}~\bibnamefont {Mercier}}, \bibinfo {author} {\bibfnamefont {C.}~\bibnamefont {Chapelier}},\ and\ \bibinfo {author} {\bibfnamefont {M.}~\bibnamefont {Hofheinz}},\ }\bibfield  {title} {\bibinfo {title} {Superconducting properties of {NbTiN} thin films deposited by high-temperature chemical vapor deposition},\ }\href {https://doi.org/10.1103/PhysRevB.97.144518} {\bibfield  {journal} {\bibinfo  {journal} {Phys. Rev. B}\ }\textbf {\bibinfo {volume} {97}},\ \bibinfo {pages} {144518} (\bibinfo {year} {2018})}\BibitemShut {NoStop}%
\bibitem [{\citenamefont {Cyberey}\ \emph {et~al.}(2019)\citenamefont {Cyberey}, \citenamefont {Farrahi}, \citenamefont {Lu}, \citenamefont {Kerr}, \citenamefont {Weikle},\ and\ \citenamefont {Lichtenberger}}]{2019-Cyberey-IEEETAS-29-1}%
  \BibitemOpen
  \bibfield  {author} {\bibinfo {author} {\bibfnamefont {M.}~\bibnamefont {Cyberey}}, \bibinfo {author} {\bibfnamefont {T.}~\bibnamefont {Farrahi}}, \bibinfo {author} {\bibfnamefont {J.}~\bibnamefont {Lu}}, \bibinfo {author} {\bibfnamefont {A.}~\bibnamefont {Kerr}}, \bibinfo {author} {\bibfnamefont {R.~M.}\ \bibnamefont {Weikle}},\ and\ \bibinfo {author} {\bibfnamefont {A.~W.}\ \bibnamefont {Lichtenberger}},\ }\bibfield  {title} {\bibinfo {title} {{NbTiN/AlN/NbTiN SIS Junctions Realized by Reactive Bias Target Ion Beam Deposition}},\ }\href {https://doi.org/10.1109/TASC.2018.2884967} {\bibfield  {journal} {\bibinfo  {journal} {IEEE Trans. Appl. Supercond.}\ }\textbf {\bibinfo {volume} {29}},\ \bibinfo {pages} {1} (\bibinfo {year} {2019})}\BibitemShut {NoStop}%
\bibitem [{\citenamefont {Cyberey}\ \emph {et~al.}(2024)\citenamefont {Cyberey}, \citenamefont {Hinton}, \citenamefont {Moore}, \citenamefont {Weikle},\ and\ \citenamefont {Lichtenberger}}]{2024-Cyberey-IEETAS-34-1}%
  \BibitemOpen
  \bibfield  {author} {\bibinfo {author} {\bibfnamefont {M.}~\bibnamefont {Cyberey}}, \bibinfo {author} {\bibfnamefont {S.}~\bibnamefont {Hinton}}, \bibinfo {author} {\bibfnamefont {C.}~\bibnamefont {Moore}}, \bibinfo {author} {\bibfnamefont {R.~M.}\ \bibnamefont {Weikle}},\ and\ \bibinfo {author} {\bibfnamefont {A.~W.}\ \bibnamefont {Lichtenberger}},\ }\bibfield  {title} {\bibinfo {title} {{SuperGaN: Synthesis of NbTiN/GaN/NbTiN Tunnel Junctions}},\ }\href {https://doi.org/10.1109/TASC.2023.3337764} {\bibfield  {journal} {\bibinfo  {journal} {IEEE Trans. Appl. Supercond.}\ }\textbf {\bibinfo {volume} {34}},\ \bibinfo {pages} {1} (\bibinfo {year} {2024})}\BibitemShut {NoStop}%
\bibitem [{\citenamefont {Zhukova}\ \emph {et~al.}(2024)\citenamefont {Zhukova}, \citenamefont {Gorshunov}, \citenamefont {Kadyrov}, \citenamefont {Zhivetev}, \citenamefont {Terentiev}, \citenamefont {Chekushkin}, \citenamefont {Khan}, \citenamefont {Khudchenko}, \citenamefont {Kinev},\ and\ \citenamefont {Koshelets}}]{2024-Zhukova-IEETAS-34-1}%
  \BibitemOpen
  \bibfield  {author} {\bibinfo {author} {\bibfnamefont {E.~S.}\ \bibnamefont {Zhukova}}, \bibinfo {author} {\bibfnamefont {B.~P.}\ \bibnamefont {Gorshunov}}, \bibinfo {author} {\bibfnamefont {L.~S.}\ \bibnamefont {Kadyrov}}, \bibinfo {author} {\bibfnamefont {K.~V.}\ \bibnamefont {Zhivetev}}, \bibinfo {author} {\bibfnamefont {A.~V.}\ \bibnamefont {Terentiev}}, \bibinfo {author} {\bibfnamefont {A.~M.}\ \bibnamefont {Chekushkin}}, \bibinfo {author} {\bibfnamefont {F.~V.}\ \bibnamefont {Khan}}, \bibinfo {author} {\bibfnamefont {A.~V.}\ \bibnamefont {Khudchenko}}, \bibinfo {author} {\bibfnamefont {N.~V.}\ \bibnamefont {Kinev}},\ and\ \bibinfo {author} {\bibfnamefont {V.~P.}\ \bibnamefont {Koshelets}},\ }\bibfield  {title} {\bibinfo {title} {{Impact of the Buffer Layers and Anodization on Properties of NbTiN Films for THz Receivers}},\ }\href {https://doi.org/10.1109/TASC.2024.3353139} {\bibfield  {journal} {\bibinfo  {journal} {IEEE Trans. Appl. Supercond.}\ }\textbf {\bibinfo {volume} {34}},\ \bibinfo {pages} {1} (\bibinfo {year} {2024})}\BibitemShut {NoStop}%
\bibitem [{\citenamefont {Uzawa}\ \emph {et~al.}(2015)\citenamefont {Uzawa}, \citenamefont {Fujii}, \citenamefont {Gonzalez}, \citenamefont {Kaneko}, \citenamefont {Kroug}, \citenamefont {Kojima}, \citenamefont {Miyachi}, \citenamefont {Makise}, \citenamefont {Saito}, \citenamefont {Terai},\ and\ \citenamefont {Wang}}]{2015-Uzawa-IEEETAS-25-1}%
  \BibitemOpen
  \bibfield  {author} {\bibinfo {author} {\bibfnamefont {Y.}~\bibnamefont {Uzawa}}, \bibinfo {author} {\bibfnamefont {Y.}~\bibnamefont {Fujii}}, \bibinfo {author} {\bibfnamefont {A.}~\bibnamefont {Gonzalez}}, \bibinfo {author} {\bibfnamefont {K.}~\bibnamefont {Kaneko}}, \bibinfo {author} {\bibfnamefont {M.}~\bibnamefont {Kroug}}, \bibinfo {author} {\bibfnamefont {T.}~\bibnamefont {Kojima}}, \bibinfo {author} {\bibfnamefont {A.}~\bibnamefont {Miyachi}}, \bibinfo {author} {\bibfnamefont {K.}~\bibnamefont {Makise}}, \bibinfo {author} {\bibfnamefont {S.}~\bibnamefont {Saito}}, \bibinfo {author} {\bibfnamefont {H.}~\bibnamefont {Terai}},\ and\ \bibinfo {author} {\bibfnamefont {Z.}~\bibnamefont {Wang}},\ }\bibfield  {title} {\bibinfo {title} {{Tuning Circuit Material for Mass-Produced Terahertz SIS Receivers}},\ }\href {https://doi.org/10.1109/TASC.2014.2386211} {\bibfield  {journal} {\bibinfo  {journal} {IEEE Trans. Appl. Supercond.}\ }\textbf {\bibinfo {volume} {25}},\ \bibinfo {pages} {1} (\bibinfo {year} {2015})}\BibitemShut {NoStop}%
\bibitem [{\citenamefont {Westig}\ \emph {et~al.}(2013)\citenamefont {Westig}, \citenamefont {Selig}, \citenamefont {Jacobs}, \citenamefont {Klapwijk},\ and\ \citenamefont {Honingh}}]{2013-Westig-JAP-114-124504}%
  \BibitemOpen
  \bibfield  {author} {\bibinfo {author} {\bibfnamefont {M.~P.}\ \bibnamefont {Westig}}, \bibinfo {author} {\bibfnamefont {S.}~\bibnamefont {Selig}}, \bibinfo {author} {\bibfnamefont {K.}~\bibnamefont {Jacobs}}, \bibinfo {author} {\bibfnamefont {T.~M.}\ \bibnamefont {Klapwijk}},\ and\ \bibinfo {author} {\bibfnamefont {C.~E.}\ \bibnamefont {Honingh}},\ }\bibfield  {title} {\bibinfo {title} {{Improved Nb SIS devices for heterodyne mixers between 700 GHz and 1.3 THz with NbTiN transmission lines using a normal metal energy relaxation layer}},\ }\href {https://doi.org/10.1063/1.4822167} {\bibfield  {journal} {\bibinfo  {journal} {J. Appl. Phys.}\ }\textbf {\bibinfo {volume} {114}},\ \bibinfo {pages} {124504} (\bibinfo {year} {2013})}\BibitemShut {NoStop}%
\bibitem [{\citenamefont {Khan}\ \emph {et~al.}(2023)\citenamefont {Khan}, \citenamefont {Zhukova}, \citenamefont {Gorshunov}, \citenamefont {Kadyrov}, \citenamefont {Chekushkin}, \citenamefont {Khudchenko},\ and\ \citenamefont {Koshelets}}]{2023-Fedor-IEETTST-13-627}%
  \BibitemOpen
  \bibfield  {author} {\bibinfo {author} {\bibfnamefont {F.~V.}\ \bibnamefont {Khan}}, \bibinfo {author} {\bibfnamefont {E.~S.}\ \bibnamefont {Zhukova}}, \bibinfo {author} {\bibfnamefont {B.~P.}\ \bibnamefont {Gorshunov}}, \bibinfo {author} {\bibfnamefont {L.~S.}\ \bibnamefont {Kadyrov}}, \bibinfo {author} {\bibfnamefont {A.~M.}\ \bibnamefont {Chekushkin}}, \bibinfo {author} {\bibfnamefont {A.~V.}\ \bibnamefont {Khudchenko}},\ and\ \bibinfo {author} {\bibfnamefont {V.~P.}\ \bibnamefont {Koshelets}},\ }\bibfield  {title} {\bibinfo {title} {{Characterization of Microwave Properties of Superconducting NbTiN Films Using TDS}},\ }\href {https://doi.org/10.1109/TTHZ.2023.3321252} {\bibfield  {journal} {\bibinfo  {journal} {IEEE Trans. Terahertz Sci. Technol.}\ }\textbf {\bibinfo {volume} {13}},\ \bibinfo {pages} {627} (\bibinfo {year} {2023})}\BibitemShut {NoStop}%
\bibitem [{\citenamefont {Valente-Feliciano}(2016)}]{2016-Anne-Marie-SST-29-113002}%
  \BibitemOpen
  \bibfield  {author} {\bibinfo {author} {\bibfnamefont {A.-M.}\ \bibnamefont {Valente-Feliciano}},\ }\bibfield  {title} {\bibinfo {title} {Superconducting {RF} materials other than bulk niobium: a review},\ }\href {https://doi.org/10.1088/0953-2048/29/11/113002} {\bibfield  {journal} {\bibinfo  {journal} {Supercond. Sci. Technol.}\ }\textbf {\bibinfo {volume} {29}},\ \bibinfo {pages} {113002} (\bibinfo {year} {2016})}\BibitemShut {NoStop}%
\bibitem [{\citenamefont {Padamsee}(2023)}]{2023-Padamsee-SRTA}%
  \BibitemOpen
  \bibfield  {author} {\bibinfo {author} {\bibfnamefont {H.}~\bibnamefont {Padamsee}},\ }\href {https://doi.org/10.1002/9783527836314} {\emph {\bibinfo {title} {Superconducting Radiofrequency Technology for Accelerators: State of the Art and Emerging Trends}}}\ (\bibinfo  {publisher} {Wiley},\ \bibinfo {address} {Weinheim},\ \bibinfo {year} {2023})\BibitemShut {NoStop}%
\bibitem [{\citenamefont {Momma}\ and\ \citenamefont {Izumi}(2011)}]{2011-Momma-JAC-44-1272}%
  \BibitemOpen
  \bibfield  {author} {\bibinfo {author} {\bibfnamefont {K.}~\bibnamefont {Momma}}\ and\ \bibinfo {author} {\bibfnamefont {F.}~\bibnamefont {Izumi}},\ }\bibfield  {title} {\bibinfo {title} {{{\it VESTA3} for three-dimensional visualization of crystal, volumetric and morphology data}},\ }\href {https://doi.org/10.1107/S0021889811038970} {\bibfield  {journal} {\bibinfo  {journal} {J. Appl. Cryst.}\ }\textbf {\bibinfo {volume} {44}},\ \bibinfo {pages} {1272} (\bibinfo {year} {2011})}\BibitemShut {NoStop}%
\bibitem [{\citenamefont {Posen}\ \emph {et~al.}(2015)\citenamefont {Posen}, \citenamefont {Valles},\ and\ \citenamefont {Liepe}}]{2015-Posen-PRL-115-047001}%
  \BibitemOpen
  \bibfield  {author} {\bibinfo {author} {\bibfnamefont {S.}~\bibnamefont {Posen}}, \bibinfo {author} {\bibfnamefont {N.}~\bibnamefont {Valles}},\ and\ \bibinfo {author} {\bibfnamefont {M.}~\bibnamefont {Liepe}},\ }\bibfield  {title} {\bibinfo {title} {Radio frequency magnetic field limits of \ch{Nb} and \ch{Nb3Sn}},\ }\href {https://doi.org/10.1103/PhysRevLett.115.047001} {\bibfield  {journal} {\bibinfo  {journal} {Phys. Rev. Lett.}\ }\textbf {\bibinfo {volume} {115}},\ \bibinfo {pages} {047001} (\bibinfo {year} {2015})}\BibitemShut {NoStop}%
\bibitem [{\citenamefont {Junginger}\ \emph {et~al.}(2017)\citenamefont {Junginger}, \citenamefont {Wasserman},\ and\ \citenamefont {Laxdal}}]{2017-Junginger-SST-30-125012}%
  \BibitemOpen
  \bibfield  {author} {\bibinfo {author} {\bibfnamefont {T.}~\bibnamefont {Junginger}}, \bibinfo {author} {\bibfnamefont {W.}~\bibnamefont {Wasserman}},\ and\ \bibinfo {author} {\bibfnamefont {R.~E.}\ \bibnamefont {Laxdal}},\ }\bibfield  {title} {\bibinfo {title} {Superheating in coated niobium},\ }\href {https://doi.org/10.1088/1361-6668/aa8e3a} {\bibfield  {journal} {\bibinfo  {journal} {Supercond. Sci. Technol.}\ }\textbf {\bibinfo {volume} {30}},\ \bibinfo {pages} {125012} (\bibinfo {year} {2017})}\BibitemShut {NoStop}%
\bibitem [{\citenamefont {Gurevich}(2006)}]{2006-Gurevich-APL-88-012511}%
  \BibitemOpen
  \bibfield  {author} {\bibinfo {author} {\bibfnamefont {A.}~\bibnamefont {Gurevich}},\ }\bibfield  {title} {\bibinfo {title} {{Enhancement of rf breakdown field of superconductors by multilayer coating}},\ }\href {https://doi.org/10.1063/1.2162264} {\bibfield  {journal} {\bibinfo  {journal} {Appl. Phys. Lett.}\ }\textbf {\bibinfo {volume} {88}},\ \bibinfo {pages} {012511} (\bibinfo {year} {2006})}\BibitemShut {NoStop}%
\bibitem [{\citenamefont {Kubo}\ \emph {et~al.}(2014)\citenamefont {Kubo}, \citenamefont {Iwashita},\ and\ \citenamefont {Saeki}}]{2014-Kubo-APL-104-032603}%
  \BibitemOpen
  \bibfield  {author} {\bibinfo {author} {\bibfnamefont {T.}~\bibnamefont {Kubo}}, \bibinfo {author} {\bibfnamefont {Y.}~\bibnamefont {Iwashita}},\ and\ \bibinfo {author} {\bibfnamefont {T.}~\bibnamefont {Saeki}},\ }\bibfield  {title} {\bibinfo {title} {Radio-frequency electromagnetic field and vortex penetration in multilayered superconductors},\ }\href {https://doi.org/10.1063/1.4862892} {\bibfield  {journal} {\bibinfo  {journal} {Appl. Phys. Lett.}\ }\textbf {\bibinfo {volume} {104}},\ \bibinfo {pages} {032603} (\bibinfo {year} {2014})}\BibitemShut {NoStop}%
\bibitem [{\citenamefont {Kubo}(2017)}]{2017-Kubo-SST-30-023001}%
  \BibitemOpen
  \bibfield  {author} {\bibinfo {author} {\bibfnamefont {T.}~\bibnamefont {Kubo}},\ }\bibfield  {title} {\bibinfo {title} {Multilayer coating for higher accelerating fields in superconducting radio-frequency cavities: a review of theoretical aspects},\ }\href {https://doi.org/10.1088/1361-6668/30/2/023001} {\bibfield  {journal} {\bibinfo  {journal} {Supercond. Sci. Technol.}\ }\textbf {\bibinfo {volume} {30}},\ \bibinfo {pages} {023001} (\bibinfo {year} {2017})}\BibitemShut {NoStop}%
\bibitem [{\citenamefont {Kubo}(2019)}]{2019-Kubo-JJAP-58-088001}%
  \BibitemOpen
  \bibfield  {author} {\bibinfo {author} {\bibfnamefont {T.}~\bibnamefont {Kubo}},\ }\bibfield  {title} {\bibinfo {title} {{Optimum multilayer coating of superconducting particle accelerator cavities and effects of thickness dependent material properties of thin films}},\ }\href {https://doi.org/10.7567/1347-4065/ab2f0a} {\bibfield  {journal} {\bibinfo  {journal} {Jpn. J. Appl. Phys.}\ }\textbf {\bibinfo {volume} {58}},\ \bibinfo {pages} {088001} (\bibinfo {year} {2019})}\BibitemShut {NoStop}%
\bibitem [{\citenamefont {Kubo}(2021)}]{2021-Kubo-SST-34-045006}%
  \BibitemOpen
  \bibfield  {author} {\bibinfo {author} {\bibfnamefont {T.}~\bibnamefont {Kubo}},\ }\bibfield  {title} {\bibinfo {title} {{Superheating fields of semi-infinite superconductors and layered superconductors in the diffusive limit: structural optimization based on the microscopic theory}},\ }\href {https://doi.org/10.1088/1361-6668/abdedd} {\bibfield  {journal} {\bibinfo  {journal} {Supercond. Sci. Technol.}\ }\textbf {\bibinfo {volume} {34}},\ \bibinfo {pages} {045006} (\bibinfo {year} {2021})}\BibitemShut {NoStop}%
\bibitem [{\citenamefont {Tan}\ \emph {et~al.}(2016)\citenamefont {Tan}, \citenamefont {Wolak}, \citenamefont {Xi}, \citenamefont {Tajima},\ and\ \citenamefont {Civale}}]{2016-Tan-SR-6-35879}%
  \BibitemOpen
  \bibfield  {author} {\bibinfo {author} {\bibfnamefont {T.}~\bibnamefont {Tan}}, \bibinfo {author} {\bibfnamefont {M.~A.}\ \bibnamefont {Wolak}}, \bibinfo {author} {\bibfnamefont {X.~X.}\ \bibnamefont {Xi}}, \bibinfo {author} {\bibfnamefont {T.}~\bibnamefont {Tajima}},\ and\ \bibinfo {author} {\bibfnamefont {L.}~\bibnamefont {Civale}},\ }\bibfield  {title} {\bibinfo {title} {Magnesium diboride coated bulk niobium: a new approach to higher acceleration gradient},\ }\href {https://doi.org/10.1038/srep35879} {\bibfield  {journal} {\bibinfo  {journal} {Sci. Rep.}\ }\textbf {\bibinfo {volume} {6}},\ \bibinfo {pages} {35879} (\bibinfo {year} {2016})}\BibitemShut {NoStop}%
\bibitem [{\citenamefont {Antoine}\ \emph {et~al.}(2019)\citenamefont {Antoine}, \citenamefont {Aburas}, \citenamefont {Four}, \citenamefont {Weiss}, \citenamefont {Iwashita}, \citenamefont {Hayano}, \citenamefont {Kato}, \citenamefont {Kubo},\ and\ \citenamefont {Saeki}}]{2019-Antoine-SST-32-085005}%
  \BibitemOpen
  \bibfield  {author} {\bibinfo {author} {\bibfnamefont {C.~Z.}\ \bibnamefont {Antoine}}, \bibinfo {author} {\bibfnamefont {M.}~\bibnamefont {Aburas}}, \bibinfo {author} {\bibfnamefont {A.}~\bibnamefont {Four}}, \bibinfo {author} {\bibfnamefont {F.}~\bibnamefont {Weiss}}, \bibinfo {author} {\bibfnamefont {Y.}~\bibnamefont {Iwashita}}, \bibinfo {author} {\bibfnamefont {H.}~\bibnamefont {Hayano}}, \bibinfo {author} {\bibfnamefont {S.}~\bibnamefont {Kato}}, \bibinfo {author} {\bibfnamefont {T.}~\bibnamefont {Kubo}},\ and\ \bibinfo {author} {\bibfnamefont {T.}~\bibnamefont {Saeki}},\ }\bibfield  {title} {\bibinfo {title} {Optimization of tailored multilayer superconductors for rf application and protection against premature vortex penetration},\ }\href {https://doi.org/10.1088/1361-6668/ab1bf1} {\bibfield  {journal} {\bibinfo  {journal} {Supercond. Sci. Technol.}\ }\textbf {\bibinfo {volume} {32}},\ \bibinfo {pages} {085005} (\bibinfo {year} {2019})}\BibitemShut {NoStop}%
\bibitem [{\citenamefont {Asaduzzaman}\ \emph {et~al.}(2023)\citenamefont {Asaduzzaman}, \citenamefont {McFadden}, \citenamefont {Valente-Feliciano}, \citenamefont {Beverstock}, \citenamefont {Suter}, \citenamefont {Salman}, \citenamefont {Prokscha},\ and\ \citenamefont {Junginger}}]{2024-Asaduzzaman-SST-37-025002}%
  \BibitemOpen
  \bibfield  {author} {\bibinfo {author} {\bibfnamefont {M.}~\bibnamefont {Asaduzzaman}}, \bibinfo {author} {\bibfnamefont {R.~M.~L.}\ \bibnamefont {McFadden}}, \bibinfo {author} {\bibfnamefont {A.-M.}\ \bibnamefont {Valente-Feliciano}}, \bibinfo {author} {\bibfnamefont {D.~R.}\ \bibnamefont {Beverstock}}, \bibinfo {author} {\bibfnamefont {A.}~\bibnamefont {Suter}}, \bibinfo {author} {\bibfnamefont {Z.}~\bibnamefont {Salman}}, \bibinfo {author} {\bibfnamefont {T.}~\bibnamefont {Prokscha}},\ and\ \bibinfo {author} {\bibfnamefont {T.}~\bibnamefont {Junginger}},\ }\bibfield  {title} {\bibinfo {title} {Evidence for current suppression in superconductor–superconductor bilayers},\ }\href {https://doi.org/10.1088/1361-6668/ad1462} {\bibfield  {journal} {\bibinfo  {journal} {Supercond. Sci. Technol.}\ }\textbf {\bibinfo {volume} {37}},\ \bibinfo {pages} {025002} (\bibinfo {year} {2023})}\BibitemShut {NoStop}%
\bibitem [{\citenamefont {Asaduzzaman}\ \emph {et~al.}(2024)\citenamefont {Asaduzzaman}, \citenamefont {McFadden}, \citenamefont {Thoeng}, \citenamefont {Laxdal},\ and\ \citenamefont {Junginger}}]{2024-Asaduzzaman-SST-37-085006}%
  \BibitemOpen
  \bibfield  {author} {\bibinfo {author} {\bibfnamefont {M.}~\bibnamefont {Asaduzzaman}}, \bibinfo {author} {\bibfnamefont {R.~M.~L.}\ \bibnamefont {McFadden}}, \bibinfo {author} {\bibfnamefont {E.}~\bibnamefont {Thoeng}}, \bibinfo {author} {\bibfnamefont {R.~E.}\ \bibnamefont {Laxdal}},\ and\ \bibinfo {author} {\bibfnamefont {T.}~\bibnamefont {Junginger}},\ }\bibfield  {title} {\bibinfo {title} {{Measurements of the first-flux-penetration field in surface-treated and coated Nb: distinguishing between near-surface pinning and an interface energy barrier}},\ }\href {https://doi.org/10.1088/1361-6668/ad54f3} {\bibfield  {journal} {\bibinfo  {journal} {Supercond. Sci. Technol.}\ }\textbf {\bibinfo {volume} {37}},\ \bibinfo {pages} {085006} (\bibinfo {year} {2024})}\BibitemShut {NoStop}%
\bibitem [{\citenamefont {Roach}\ \emph {et~al.}(2013)\citenamefont {Roach}, \citenamefont {Beringer}, \citenamefont {Li}, \citenamefont {Clavero},\ and\ \citenamefont {Lukaszew}}]{2023-Roach-IEETAS-23-8600203}%
  \BibitemOpen
  \bibfield  {author} {\bibinfo {author} {\bibfnamefont {W.~M.}\ \bibnamefont {Roach}}, \bibinfo {author} {\bibfnamefont {D.~B.}\ \bibnamefont {Beringer}}, \bibinfo {author} {\bibfnamefont {Z.}~\bibnamefont {Li}}, \bibinfo {author} {\bibfnamefont {C.}~\bibnamefont {Clavero}},\ and\ \bibinfo {author} {\bibfnamefont {R.~A.}\ \bibnamefont {Lukaszew}},\ }\bibfield  {title} {\bibinfo {title} {{Magnetic Shielding Larger Than the Lower Critical Field of Niobium in Multilayers}},\ }\href {https://doi.org/10.1109/TASC.2012.2234956} {\bibfield  {journal} {\bibinfo  {journal} {IEEE Trans. Appl. Supercond.}\ }\textbf {\bibinfo {volume} {23}},\ \bibinfo {pages} {8600203} (\bibinfo {year} {2013})}\BibitemShut {NoStop}%
\bibitem [{\citenamefont {Cody}\ and\ \citenamefont {Miller}(1968)}]{1968-Cody-PR-173-481}%
  \BibitemOpen
  \bibfield  {author} {\bibinfo {author} {\bibfnamefont {G.~D.}\ \bibnamefont {Cody}}\ and\ \bibinfo {author} {\bibfnamefont {R.~E.}\ \bibnamefont {Miller}},\ }\bibfield  {title} {\bibinfo {title} {{Magnetic Transitions of Superconducting Thin Films and Foils. {I}. Lead}},\ }\href {https://doi.org/10.1103/PhysRev.173.481} {\bibfield  {journal} {\bibinfo  {journal} {Phys. Rev.}\ }\textbf {\bibinfo {volume} {173}},\ \bibinfo {pages} {481} (\bibinfo {year} {1968})}\BibitemShut {NoStop}%
\bibitem [{\citenamefont {Bell}\ \emph {et~al.}(1968)\citenamefont {Bell}, \citenamefont {Shy}, \citenamefont {Anderson},\ and\ \citenamefont {Toth}}]{1968-Bell-JAP-39-2797}%
  \BibitemOpen
  \bibfield  {author} {\bibinfo {author} {\bibfnamefont {H.}~\bibnamefont {Bell}}, \bibinfo {author} {\bibfnamefont {Y.~M.}\ \bibnamefont {Shy}}, \bibinfo {author} {\bibfnamefont {D.~E.}\ \bibnamefont {Anderson}},\ and\ \bibinfo {author} {\bibfnamefont {L.~E.}\ \bibnamefont {Toth}},\ }\bibfield  {title} {\bibinfo {title} {{Superconducting Properties of Reactively Sputtered Thin‐Film Ternary Nitrides, Nb–Ti–N and Nb–Zr–N}},\ }\href {https://doi.org/10.1063/1.1656676} {\bibfield  {journal} {\bibinfo  {journal} {J. Appl. Phys.}\ }\textbf {\bibinfo {volume} {39}},\ \bibinfo {pages} {2797} (\bibinfo {year} {1968})}\BibitemShut {NoStop}%
\bibitem [{\citenamefont {{Di Leo}}\ \emph {et~al.}(1990)\citenamefont {{Di Leo}}, \citenamefont {Nigro}, \citenamefont {Nobile},\ and\ \citenamefont {Vaglio}}]{1990-DiLeo-JLTP-78-41}%
  \BibitemOpen
  \bibfield  {author} {\bibinfo {author} {\bibfnamefont {R.}~\bibnamefont {{Di Leo}}}, \bibinfo {author} {\bibfnamefont {A.}~\bibnamefont {Nigro}}, \bibinfo {author} {\bibfnamefont {G.}~\bibnamefont {Nobile}},\ and\ \bibinfo {author} {\bibfnamefont {R.}~\bibnamefont {Vaglio}},\ }\bibfield  {title} {\bibinfo {title} {{Niboium-titanium nitride thin films for superconducting rf accelerator cavities}},\ }\href {https://doi.org/10.1007/BF00682108} {\bibfield  {journal} {\bibinfo  {journal} {J. Low Temp. Phys.}\ }\textbf {\bibinfo {volume} {78}},\ \bibinfo {pages} {41} (\bibinfo {year} {1990})}\BibitemShut {NoStop}%
\bibitem [{\citenamefont {Burton}\ \emph {et~al.}(2016)\citenamefont {Burton}, \citenamefont {Beebe}, \citenamefont {Yang}, \citenamefont {Lukaszew}, \citenamefont {Valente-Feliciano},\ and\ \citenamefont {Reece}}]{2016-Burton-JVSTA-34-021518}%
  \BibitemOpen
  \bibfield  {author} {\bibinfo {author} {\bibfnamefont {M.~C.}\ \bibnamefont {Burton}}, \bibinfo {author} {\bibfnamefont {M.~R.}\ \bibnamefont {Beebe}}, \bibinfo {author} {\bibfnamefont {K.}~\bibnamefont {Yang}}, \bibinfo {author} {\bibfnamefont {R.~A.}\ \bibnamefont {Lukaszew}}, \bibinfo {author} {\bibfnamefont {A.-M.}\ \bibnamefont {Valente-Feliciano}},\ and\ \bibinfo {author} {\bibfnamefont {C.}~\bibnamefont {Reece}},\ }\bibfield  {title} {\bibinfo {title} {Superconducting \ch{NbTiN} thin films for superconducting radio frequency accelerator cavity applications},\ }\href {https://doi.org/10.1116/1.4941735} {\bibfield  {journal} {\bibinfo  {journal} {J. Vac. Sci. Technol. A}\ }\textbf {\bibinfo {volume} {34}},\ \bibinfo {pages} {021518} (\bibinfo {year} {2016})}\BibitemShut {NoStop}%
\bibitem [{\citenamefont {Yen}\ \emph {et~al.}(1967)\citenamefont {Yen}, \citenamefont {Toth}, \citenamefont {Shy}, \citenamefont {Anderson},\ and\ \citenamefont {Rosner}}]{1967-Yen-JAP-38-2268}%
  \BibitemOpen
  \bibfield  {author} {\bibinfo {author} {\bibfnamefont {C.~M.}\ \bibnamefont {Yen}}, \bibinfo {author} {\bibfnamefont {L.~E.}\ \bibnamefont {Toth}}, \bibinfo {author} {\bibfnamefont {Y.~M.}\ \bibnamefont {Shy}}, \bibinfo {author} {\bibfnamefont {D.~E.}\ \bibnamefont {Anderson}},\ and\ \bibinfo {author} {\bibfnamefont {L.~G.}\ \bibnamefont {Rosner}},\ }\bibfield  {title} {\bibinfo {title} {{Superconducting $H_c$‐$J_c$ and $T_c$ Measurements in the Nb–Ti–N, Nb–Hf–N, and Nb–V–N Ternary Systems}},\ }\href {https://doi.org/10.1063/1.1709868} {\bibfield  {journal} {\bibinfo  {journal} {J. Appl. Phys.}\ }\textbf {\bibinfo {volume} {38}},\ \bibinfo {pages} {2268} (\bibinfo {year} {1967})}\BibitemShut {NoStop}%
\bibitem [{\citenamefont {Pan}\ \emph {et~al.}(1983)\citenamefont {Pan}, \citenamefont {Gorishnyak}, \citenamefont {Rudenko}, \citenamefont {Shaternik}, \citenamefont {Belous}, \citenamefont {Koziychuk},\ and\ \citenamefont {Korzhinsky}}]{1983-Pan-Cryo-23-258}%
  \BibitemOpen
  \bibfield  {author} {\bibinfo {author} {\bibfnamefont {V.}~\bibnamefont {Pan}}, \bibinfo {author} {\bibfnamefont {V.}~\bibnamefont {Gorishnyak}}, \bibinfo {author} {\bibfnamefont {E.}~\bibnamefont {Rudenko}}, \bibinfo {author} {\bibfnamefont {V.}~\bibnamefont {Shaternik}}, \bibinfo {author} {\bibfnamefont {M.}~\bibnamefont {Belous}}, \bibinfo {author} {\bibfnamefont {S.}~\bibnamefont {Koziychuk}},\ and\ \bibinfo {author} {\bibfnamefont {F.}~\bibnamefont {Korzhinsky}},\ }\bibfield  {title} {\bibinfo {title} {Investigation of the properties of superconducting niobium nitride films},\ }\href {https://doi.org/10.1016/0011-2275(83)90146-7} {\bibfield  {journal} {\bibinfo  {journal} {Cryogenics}\ }\textbf {\bibinfo {volume} {23}},\ \bibinfo {pages} {258} (\bibinfo {year} {1983})}\BibitemShut {NoStop}%
\bibitem [{\citenamefont {Isagawa}(1981)}]{1981-Isagawa-JAP-52-921}%
  \BibitemOpen
  \bibfield  {author} {\bibinfo {author} {\bibfnamefont {S.}~\bibnamefont {Isagawa}},\ }\bibfield  {title} {\bibinfo {title} {{rf superconducting properties of reactively sputtered NbN}},\ }\href {https://doi.org/10.1063/1.328846} {\bibfield  {journal} {\bibinfo  {journal} {J. Appl. Phys.}\ }\textbf {\bibinfo {volume} {52}},\ \bibinfo {pages} {921} (\bibinfo {year} {1981})}\BibitemShut {NoStop}%
\bibitem [{\citenamefont {Torgovkin}\ \emph {et~al.}(2018)\citenamefont {Torgovkin}, \citenamefont {Chaudhuri}, \citenamefont {Ruhtinas}, \citenamefont {Lahtinen}, \citenamefont {Sajavaara},\ and\ \citenamefont {Maasilta}}]{2018-Torgovkin-SST-31-055017}%
  \BibitemOpen
  \bibfield  {author} {\bibinfo {author} {\bibfnamefont {A.}~\bibnamefont {Torgovkin}}, \bibinfo {author} {\bibfnamefont {S.}~\bibnamefont {Chaudhuri}}, \bibinfo {author} {\bibfnamefont {A.}~\bibnamefont {Ruhtinas}}, \bibinfo {author} {\bibfnamefont {M.}~\bibnamefont {Lahtinen}}, \bibinfo {author} {\bibfnamefont {T.}~\bibnamefont {Sajavaara}},\ and\ \bibinfo {author} {\bibfnamefont {I.~J.}\ \bibnamefont {Maasilta}},\ }\bibfield  {title} {\bibinfo {title} {High quality superconducting titanium nitride thin film growth using infrared pulsed laser deposition},\ }\href {https://doi.org/10.1088/1361-6668/aab7d6} {\bibfield  {journal} {\bibinfo  {journal} {Supercond. Sci. Technol.}\ }\textbf {\bibinfo {volume} {31}},\ \bibinfo {pages} {055017} (\bibinfo {year} {2018})}\BibitemShut {NoStop}%
\bibitem [{\citenamefont {Vissers}\ \emph {et~al.}(2013)\citenamefont {Vissers}, \citenamefont {Gao}, \citenamefont {Kline}, \citenamefont {Sandberg}, \citenamefont {Weides}, \citenamefont {Wisbey},\ and\ \citenamefont {Pappas}}]{2013-Michael-TSF-548-485}%
  \BibitemOpen
  \bibfield  {author} {\bibinfo {author} {\bibfnamefont {M.~R.}\ \bibnamefont {Vissers}}, \bibinfo {author} {\bibfnamefont {J.}~\bibnamefont {Gao}}, \bibinfo {author} {\bibfnamefont {J.~S.}\ \bibnamefont {Kline}}, \bibinfo {author} {\bibfnamefont {M.}~\bibnamefont {Sandberg}}, \bibinfo {author} {\bibfnamefont {M.~P.}\ \bibnamefont {Weides}}, \bibinfo {author} {\bibfnamefont {D.~S.}\ \bibnamefont {Wisbey}},\ and\ \bibinfo {author} {\bibfnamefont {D.~P.}\ \bibnamefont {Pappas}},\ }\bibfield  {title} {\bibinfo {title} {Characterization and in-situ monitoring of sub-stoichiometric adjustable superconducting critical temperature titanium nitride growth},\ }\href {https://doi.org/10.1016/j.tsf.2013.07.046} {\bibfield  {journal} {\bibinfo  {journal} {Thin Solid Films}\ }\textbf {\bibinfo {volume} {548}},\ \bibinfo {pages} {485} (\bibinfo {year} {2013})}\BibitemShut {NoStop}%
\bibitem [{\citenamefont {Yu}\ \emph {et~al.}(2005)\citenamefont {Yu}, \citenamefont {Singh}, \citenamefont {Liu}, \citenamefont {Wu}, \citenamefont {Hu}, \citenamefont {Durand}, \citenamefont {Bulman}, \citenamefont {Rowell},\ and\ \citenamefont {Newman}}]{2005-LeiYu-IEETAS-15-44}%
  \BibitemOpen
  \bibfield  {author} {\bibinfo {author} {\bibfnamefont {L.}~\bibnamefont {Yu}}, \bibinfo {author} {\bibfnamefont {R.~K.}\ \bibnamefont {Singh}}, \bibinfo {author} {\bibfnamefont {H.}~\bibnamefont {Liu}}, \bibinfo {author} {\bibfnamefont {S.~Y.}\ \bibnamefont {Wu}}, \bibinfo {author} {\bibfnamefont {R.}~\bibnamefont {Hu}}, \bibinfo {author} {\bibfnamefont {D.}~\bibnamefont {Durand}}, \bibinfo {author} {\bibfnamefont {J.}~\bibnamefont {Bulman}}, \bibinfo {author} {\bibfnamefont {J.~M.}\ \bibnamefont {Rowell}},\ and\ \bibinfo {author} {\bibfnamefont {N.}~\bibnamefont {Newman}},\ }\bibfield  {title} {\bibinfo {title} {Fabrication of niobium titanium nitride thin films with high superconducting transition temperatures and short penetration lengths},\ }\href {https://doi.org/10.1109/TASC.2005.844126} {\bibfield  {journal} {\bibinfo  {journal} {IEEE Trans. Appl. Supercond.}\ }\textbf {\bibinfo {volume} {15}},\ \bibinfo {pages} {44} (\bibinfo {year} {2005})}\BibitemShut {NoStop}%
\bibitem [{\citenamefont {Khan}\ \emph {et~al.}(2022)\citenamefont {Khan}, \citenamefont {Khudchenko}, \citenamefont {Chekushkin},\ and\ \citenamefont {Koshelets}}]{2022-Khan-IEEETAS-32-1}%
  \BibitemOpen
  \bibfield  {author} {\bibinfo {author} {\bibfnamefont {F.}~\bibnamefont {Khan}}, \bibinfo {author} {\bibfnamefont {A.~V.}\ \bibnamefont {Khudchenko}}, \bibinfo {author} {\bibfnamefont {A.~M.}\ \bibnamefont {Chekushkin}},\ and\ \bibinfo {author} {\bibfnamefont {V.~P.}\ \bibnamefont {Koshelets}},\ }\bibfield  {title} {\bibinfo {title} {{Characterization of the Parameters of Superconducting NbN and NbTiN Films Using Parallel Plate Resonator}},\ }\href {https://doi.org/10.1109/TASC.2022.3148687} {\bibfield  {journal} {\bibinfo  {journal} {IEEE Trans. Appl. Supercond.}\ }\textbf {\bibinfo {volume} {32}},\ \bibinfo {pages} {1} (\bibinfo {year} {2022})}\BibitemShut {NoStop}%
\bibitem [{\citenamefont {Junginger}\ \emph {et~al.}(2018)\citenamefont {Junginger}, \citenamefont {Prokscha}, \citenamefont {Salman}, \citenamefont {Suter},\ and\ \citenamefont {{Valente-Feliciano}}}]{2018-Junginger-IPAC-3921}%
  \BibitemOpen
  \bibfield  {author} {\bibinfo {author} {\bibfnamefont {T.}~\bibnamefont {Junginger}}, \bibinfo {author} {\bibfnamefont {T.}~\bibnamefont {Prokscha}}, \bibinfo {author} {\bibfnamefont {Z.}~\bibnamefont {Salman}}, \bibinfo {author} {\bibfnamefont {A.}~\bibnamefont {Suter}},\ and\ \bibinfo {author} {\bibfnamefont {A.-M.}\ \bibnamefont {{Valente-Feliciano}}},\ }\bibfield  {title} {\bibinfo {title} {Critical fields of {SRF} materials},\ }in\ \href {https://doi.org/10.18429/JACoW-IPAC2018-THPAL118} {\emph {\bibinfo {booktitle} {Proceedings of {IPAC'18}}}},\ \bibinfo {series and number} {\bibinfo {series} {International Particle Accelerator Conference}\ No.~\bibinfo {number} {9}},\ \bibinfo {organization} {TRIUMF}\ (\bibinfo  {publisher} {JACoW Publishing},\ \bibinfo {address} {Geneva, Switzerland},\ \bibinfo {year} {2018})\ pp.\ \bibinfo {pages} {3921--3924}\BibitemShut {NoStop}%
\bibitem [{\citenamefont {Hong}\ \emph {et~al.}(2013)\citenamefont {Hong}, \citenamefont {Choi}, \citenamefont {Ik~Sim}, \citenamefont {Ha}, \citenamefont {Cheol~Park}, \citenamefont {Yamamori},\ and\ \citenamefont {Hoon~Kim}}]{2013-Hong-JAP-114-243905}%
  \BibitemOpen
  \bibfield  {author} {\bibinfo {author} {\bibfnamefont {T.}~\bibnamefont {Hong}}, \bibinfo {author} {\bibfnamefont {K.}~\bibnamefont {Choi}}, \bibinfo {author} {\bibfnamefont {K.}~\bibnamefont {Ik~Sim}}, \bibinfo {author} {\bibfnamefont {T.}~\bibnamefont {Ha}}, \bibinfo {author} {\bibfnamefont {B.}~\bibnamefont {Cheol~Park}}, \bibinfo {author} {\bibfnamefont {H.}~\bibnamefont {Yamamori}},\ and\ \bibinfo {author} {\bibfnamefont {J.}~\bibnamefont {Hoon~Kim}},\ }\bibfield  {title} {\bibinfo {title} {{Terahertz electrodynamics and superconducting energy gap of NbTiN}},\ }\href {https://doi.org/10.1063/1.4856995} {\bibfield  {journal} {\bibinfo  {journal} {J. Appl. Phys.}\ }\textbf {\bibinfo {volume} {114}},\ \bibinfo {pages} {243905} (\bibinfo {year} {2013})}\BibitemShut {NoStop}%
\bibitem [{\citenamefont {Sidorova}\ \emph {et~al.}(2021)\citenamefont {Sidorova}, \citenamefont {Semenov}, \citenamefont {H\"ubers}, \citenamefont {Gyger}, \citenamefont {Steinhauer}, \citenamefont {Zhang},\ and\ \citenamefont {Schilling}}]{2021-Sidorova-PRB-104-184514}%
  \BibitemOpen
  \bibfield  {author} {\bibinfo {author} {\bibfnamefont {M.}~\bibnamefont {Sidorova}}, \bibinfo {author} {\bibfnamefont {A.~D.}\ \bibnamefont {Semenov}}, \bibinfo {author} {\bibfnamefont {H.-W.}\ \bibnamefont {H\"ubers}}, \bibinfo {author} {\bibfnamefont {S.}~\bibnamefont {Gyger}}, \bibinfo {author} {\bibfnamefont {S.}~\bibnamefont {Steinhauer}}, \bibinfo {author} {\bibfnamefont {X.}~\bibnamefont {Zhang}},\ and\ \bibinfo {author} {\bibfnamefont {A.}~\bibnamefont {Schilling}},\ }\bibfield  {title} {\bibinfo {title} {{Magnetoconductance and photoresponse properties of disordered NbTiN films}},\ }\href {https://doi.org/10.1103/PhysRevB.104.184514} {\bibfield  {journal} {\bibinfo  {journal} {Phys. Rev. B}\ }\textbf {\bibinfo {volume} {104}},\ \bibinfo {pages} {184514} (\bibinfo {year} {2021})}\BibitemShut {NoStop}%
\bibitem [{\citenamefont {Yu}\ \emph {et~al.}(2002)\citenamefont {Yu}, \citenamefont {Newman},\ and\ \citenamefont {Rowell}}]{2002-Lei-IEEETAS-12-1795}%
  \BibitemOpen
  \bibfield  {author} {\bibinfo {author} {\bibfnamefont {L.}~\bibnamefont {Yu}}, \bibinfo {author} {\bibfnamefont {N.}~\bibnamefont {Newman}},\ and\ \bibinfo {author} {\bibfnamefont {J.~M.}\ \bibnamefont {Rowell}},\ }\bibfield  {title} {\bibinfo {title} {Measurement of the coherence length of sputtered \ch{Nb_{0.62}Ti_{0.38}N} thin films},\ }\href {https://doi.org/10.1109/TASC.2002.1020339} {\bibfield  {journal} {\bibinfo  {journal} {IEEE Trans. Appl. Supercond.}\ }\textbf {\bibinfo {volume} {12}},\ \bibinfo {pages} {1795} (\bibinfo {year} {2002})}\BibitemShut {NoStop}%
\bibitem [{\citenamefont {González Díaz-Palacio}\ \emph {et~al.}(2023)\citenamefont {González Díaz-Palacio}, \citenamefont {Wenskat}, \citenamefont {Deyu}, \citenamefont {Hillert}, \citenamefont {Blick},\ and\ \citenamefont {Zierold}}]{2023-Gonzalez-JAP-134-035301}%
  \BibitemOpen
  \bibfield  {author} {\bibinfo {author} {\bibfnamefont {I.}~\bibnamefont {González Díaz-Palacio}}, \bibinfo {author} {\bibfnamefont {M.}~\bibnamefont {Wenskat}}, \bibinfo {author} {\bibfnamefont {G.~K.}\ \bibnamefont {Deyu}}, \bibinfo {author} {\bibfnamefont {W.}~\bibnamefont {Hillert}}, \bibinfo {author} {\bibfnamefont {R.~H.}\ \bibnamefont {Blick}},\ and\ \bibinfo {author} {\bibfnamefont {R.}~\bibnamefont {Zierold}},\ }\bibfield  {title} {\bibinfo {title} {{Thermal annealing of superconducting niobium titanium nitride thin films deposited by plasma-enhanced atomic layer deposition}},\ }\href {https://doi.org/10.1063/5.0155557} {\bibfield  {journal} {\bibinfo  {journal} {J. Appl. Phys.}\ }\textbf {\bibinfo {volume} {134}},\ \bibinfo {pages} {035301} (\bibinfo {year} {2023})}\BibitemShut {NoStop}%
\bibitem [{\citenamefont {Valente-Feliciano}\ \emph {et~al.}(2015)\citenamefont {Valente-Feliciano}, \citenamefont {Burton}, \citenamefont {Eremeev}, \citenamefont {Lukaszew}, \citenamefont {Reece},\ and\ \citenamefont {Spradlin}}]{2015-Valente-Feliciano-SRF-TUBA08}%
  \BibitemOpen
  \bibfield  {author} {\bibinfo {author} {\bibfnamefont {A.-M.}\ \bibnamefont {Valente-Feliciano}}, \bibinfo {author} {\bibfnamefont {M.}~\bibnamefont {Burton}}, \bibinfo {author} {\bibfnamefont {G.}~\bibnamefont {Eremeev}}, \bibinfo {author} {\bibfnamefont {R.}~\bibnamefont {Lukaszew}}, \bibinfo {author} {\bibfnamefont {C.}~\bibnamefont {Reece}},\ and\ \bibinfo {author} {\bibfnamefont {J.}~\bibnamefont {Spradlin}},\ }\bibfield  {title} {\bibinfo {title} {{G}rowth and {C}haracterization of {M}ulti{-L}ayer {N}b{T}i{N} {F}ilms},\ }in\ \href {https://doi.org/10.18429/JACoW-SRF2015-TUBA08} {\emph {\bibinfo {booktitle} {Proc. of International Conference on RF Superconductivity (SRF2015), Whistler, BC, Canada, Sept. 13-18, 2015}}},\ \bibinfo {series and number} {\bibinfo {series} {International Conference on RF Superconductivity}\ No.~\bibinfo {number} {17}}\ (\bibinfo  {publisher} {JACoW},\ \bibinfo {address} {Geneva, Switzerland},\ \bibinfo {year} {2015})\ pp.\ \bibinfo {pages} {516--520}\BibitemShut {NoStop}%
\bibitem [{\citenamefont {{Hechler}}\ \emph {et~al.}(1969)\citenamefont {{Hechler}}, \citenamefont {{Horn}}, \citenamefont {{Otto}},\ and\ \citenamefont {{Saur}}}]{1969-Hechler-JLTP-1-29}%
  \BibitemOpen
  \bibfield  {author} {\bibinfo {author} {\bibfnamefont {K.}~\bibnamefont {{Hechler}}}, \bibinfo {author} {\bibfnamefont {G.}~\bibnamefont {{Horn}}}, \bibinfo {author} {\bibfnamefont {G.}~\bibnamefont {{Otto}}},\ and\ \bibinfo {author} {\bibfnamefont {E.}~\bibnamefont {{Saur}}},\ }\bibfield  {title} {\bibinfo {title} {{Measurements of critical data for some type II superconductors and comparison with theory}},\ }\href {https://doi.org/10.1007/BF00628332} {\bibfield  {journal} {\bibinfo  {journal} {J. Low Temp. Phys.}\ }\textbf {\bibinfo {volume} {1}},\ \bibinfo {pages} {29} (\bibinfo {year} {1969})}\BibitemShut {NoStop}%
\bibitem [{\citenamefont {Nieto}\ \emph {et~al.}(2023)\citenamefont {Nieto}, \citenamefont {Hofer}, \citenamefont {Sirena},\ and\ \citenamefont {Haberkorn}}]{2023-Rezinovsky-PC-607-1354241}%
  \BibitemOpen
  \bibfield  {author} {\bibinfo {author} {\bibfnamefont {S.~R.}\ \bibnamefont {Nieto}}, \bibinfo {author} {\bibfnamefont {J.}~\bibnamefont {Hofer}}, \bibinfo {author} {\bibfnamefont {M.}~\bibnamefont {Sirena}},\ and\ \bibinfo {author} {\bibfnamefont {N.}~\bibnamefont {Haberkorn}},\ }\bibfield  {title} {\bibinfo {title} {{Flexible NbTiN thin films for superconducting electronics}},\ }\href {https://doi.org/10.1016/j.physc.2023.1354241} {\bibfield  {journal} {\bibinfo  {journal} {Physica C}\ }\textbf {\bibinfo {volume} {607}},\ \bibinfo {pages} {1354241} (\bibinfo {year} {2023})}\BibitemShut {NoStop}%
\bibitem [{\citenamefont {Groll}\ \emph {et~al.}(2014)\citenamefont {Groll}, \citenamefont {Klug}, \citenamefont {Cao}, \citenamefont {Altin}, \citenamefont {Claus}, \citenamefont {Becker}, \citenamefont {Zasadzinski}, \citenamefont {Pellin},\ and\ \citenamefont {Proslier}}]{2014-Groll-APL-104-092602}%
  \BibitemOpen
  \bibfield  {author} {\bibinfo {author} {\bibfnamefont {N.~R.}\ \bibnamefont {Groll}}, \bibinfo {author} {\bibfnamefont {J.~A.}\ \bibnamefont {Klug}}, \bibinfo {author} {\bibfnamefont {C.}~\bibnamefont {Cao}}, \bibinfo {author} {\bibfnamefont {S.}~\bibnamefont {Altin}}, \bibinfo {author} {\bibfnamefont {H.}~\bibnamefont {Claus}}, \bibinfo {author} {\bibfnamefont {N.~G.}\ \bibnamefont {Becker}}, \bibinfo {author} {\bibfnamefont {J.~F.}\ \bibnamefont {Zasadzinski}}, \bibinfo {author} {\bibfnamefont {M.~J.}\ \bibnamefont {Pellin}},\ and\ \bibinfo {author} {\bibfnamefont {T.}~\bibnamefont {Proslier}},\ }\bibfield  {title} {\bibinfo {title} {{{Tunneling spectroscopy of superconducting MoN and NbTiN grown by atomic layer deposition}}},\ }\href {https://doi.org/10.1063/1.4867880} {\bibfield  {journal} {\bibinfo  {journal} {Appl. Phys. Lett.}\ }\textbf {\bibinfo {volume} {104}},\ \bibinfo {pages} {092602} (\bibinfo {year} {2014})}\BibitemShut {NoStop}%
\bibitem [{\citenamefont {Lap}\ \emph {et~al.}(2021)\citenamefont {Lap}, \citenamefont {Khudchenko}, \citenamefont {Hesper}, \citenamefont {Rudakov}, \citenamefont {Dmitriev}, \citenamefont {Khan}, \citenamefont {Koshelets},\ and\ \citenamefont {Baryshev}}]{2021-Lap-APL-119-152601}%
  \BibitemOpen
  \bibfield  {author} {\bibinfo {author} {\bibfnamefont {B.~N.~R.}\ \bibnamefont {Lap}}, \bibinfo {author} {\bibfnamefont {A.}~\bibnamefont {Khudchenko}}, \bibinfo {author} {\bibfnamefont {R.}~\bibnamefont {Hesper}}, \bibinfo {author} {\bibfnamefont {K.~I.}\ \bibnamefont {Rudakov}}, \bibinfo {author} {\bibfnamefont {P.}~\bibnamefont {Dmitriev}}, \bibinfo {author} {\bibfnamefont {F.}~\bibnamefont {Khan}}, \bibinfo {author} {\bibfnamefont {V.~P.}\ \bibnamefont {Koshelets}},\ and\ \bibinfo {author} {\bibfnamefont {A.~M.}\ \bibnamefont {Baryshev}},\ }\bibfield  {title} {\bibinfo {title} {{Characterization of superconducting NbTiN films using a dispersive Fourier transform spectrometer}},\ }\href {https://doi.org/10.1063/5.0066371} {\bibfield  {journal} {\bibinfo  {journal} {Appl. Phys. Lett.}\ }\textbf {\bibinfo {volume} {119}},\ \bibinfo {pages} {152601} (\bibinfo {year} {2021})}\BibitemShut {NoStop}%
\bibitem [{\citenamefont {Barends}\ \emph {et~al.}(2010)\citenamefont {Barends}, \citenamefont {Vercruyssen}, \citenamefont {Endo}, \citenamefont {de~Visser}, \citenamefont {Zijlstra}, \citenamefont {Klapwijk},\ and\ \citenamefont {Baselmans}}]{2010-Barends-APL-97-033507}%
  \BibitemOpen
  \bibfield  {author} {\bibinfo {author} {\bibfnamefont {R.}~\bibnamefont {Barends}}, \bibinfo {author} {\bibfnamefont {N.}~\bibnamefont {Vercruyssen}}, \bibinfo {author} {\bibfnamefont {A.}~\bibnamefont {Endo}}, \bibinfo {author} {\bibfnamefont {P.~J.}\ \bibnamefont {de~Visser}}, \bibinfo {author} {\bibfnamefont {T.}~\bibnamefont {Zijlstra}}, \bibinfo {author} {\bibfnamefont {T.~M.}\ \bibnamefont {Klapwijk}},\ and\ \bibinfo {author} {\bibfnamefont {J.~J.~A.}\ \bibnamefont {Baselmans}},\ }\bibfield  {title} {\bibinfo {title} {Reduced frequency noise in superconducting resonators},\ }\href {https://doi.org/10.1063/1.3467052} {\bibfield  {journal} {\bibinfo  {journal} {Appl. Phys. Lett.}\ }\textbf {\bibinfo {volume} {97}},\ \bibinfo {pages} {033507} (\bibinfo {year} {2010})}\BibitemShut {NoStop}%
\bibitem [{\citenamefont {Driessen}\ \emph {et~al.}(2012)\citenamefont {Driessen}, \citenamefont {Coumou}, \citenamefont {Tromp}, \citenamefont {de~Visser},\ and\ \citenamefont {Klapwijk}}]{2012-Driessen-PRL-109-107003}%
  \BibitemOpen
  \bibfield  {author} {\bibinfo {author} {\bibfnamefont {E.~F.~C.}\ \bibnamefont {Driessen}}, \bibinfo {author} {\bibfnamefont {P.~C. J.~J.}\ \bibnamefont {Coumou}}, \bibinfo {author} {\bibfnamefont {R.~R.}\ \bibnamefont {Tromp}}, \bibinfo {author} {\bibfnamefont {P.~J.}\ \bibnamefont {de~Visser}},\ and\ \bibinfo {author} {\bibfnamefont {T.~M.}\ \bibnamefont {Klapwijk}},\ }\bibfield  {title} {\bibinfo {title} {{Strongly Disordered TiN and NbTiN $s$-Wave Superconductors Probed by Microwave Electrodynamics}},\ }\href {https://doi.org/10.1103/PhysRevLett.109.107003} {\bibfield  {journal} {\bibinfo  {journal} {Phys. Rev. Lett.}\ }\textbf {\bibinfo {volume} {109}},\ \bibinfo {pages} {107003} (\bibinfo {year} {2012})}\BibitemShut {NoStop}%
\bibitem [{\citenamefont {Kalboussi}(2023)}]{2023-Kalboussi-thesis}%
  \BibitemOpen
  \bibfield  {author} {\bibinfo {author} {\bibfnamefont {Y.}~\bibnamefont {Kalboussi}},\ }\emph {\bibinfo {title} {{Nano hetero-structures for improving performances of superconductors under high fields}}},\ \href {https://theses.hal.science/tel-04116992} {Ph.D. thesis},\ \bibinfo  {school} {{Universit{\'e} Paris-Saclay}} (\bibinfo {year} {2023})\BibitemShut {NoStop}%
\bibitem [{\citenamefont {Proslier}\ \emph {et~al.}(2011)\citenamefont {Proslier}, \citenamefont {Klug}, \citenamefont {Becker}, \citenamefont {Elam},\ and\ \citenamefont {Pellin}}]{2011-Proslier-ET-41-237}%
  \BibitemOpen
  \bibfield  {author} {\bibinfo {author} {\bibfnamefont {T.}~\bibnamefont {Proslier}}, \bibinfo {author} {\bibfnamefont {J.}~\bibnamefont {Klug}}, \bibinfo {author} {\bibfnamefont {N.~C.}\ \bibnamefont {Becker}}, \bibinfo {author} {\bibfnamefont {J.~W.}\ \bibnamefont {Elam}},\ and\ \bibinfo {author} {\bibfnamefont {M.}~\bibnamefont {Pellin}},\ }\bibfield  {title} {\bibinfo {title} {{(Invited) Atomic Layer Deposition of Superconductors}},\ }\href {https://doi.org/10.1149/1.3633673} {\bibfield  {journal} {\bibinfo  {journal} {ECS Trans.}\ }\textbf {\bibinfo {volume} {41}},\ \bibinfo {pages} {237} (\bibinfo {year} {2011})}\BibitemShut {NoStop}%
\bibitem [{\citenamefont {Kalboussi}\ \emph {et~al.}(2022)\citenamefont {Kalboussi}, \citenamefont {Antoine}, \citenamefont {Bira}, \citenamefont {Delatte}, \citenamefont {Dragoe}, \citenamefont {Leroy}, \citenamefont {Longuevergne}, \citenamefont {Proslier},\ and\ \citenamefont {Tusseau-Nenez}}]{2021-Kalboussi-SRF-2021}%
  \BibitemOpen
  \bibfield  {author} {\bibinfo {author} {\bibfnamefont {Y.}~\bibnamefont {Kalboussi}}, \bibinfo {author} {\bibfnamefont {C.}~\bibnamefont {Antoine}}, \bibinfo {author} {\bibfnamefont {S.}~\bibnamefont {Bira}}, \bibinfo {author} {\bibfnamefont {B.}~\bibnamefont {Delatte}}, \bibinfo {author} {\bibfnamefont {D.}~\bibnamefont {Dragoe}}, \bibinfo {author} {\bibfnamefont {J.}~\bibnamefont {Leroy}}, \bibinfo {author} {\bibfnamefont {D.}~\bibnamefont {Longuevergne}}, \bibinfo {author} {\bibfnamefont {T.}~\bibnamefont {Proslier}},\ and\ \bibinfo {author} {\bibfnamefont {S.}~\bibnamefont {Tusseau-Nenez}},\ }\bibfield  {title} {\bibinfo {title} {{Material Engineering of ALD- Deposited Multilayer to Improve the Superconducting Performances of RF Cavities Under Intense Fields}}\ }(\bibinfo  {publisher} {JACoW Publishing, Geneva, Switzerland},\ \bibinfo {year} {2022})\ \bibinfo {note} {presented at SRF'21 in East Lansing, MI, USA, unpublished}\BibitemShut {NoStop}%
\bibitem [{\citenamefont {Abrikosov}(1957)}]{1957-Abrikosov-SPJ-5-1174}%
  \BibitemOpen
  \bibfield  {author} {\bibinfo {author} {\bibfnamefont {A.~A.}\ \bibnamefont {Abrikosov}},\ }\bibfield  {title} {\bibinfo {title} {On the magnetic properties of superconductors of the second group},\ }\href@noop {} {\bibfield  {journal} {\bibinfo  {journal} {Sov. Phys. JETP}\ }\textbf {\bibinfo {volume} {5}},\ \bibinfo {pages} {1174} (\bibinfo {year} {1957})}\BibitemShut {NoStop}%
\bibitem [{\citenamefont {Brandt}(1995)}]{1995-Brandt-RPP-58-1465}%
  \BibitemOpen
  \bibfield  {author} {\bibinfo {author} {\bibfnamefont {E.~H.}\ \bibnamefont {Brandt}},\ }\bibfield  {title} {\bibinfo {title} {The flux-line lattice in superconductors},\ }\href {https://doi.org/10.1088/0034-4885/58/11/003} {\bibfield  {journal} {\bibinfo  {journal} {Rep. Prog. Phys.}\ }\textbf {\bibinfo {volume} {58}},\ \bibinfo {pages} {1465} (\bibinfo {year} {1995})}\BibitemShut {NoStop}%
\bibitem [{\citenamefont {MacLaughlin}(1976)}]{1976-MacLaughlin-SSP-31-1}%
  \BibitemOpen
  \bibfield  {author} {\bibinfo {author} {\bibfnamefont {D.~E.}\ \bibnamefont {MacLaughlin}},\ }\bibfield  {title} {\bibinfo {title} {{Magnetic Resonance in the Superconducting State}}\ }(\bibinfo  {publisher} {Academic Press},\ \bibinfo {year} {1976})\ pp.\ \bibinfo {pages} {1--69}\BibitemShut {NoStop}%
\bibitem [{\citenamefont {Walstedt}(2008)}]{2008-Walstedt-NMRPHTM-2-13}%
  \BibitemOpen
  \bibfield  {author} {\bibinfo {author} {\bibfnamefont {R.~E.}\ \bibnamefont {Walstedt}},\ }\bibinfo {title} {Introduction to {NMR} studies of metals, metallic compounds, and superconductors},\ in\ \href {https://doi.org/10.1007/978-3-540-75565-4_2} {\emph {\bibinfo {booktitle} {The {NMR} Probe of High-{$T_{c}$} Materials}}},\ \bibinfo {series} {Springer Tracts in Modern Physics}, Vol.\ \bibinfo {volume} {228}\ (\bibinfo  {publisher} {Springer},\ \bibinfo {address} {Berlin},\ \bibinfo {year} {2008})\ Chap.~\bibinfo {chapter} {2}, pp.\ \bibinfo {pages} {13--65}\BibitemShut {NoStop}%
\bibitem [{\citenamefont {Sonier}\ \emph {et~al.}(2000)\citenamefont {Sonier}, \citenamefont {Brewer},\ and\ \citenamefont {Kiefl}}]{2000-Sonier-RMP-72-769}%
  \BibitemOpen
  \bibfield  {author} {\bibinfo {author} {\bibfnamefont {J.~E.}\ \bibnamefont {Sonier}}, \bibinfo {author} {\bibfnamefont {J.~H.}\ \bibnamefont {Brewer}},\ and\ \bibinfo {author} {\bibfnamefont {R.~F.}\ \bibnamefont {Kiefl}},\ }\bibfield  {title} {\bibinfo {title} {{\ensuremath{\mu}SR studies of the vortex state in type-II superconductors}},\ }\href {https://doi.org/10.1103/RevModPhys.72.769} {\bibfield  {journal} {\bibinfo  {journal} {Rev. Mod. Phys.}\ }\textbf {\bibinfo {volume} {72}},\ \bibinfo {pages} {769} (\bibinfo {year} {2000})}\BibitemShut {NoStop}%
\bibitem [{\citenamefont {Amato}\ and\ \citenamefont {Morenzoni}(2024)}]{2024-Amato-musr-book-new}%
  \BibitemOpen
  \bibfield  {author} {\bibinfo {author} {\bibfnamefont {A.}~\bibnamefont {Amato}}\ and\ \bibinfo {author} {\bibfnamefont {E.}~\bibnamefont {Morenzoni}},\ }\href {https://doi.org/10.1007/978-3-031-44959-8} {\emph {\bibinfo {title} {Introduction to Muon Spin Spectroscopy: Applications to Solid State and Material Sciences}}},\ \bibinfo {edition} {1st}\ ed.,\ Lecture Notes in Physics\ (\bibinfo  {publisher} {Springer Cham},\ \bibinfo {year} {2024})\BibitemShut {NoStop}%
\bibitem [{\citenamefont {Morenzoni}\ \emph {et~al.}(2004)\citenamefont {Morenzoni}, \citenamefont {Prokscha}, \citenamefont {Suter}, \citenamefont {Luetkens},\ and\ \citenamefont {Khasanov}}]{2004-Morenzoni-JPCM-16-S4583}%
  \BibitemOpen
  \bibfield  {author} {\bibinfo {author} {\bibfnamefont {E.}~\bibnamefont {Morenzoni}}, \bibinfo {author} {\bibfnamefont {T.}~\bibnamefont {Prokscha}}, \bibinfo {author} {\bibfnamefont {A.}~\bibnamefont {Suter}}, \bibinfo {author} {\bibfnamefont {H.}~\bibnamefont {Luetkens}},\ and\ \bibinfo {author} {\bibfnamefont {R.}~\bibnamefont {Khasanov}},\ }\bibfield  {title} {\bibinfo {title} {Nano-scale thin film investigations with slow polarized muons},\ }\href {https://doi.org/10.1088/0953-8984/16/40/010} {\bibfield  {journal} {\bibinfo  {journal} {J. Phys.: Condens. Matter}\ }\textbf {\bibinfo {volume} {16}},\ \bibinfo {pages} {S4583} (\bibinfo {year} {2004})}\BibitemShut {NoStop}%
\bibitem [{\citenamefont {Blundell}\ \emph {et~al.}(2021)\citenamefont {Blundell}, \citenamefont {De~Renzi}, \citenamefont {Lancaster},\ and\ \citenamefont {Pratt}}]{2021-Blundell-Book-OUP}%
  \BibitemOpen
  \bibfield  {author} {\bibinfo {author} {\bibfnamefont {S.~J.}\ \bibnamefont {Blundell}}, \bibinfo {author} {\bibfnamefont {R.}~\bibnamefont {De~Renzi}}, \bibinfo {author} {\bibfnamefont {T.}~\bibnamefont {Lancaster}},\ and\ \bibinfo {author} {\bibfnamefont {F.~L.}\ \bibnamefont {Pratt}},\ }\bibfield  {title} {\bibinfo {title} {{Low energy $\mu${SR}}},\ }in\ \href {https://doi.org/10.1093/oso/9780198858959.003.0018} {\emph {\bibinfo {booktitle} {{Muon Spectroscopy: An Introduction}}}}\ (\bibinfo  {publisher} {Oxford University Press},\ \bibinfo {year} {2021})\BibitemShut {NoStop}%
\bibitem [{\citenamefont {MacFarlane}(2015)}]{2015-MacFarlane-SSNMR-68-1}%
  \BibitemOpen
  \bibfield  {author} {\bibinfo {author} {\bibfnamefont {W.}~\bibnamefont {MacFarlane}},\ }\bibfield  {title} {\bibinfo {title} {{Implanted-ion $\beta$NMR: A new probe for nanoscience}},\ }\href {https://doi.org/10.1016/j.ssnmr.2015.02.004} {\bibfield  {journal} {\bibinfo  {journal} {Solid State Nucl. Magn. Reson.}\ }\textbf {\bibinfo {volume} {68-69}},\ \bibinfo {pages} {1} (\bibinfo {year} {2015})}\BibitemShut {NoStop}%
\bibitem [{\citenamefont {MacFarlane}(2022)}]{2022-MacFarlane-ZPC-236-757}%
  \BibitemOpen
  \bibfield  {author} {\bibinfo {author} {\bibfnamefont {W.~A.}\ \bibnamefont {MacFarlane}},\ }\bibfield  {title} {\bibinfo {title} {{Status and progress of ion-implanted $\beta$NMR at TRIUMF}},\ }\href {https://doi.org/doi:10.1515/zpch-2021-3154} {\bibfield  {journal} {\bibinfo  {journal} {Z. Phys. Chem.}\ }\textbf {\bibinfo {volume} {236}},\ \bibinfo {pages} {757} (\bibinfo {year} {2022})}\BibitemShut {NoStop}%
\bibitem [{\citenamefont {Korringa}(1950)}]{1950-Korringa-Physica-16-601}%
  \BibitemOpen
  \bibfield  {author} {\bibinfo {author} {\bibfnamefont {J.}~\bibnamefont {Korringa}},\ }\bibfield  {title} {\bibinfo {title} {Nuclear magnetic relaxation and resonnance line shift in metals},\ }\href {https://doi.org/10.1016/0031-8914(50)90105-4} {\bibfield  {journal} {\bibinfo  {journal} {Physica}\ }\textbf {\bibinfo {volume} {16}},\ \bibinfo {pages} {601} (\bibinfo {year} {1950})}\BibitemShut {NoStop}%
\bibitem [{\citenamefont {Hebel}\ and\ \citenamefont {Slichter}(1957)}]{1957-Hebel-PR-107-901}%
  \BibitemOpen
  \bibfield  {author} {\bibinfo {author} {\bibfnamefont {L.~C.}\ \bibnamefont {Hebel}}\ and\ \bibinfo {author} {\bibfnamefont {C.~P.}\ \bibnamefont {Slichter}},\ }\bibfield  {title} {\bibinfo {title} {Nuclear relaxation in superconducting aluminum},\ }\href {https://doi.org/10.1103/PhysRev.107.901} {\bibfield  {journal} {\bibinfo  {journal} {Phys. Rev.}\ }\textbf {\bibinfo {volume} {107}},\ \bibinfo {pages} {901} (\bibinfo {year} {1957})}\BibitemShut {NoStop}%
\bibitem [{\citenamefont {Hebel}\ and\ \citenamefont {Slichter}(1959)}]{1959-Hebel-PR-113-1504}%
  \BibitemOpen
  \bibfield  {author} {\bibinfo {author} {\bibfnamefont {L.~C.}\ \bibnamefont {Hebel}}\ and\ \bibinfo {author} {\bibfnamefont {C.~P.}\ \bibnamefont {Slichter}},\ }\bibfield  {title} {\bibinfo {title} {{Nuclear Spin Relaxation in Normal and Superconducting Aluminum}},\ }\href {https://doi.org/10.1103/PhysRev.113.1504} {\bibfield  {journal} {\bibinfo  {journal} {Phys. Rev.}\ }\textbf {\bibinfo {volume} {113}},\ \bibinfo {pages} {1504} (\bibinfo {year} {1959})}\BibitemShut {NoStop}%
\bibitem [{\citenamefont {Ziegler}\ \emph {et~al.}(2008)\citenamefont {Ziegler}, \citenamefont {Biersack},\ and\ \citenamefont {Ziegler}}]{2008-Ziegler-book-srim}%
  \BibitemOpen
  \bibfield  {author} {\bibinfo {author} {\bibfnamefont {J.~F.}\ \bibnamefont {Ziegler}}, \bibinfo {author} {\bibfnamefont {J.~P.}\ \bibnamefont {Biersack}},\ and\ \bibinfo {author} {\bibfnamefont {M.~D.}\ \bibnamefont {Ziegler}},\ }\href {http://www.srim.org/} {\emph {\bibinfo {title} {{SRIM} --- The Stopping and Range of Ions in Matter}}},\ \bibinfo {edition} {7th}\ ed.\ (\bibinfo  {publisher} {SRIM Co.},\ \bibinfo {address} {Chester},\ \bibinfo {year} {2008})\BibitemShut {NoStop}%
\bibitem [{\citenamefont {Levy}\ \emph {et~al.}(2013)\citenamefont {Levy}, \citenamefont {PPearson}, \citenamefont {Kiefl}, \citenamefont {Mané}, \citenamefont {Morris},\ and\ \citenamefont {Voss}}]{2013-Levy-HI-225-165}%
  \BibitemOpen
  \bibfield  {author} {\bibinfo {author} {\bibfnamefont {C.~D.~P.}\ \bibnamefont {Levy}}, \bibinfo {author} {\bibfnamefont {M.~R.}\ \bibnamefont {PPearson}}, \bibinfo {author} {\bibfnamefont {R.~F.}\ \bibnamefont {Kiefl}}, \bibinfo {author} {\bibfnamefont {E.}~\bibnamefont {Mané}}, \bibinfo {author} {\bibfnamefont {G.~D.}\ \bibnamefont {Morris}},\ and\ \bibinfo {author} {\bibfnamefont {A.}~\bibnamefont {Voss}},\ }\bibfield  {title} {\bibinfo {title} {{Laser polarization facility}},\ }\href {https://doi.org/10.1007/s10751-013-0896-4} {\bibfield  {journal} {\bibinfo  {journal} {Hyperfine Interact.}\ }\textbf {\bibinfo {volume} {225}},\ \bibinfo {pages} {165} (\bibinfo {year} {2013})}\BibitemShut {NoStop}%
\bibitem [{\citenamefont {Levy}\ \emph {et~al.}(2003)\citenamefont {Levy}, \citenamefont {Hatakeyama}, \citenamefont {Hirayama}, \citenamefont {Kiefl}, \citenamefont {Baartman}, \citenamefont {Behr}, \citenamefont {Izumi}, \citenamefont {Melconian}, \citenamefont {Morris}, \citenamefont {Nussbaumer}, \citenamefont {Olivo}, \citenamefont {Pearson}, \citenamefont {Poutissou},\ and\ \citenamefont {Wight}}]{2003-Levy-NIMPRS-204-689}%
  \BibitemOpen
  \bibfield  {author} {\bibinfo {author} {\bibfnamefont {C.}~\bibnamefont {Levy}}, \bibinfo {author} {\bibfnamefont {A.}~\bibnamefont {Hatakeyama}}, \bibinfo {author} {\bibfnamefont {Y.}~\bibnamefont {Hirayama}}, \bibinfo {author} {\bibfnamefont {R.}~\bibnamefont {Kiefl}}, \bibinfo {author} {\bibfnamefont {R.}~\bibnamefont {Baartman}}, \bibinfo {author} {\bibfnamefont {J.}~\bibnamefont {Behr}}, \bibinfo {author} {\bibfnamefont {H.}~\bibnamefont {Izumi}}, \bibinfo {author} {\bibfnamefont {D.}~\bibnamefont {Melconian}}, \bibinfo {author} {\bibfnamefont {G.}~\bibnamefont {Morris}}, \bibinfo {author} {\bibfnamefont {R.}~\bibnamefont {Nussbaumer}}, \bibinfo {author} {\bibfnamefont {M.}~\bibnamefont {Olivo}}, \bibinfo {author} {\bibfnamefont {M.}~\bibnamefont {Pearson}}, \bibinfo {author} {\bibfnamefont {R.}~\bibnamefont {Poutissou}},\ and\ \bibinfo {author} {\bibfnamefont {G.}~\bibnamefont {Wight}},\ }\bibfield  {title} {\bibinfo {title} {Polarized radioactive beam at {ISAC}},\ }\href {https://doi.org/10.1016/S0168-583X(03)00485-3} {\bibfield  {journal} {\bibinfo  {journal} {Nucl. Instrum. Methods Phys. Res., Sect. B}\ }\textbf {\bibinfo {volume} {204}},\ \bibinfo {pages} {689} (\bibinfo {year} {2003})}\BibitemShut {NoStop}%
\bibitem [{\citenamefont {{Morris}}(2014)}]{2014-Morris-HI-225-173}%
  \BibitemOpen
  \bibfield  {author} {\bibinfo {author} {\bibfnamefont {G.~D.}\ \bibnamefont {{Morris}}},\ }\bibfield  {title} {\bibinfo {title} {{{\ensuremath{\beta}}-NMR}},\ }\href {https://doi.org/10.1007/s10751-013-0894-6} {\bibfield  {journal} {\bibinfo  {journal} {Hyperfine Interact.}\ }\textbf {\bibinfo {volume} {225}},\ \bibinfo {pages} {173} (\bibinfo {year} {2014})}\BibitemShut {NoStop}%
\bibitem [{Note1()}]{Note1}%
  \BibitemOpen
  \bibinfo {note} {Forming the asymmetry in this manner has the advantage of implicitly removing select detection systematics (e.g., different detector efficiencies).}\BibitemShut {Stop}%
\bibitem [{\citenamefont {{Minamisono}}\ \emph {et~al.}(1993)\citenamefont {{Minamisono}}, \citenamefont {{Ohtsubo}}, \citenamefont {{Fukuda}}, \citenamefont {{Minami}}, \citenamefont {{Nakayama}}, \citenamefont {{Fukuda}}, \citenamefont {{Matsuta}},\ and\ \citenamefont {{Nojiri}}}]{1993-Minamisono-HI-80-1315}%
  \BibitemOpen
  \bibfield  {author} {\bibinfo {author} {\bibfnamefont {T.}~\bibnamefont {{Minamisono}}}, \bibinfo {author} {\bibfnamefont {T.}~\bibnamefont {{Ohtsubo}}}, \bibinfo {author} {\bibfnamefont {S.}~\bibnamefont {{Fukuda}}}, \bibinfo {author} {\bibfnamefont {I.}~\bibnamefont {{Minami}}}, \bibinfo {author} {\bibfnamefont {Y.}~\bibnamefont {{Nakayama}}}, \bibinfo {author} {\bibfnamefont {M.}~\bibnamefont {{Fukuda}}}, \bibinfo {author} {\bibfnamefont {K.}~\bibnamefont {{Matsuta}}},\ and\ \bibinfo {author} {\bibfnamefont {Y.}~\bibnamefont {{Nojiri}}},\ }\bibfield  {title} {\bibinfo {title} {{{New nuclear quadrupole resonance technique in {\ensuremath{\beta}}-NMR}}},\ }\href {https://doi.org/10.1007/BF00567497} {\bibfield  {journal} {\bibinfo  {journal} {Hyperfine Interact.}\ }\textbf {\bibinfo {volume} {80}},\ \bibinfo {pages} {1315} (\bibinfo {year} {1993})}\BibitemShut {NoStop}%
\bibitem [{\citenamefont {Adelman}\ \emph {et~al.}(2022)\citenamefont {Adelman}, \citenamefont {Fujimoto}, \citenamefont {Dehn}, \citenamefont {Dunsiger}, \citenamefont {Karner}, \citenamefont {Levy}, \citenamefont {Li}, \citenamefont {McKenzie}, \citenamefont {McFadden}, \citenamefont {Morris}, \citenamefont {Pearson}, \citenamefont {Stachura}, \citenamefont {Thoeng}, \citenamefont {Ticknor}, \citenamefont {Ohashi}, \citenamefont {Kojima},\ and\ \citenamefont {MacFarlane}}]{2022-Adelman-PRB-106-035205}%
  \BibitemOpen
  \bibfield  {author} {\bibinfo {author} {\bibfnamefont {J.~R.}\ \bibnamefont {Adelman}}, \bibinfo {author} {\bibfnamefont {D.}~\bibnamefont {Fujimoto}}, \bibinfo {author} {\bibfnamefont {M.~H.}\ \bibnamefont {Dehn}}, \bibinfo {author} {\bibfnamefont {S.~R.}\ \bibnamefont {Dunsiger}}, \bibinfo {author} {\bibfnamefont {V.~L.}\ \bibnamefont {Karner}}, \bibinfo {author} {\bibfnamefont {C.~D.~P.}\ \bibnamefont {Levy}}, \bibinfo {author} {\bibfnamefont {R.}~\bibnamefont {Li}}, \bibinfo {author} {\bibfnamefont {I.}~\bibnamefont {McKenzie}}, \bibinfo {author} {\bibfnamefont {R.~M.~L.}\ \bibnamefont {McFadden}}, \bibinfo {author} {\bibfnamefont {G.~D.}\ \bibnamefont {Morris}}, \bibinfo {author} {\bibfnamefont {M.~R.}\ \bibnamefont {Pearson}}, \bibinfo {author} {\bibfnamefont {M.}~\bibnamefont {Stachura}}, \bibinfo {author} {\bibfnamefont {E.}~\bibnamefont {Thoeng}}, \bibinfo {author} {\bibfnamefont {J.~O.}\ \bibnamefont {Ticknor}}, \bibinfo {author} {\bibfnamefont {N.}~\bibnamefont {Ohashi}}, \bibinfo {author} {\bibfnamefont {K.~M.}\ \bibnamefont {Kojima}},\ and\ \bibinfo {author} {\bibfnamefont {W.~A.}\ \bibnamefont {MacFarlane}},\ }\bibfield  {title} {\bibinfo {title} {{Nuclear magnetic resonance of $^{8}\mathrm{Li}$ ions implanted in ZnO}},\ }\href {https://doi.org/10.1103/PhysRevB.106.035205} {\bibfield  {journal} {\bibinfo  {journal} {Phys. Rev. B}\ }\textbf {\bibinfo {volume} {106}},\ \bibinfo {pages} {035205} (\bibinfo {year} {2022})}\BibitemShut {NoStop}%
\bibitem [{\citenamefont {Ciovati}\ \emph {et~al.}(2011)\citenamefont {Ciovati}, \citenamefont {Tian},\ and\ \citenamefont {Corcoran}}]{2011-Ciovati-JAE-41-721}%
  \BibitemOpen
  \bibfield  {author} {\bibinfo {author} {\bibfnamefont {G.}~\bibnamefont {Ciovati}}, \bibinfo {author} {\bibfnamefont {H.}~\bibnamefont {Tian}},\ and\ \bibinfo {author} {\bibfnamefont {S.~G.}\ \bibnamefont {Corcoran}},\ }\bibfield  {title} {\bibinfo {title} {Buffered electrochemical polishing of niobium},\ }\href {https://doi.org/10.1007/s10800-011-0286-z} {\bibfield  {journal} {\bibinfo  {journal} {J. Appl. Electrochem.}\ }\textbf {\bibinfo {volume} {41}},\ \bibinfo {pages} {721} (\bibinfo {year} {2011})}\BibitemShut {NoStop}%
\bibitem [{\citenamefont {Miikkulainen}\ \emph {et~al.}(2013)\citenamefont {Miikkulainen}, \citenamefont {Leskelä}, \citenamefont {Ritala},\ and\ \citenamefont {Puurunen}}]{2013-Miikkulainen-JAP-113-021301}%
  \BibitemOpen
  \bibfield  {author} {\bibinfo {author} {\bibfnamefont {V.}~\bibnamefont {Miikkulainen}}, \bibinfo {author} {\bibfnamefont {M.}~\bibnamefont {Leskelä}}, \bibinfo {author} {\bibfnamefont {M.}~\bibnamefont {Ritala}},\ and\ \bibinfo {author} {\bibfnamefont {R.~L.}\ \bibnamefont {Puurunen}},\ }\bibfield  {title} {\bibinfo {title} {{Crystallinity of inorganic films grown by atomic layer deposition: Overview and general trends}},\ }\href {https://doi.org/10.1063/1.4757907} {\bibfield  {journal} {\bibinfo  {journal} {J. Appl. Phys.}\ }\textbf {\bibinfo {volume} {113}},\ \bibinfo {pages} {021301} (\bibinfo {year} {2013})}\BibitemShut {NoStop}%
\bibitem [{\citenamefont {Rontu}\ \emph {et~al.}(2018)\citenamefont {Rontu}, \citenamefont {Sippola}, \citenamefont {Broas}, \citenamefont {Ross}, \citenamefont {Sajavaara}, \citenamefont {Lipsanen}, \citenamefont {Paulasto-Kröckel},\ and\ \citenamefont {Franssila}}]{2018-Rontu-JVSTA-36-021508}%
  \BibitemOpen
  \bibfield  {author} {\bibinfo {author} {\bibfnamefont {V.}~\bibnamefont {Rontu}}, \bibinfo {author} {\bibfnamefont {P.}~\bibnamefont {Sippola}}, \bibinfo {author} {\bibfnamefont {M.}~\bibnamefont {Broas}}, \bibinfo {author} {\bibfnamefont {G.}~\bibnamefont {Ross}}, \bibinfo {author} {\bibfnamefont {T.}~\bibnamefont {Sajavaara}}, \bibinfo {author} {\bibfnamefont {H.}~\bibnamefont {Lipsanen}}, \bibinfo {author} {\bibfnamefont {M.}~\bibnamefont {Paulasto-Kröckel}},\ and\ \bibinfo {author} {\bibfnamefont {S.}~\bibnamefont {Franssila}},\ }\bibfield  {title} {\bibinfo {title} {{Atomic layer deposition of AlN from AlCl$_3$ using NH$_3$ and Ar/NH$_3$ plasma}},\ }\href {https://doi.org/10.1116/1.5003381} {\bibfield  {journal} {\bibinfo  {journal} {J. Vac. Sci. Technol. A}\ }\textbf {\bibinfo {volume} {36}},\ \bibinfo {pages} {021508} (\bibinfo {year} {2018})}\BibitemShut {NoStop}%
\bibitem [{\citenamefont {Groll}\ \emph {et~al.}(2015)\citenamefont {Groll}, \citenamefont {Pellin}, \citenamefont {Zasadzinksi},\ and\ \citenamefont {Proslier}}]{2015-Groll-RSI-86-095111}%
  \BibitemOpen
  \bibfield  {author} {\bibinfo {author} {\bibfnamefont {N.}~\bibnamefont {Groll}}, \bibinfo {author} {\bibfnamefont {M.~J.}\ \bibnamefont {Pellin}}, \bibinfo {author} {\bibfnamefont {J.~F.}\ \bibnamefont {Zasadzinksi}},\ and\ \bibinfo {author} {\bibfnamefont {T.}~\bibnamefont {Proslier}},\ }\bibfield  {title} {\bibinfo {title} {Point contact tunneling spectroscopy apparatus for large scale mapping of surface superconducting properties},\ }\href {https://doi.org/10.1063/1.4931066} {\bibfield  {journal} {\bibinfo  {journal} {Rev. Sci. Instrum.}\ }\textbf {\bibinfo {volume} {86}},\ \bibinfo {pages} {095111} (\bibinfo {year} {2015})}\BibitemShut {NoStop}%
\bibitem [{\citenamefont {Kalboussi}\ \emph {et~al.}(2025)\citenamefont {Kalboussi}, \citenamefont {Curci}, \citenamefont {Miserque}, \citenamefont {Troadec}, \citenamefont {Brun}, \citenamefont {Walls}, \citenamefont {Jullien}, \citenamefont {Eozenou}, \citenamefont {Baudrier}, \citenamefont {Maurice}, \citenamefont {Bertrand}, \citenamefont {Sahuquet},\ and\ \citenamefont {Proslier}}]{2025-Kalboussi-PRA-23-044023}%
  \BibitemOpen
  \bibfield  {author} {\bibinfo {author} {\bibfnamefont {Y.}~\bibnamefont {Kalboussi}}, \bibinfo {author} {\bibfnamefont {I.}~\bibnamefont {Curci}}, \bibinfo {author} {\bibfnamefont {F.}~\bibnamefont {Miserque}}, \bibinfo {author} {\bibfnamefont {D.}~\bibnamefont {Troadec}}, \bibinfo {author} {\bibfnamefont {N.}~\bibnamefont {Brun}}, \bibinfo {author} {\bibfnamefont {M.}~\bibnamefont {Walls}}, \bibinfo {author} {\bibfnamefont {G.}~\bibnamefont {Jullien}}, \bibinfo {author} {\bibfnamefont {F.}~\bibnamefont {Eozenou}}, \bibinfo {author} {\bibfnamefont {M.}~\bibnamefont {Baudrier}}, \bibinfo {author} {\bibfnamefont {L.}~\bibnamefont {Maurice}}, \bibinfo {author} {\bibfnamefont {Q.}~\bibnamefont {Bertrand}}, \bibinfo {author} {\bibfnamefont {P.}~\bibnamefont {Sahuquet}},\ and\ \bibinfo {author} {\bibfnamefont {T.}~\bibnamefont {Proslier}},\ }\bibfield  {title} {\bibinfo {title} {{Crystallinity in niobium oxides: A pathway to mitigate two-level-system defects in niobium three-dimensional resonators for quantum applications}},\ }\href {https://doi.org/10.1103/PhysRevApplied.23.044023} {\bibfield  {journal} {\bibinfo  {journal} {Phys. Rev. Appl.}\ }\textbf {\bibinfo {volume} {23}},\ \bibinfo {pages} {044023} (\bibinfo {year} {2025})}\BibitemShut {NoStop}%
\bibitem [{\citenamefont {Dynes}\ \emph {et~al.}(1978)\citenamefont {Dynes}, \citenamefont {Narayanamurti},\ and\ \citenamefont {Garno}}]{1978-Dynes-PRL-41-1509}%
  \BibitemOpen
  \bibfield  {author} {\bibinfo {author} {\bibfnamefont {R.~C.}\ \bibnamefont {Dynes}}, \bibinfo {author} {\bibfnamefont {V.}~\bibnamefont {Narayanamurti}},\ and\ \bibinfo {author} {\bibfnamefont {J.~P.}\ \bibnamefont {Garno}},\ }\bibfield  {title} {\bibinfo {title} {{Direct Measurement of Quasiparticle-Lifetime Broadening in a Strong-Coupled Superconductor}},\ }\href {https://doi.org/10.1103/PhysRevLett.41.1509} {\bibfield  {journal} {\bibinfo  {journal} {Phys. Rev. Lett.}\ }\textbf {\bibinfo {volume} {41}},\ \bibinfo {pages} {1509} (\bibinfo {year} {1978})}\BibitemShut {NoStop}%
\bibitem [{sup()}]{supp}%
  \BibitemOpen
  \href@noop {} {}\bibinfo {note} {See Supplemental Material at [URL will be inserted by publisher] for structural (i.e., \gls{xrr}, \gls{gixrd}) and superconducting (i.e., \gls{vsm}, \gls{pct}) characterizations of the thin films.}\BibitemShut {Stop}%
\bibitem [{Note2()}]{Note2}%
  \BibitemOpen
  \bibinfo {note} {To confirm the absence of any (static) non-zero \glspl {efg}, we also performed frequency comb measurements (see, e.g., Refs.~\cite {1993-Minamisono-HI-80-1315,2022-Adelman-PRB-106-035205}), which greatly amplifies the sensitivity to such features. No evidence for any finite \glspl {efg} was found at all measured temperatures.}\BibitemShut {Stop}%
\bibitem [{\citenamefont {MacFarlane}\ \emph {et~al.}(2014{\natexlab{a}})\citenamefont {MacFarlane}, \citenamefont {Parolin}, \citenamefont {Cortie}, \citenamefont {Chow}, \citenamefont {Hossain}, \citenamefont {Kiefl}, \citenamefont {Levy}, \citenamefont {McFadden}, \citenamefont {Morris}, \citenamefont {Pearson}, \citenamefont {Saadaoui}, \citenamefont {Salman}, \citenamefont {Song},\ and\ \citenamefont {Wang}}]{2014-MacFarlane-JPCS-551-012033}%
  \BibitemOpen
  \bibfield  {author} {\bibinfo {author} {\bibfnamefont {W.~A.}\ \bibnamefont {MacFarlane}}, \bibinfo {author} {\bibfnamefont {T.~J.}\ \bibnamefont {Parolin}}, \bibinfo {author} {\bibfnamefont {D.~L.}\ \bibnamefont {Cortie}}, \bibinfo {author} {\bibfnamefont {K.~H.}\ \bibnamefont {Chow}}, \bibinfo {author} {\bibfnamefont {M.~D.}\ \bibnamefont {Hossain}}, \bibinfo {author} {\bibfnamefont {R.~F.}\ \bibnamefont {Kiefl}}, \bibinfo {author} {\bibfnamefont {C.~D.~P.}\ \bibnamefont {Levy}}, \bibinfo {author} {\bibfnamefont {R.~M.~L.}\ \bibnamefont {McFadden}}, \bibinfo {author} {\bibfnamefont {G.~D.}\ \bibnamefont {Morris}}, \bibinfo {author} {\bibfnamefont {M.~R.}\ \bibnamefont {Pearson}}, \bibinfo {author} {\bibfnamefont {H.}~\bibnamefont {Saadaoui}}, \bibinfo {author} {\bibfnamefont {Z.}~\bibnamefont {Salman}}, \bibinfo {author} {\bibfnamefont {Q.}~\bibnamefont {Song}},\ and\ \bibinfo {author} {\bibfnamefont {D.}~\bibnamefont {Wang}},\ }\bibfield  {title} {\bibinfo {title} {{$^8$Li$^+$ $\beta$-NMR in the Cubic Insulator MgO}},\ }\href {https://doi.org/10.1088/1742-6596/551/1/012033} {\bibfield  {journal} {\bibinfo  {journal} {J. Phys. Conf. Ser.}\ }\textbf {\bibinfo {volume} {551}},\ \bibinfo {pages} {012033} (\bibinfo {year} {2014}{\natexlab{a}})}\BibitemShut {NoStop}%
\bibitem [{\citenamefont {Brandt}(2003)}]{2003-Brandt-PRB-68-054506}%
  \BibitemOpen
  \bibfield  {author} {\bibinfo {author} {\bibfnamefont {E.~H.}\ \bibnamefont {Brandt}},\ }\bibfield  {title} {\bibinfo {title} {Properties of the ideal {Ginzburg-Landau} vortex lattice},\ }\href {https://doi.org/10.1103/PhysRevB.68.054506} {\bibfield  {journal} {\bibinfo  {journal} {Phys. Rev. B}\ }\textbf {\bibinfo {volume} {68}},\ \bibinfo {pages} {054506} (\bibinfo {year} {2003})}\BibitemShut {NoStop}%
\bibitem [{\citenamefont {Brandt}(1988)}]{1988-Brandt-PRB-37-2349}%
  \BibitemOpen
  \bibfield  {author} {\bibinfo {author} {\bibfnamefont {E.~H.}\ \bibnamefont {Brandt}},\ }\bibfield  {title} {\bibinfo {title} {Flux distribution and penetration depth measured by muon spin rotation in high-{$T_{c}$} superconductors},\ }\href {https://doi.org/10.1103/PhysRevB.37.2349} {\bibfield  {journal} {\bibinfo  {journal} {Phys. Rev. B}\ }\textbf {\bibinfo {volume} {37}},\ \bibinfo {pages} {2349} (\bibinfo {year} {1988})}\BibitemShut {NoStop}%
\bibitem [{\citenamefont {Pearl}(1964)}]{1964-Pearl-APL-5-65}%
  \BibitemOpen
  \bibfield  {author} {\bibinfo {author} {\bibfnamefont {J.}~\bibnamefont {Pearl}},\ }\bibfield  {title} {\bibinfo {title} {{Current distribution in superconducting films carrying quantized fluxoids}},\ }\href {https://doi.org/10.1063/1.1754056} {\bibfield  {journal} {\bibinfo  {journal} {Appl. Phys. Lett.}\ }\textbf {\bibinfo {volume} {5}},\ \bibinfo {pages} {65} (\bibinfo {year} {1964})}\BibitemShut {NoStop}%
\bibitem [{Note3()}]{Note3}%
  \BibitemOpen
  \bibinfo {note} {The form of \protect \Cref {eq:lambda-T-dependence} closely approximates the $T$-dependence predicted by \gls {bcs} theory (see, e.g.,~\cite {2024-Amato-musr-book-new}).}\BibitemShut {Stop}%
\bibitem [{Note4()}]{Note4}%
  \BibitemOpen
  \bibinfo {note} {Note that the exponent in the denominator of \protect \Cref {eq:lambda-T-dependence} is 2, unlike its more common value of 4 found in the two-fluid model. This choice is intentional, as it better describes $\lambda (T)$ in \ch {Nb_{1-x}Ti_xN}~\cite {2013-Hong-JAP-114-243905} and \ch {NbTi}~\cite {2024-Yeonkyu-PS-99-065963}.}\BibitemShut {Stop}%
\bibitem [{\citenamefont {Werthamer}\ \emph {et~al.}(1966)\citenamefont {Werthamer}, \citenamefont {Helfand},\ and\ \citenamefont {Hohenberg}}]{1996-Werthamer-PR-147-295}%
  \BibitemOpen
  \bibfield  {author} {\bibinfo {author} {\bibfnamefont {N.~R.}\ \bibnamefont {Werthamer}}, \bibinfo {author} {\bibfnamefont {E.}~\bibnamefont {Helfand}},\ and\ \bibinfo {author} {\bibfnamefont {P.~C.}\ \bibnamefont {Hohenberg}},\ }\bibfield  {title} {\bibinfo {title} {{Temperature and Purity Dependence of the Superconducting Critical Field, ${H}_{c2}$. III. Electron Spin and Spin-Orbit Effects}},\ }\href {https://doi.org/10.1103/PhysRev.147.295} {\bibfield  {journal} {\bibinfo  {journal} {Phys. Rev.}\ }\textbf {\bibinfo {volume} {147}},\ \bibinfo {pages} {295} (\bibinfo {year} {1966})}\BibitemShut {NoStop}%
\bibitem [{\citenamefont {Baumgartner}\ \emph {et~al.}(2013)\citenamefont {Baumgartner}, \citenamefont {Eisterer}, \citenamefont {Weber}, \citenamefont {Flükiger}, \citenamefont {Scheuerlein},\ and\ \citenamefont {Bottura}}]{2014-Baumgartner-SST-27-015005}%
  \BibitemOpen
  \bibfield  {author} {\bibinfo {author} {\bibfnamefont {T.}~\bibnamefont {Baumgartner}}, \bibinfo {author} {\bibfnamefont {M.}~\bibnamefont {Eisterer}}, \bibinfo {author} {\bibfnamefont {H.~W.}\ \bibnamefont {Weber}}, \bibinfo {author} {\bibfnamefont {R.}~\bibnamefont {Flükiger}}, \bibinfo {author} {\bibfnamefont {C.}~\bibnamefont {Scheuerlein}},\ and\ \bibinfo {author} {\bibfnamefont {L.}~\bibnamefont {Bottura}},\ }\bibfield  {title} {\bibinfo {title} {{Effects of neutron irradiation on pinning force scaling in state-of-the-art \ch{Nb3Sn} wires}},\ }\href {https://doi.org/10.1088/0953-2048/27/1/015005} {\bibfield  {journal} {\bibinfo  {journal} {Supercond. Sci. Technol.}\ }\textbf {\bibinfo {volume} {27}},\ \bibinfo {pages} {015005} (\bibinfo {year} {2013})}\BibitemShut {NoStop}%
\bibitem [{\citenamefont {Lee}\ \emph {et~al.}(2024)\citenamefont {Lee}, \citenamefont {Yun}, \citenamefont {Lee}, \citenamefont {Sirena}, \citenamefont {Kim},\ and\ \citenamefont {Haberkorn}}]{2024-Yeonkyu-PS-99-065963}%
  \BibitemOpen
  \bibfield  {author} {\bibinfo {author} {\bibfnamefont {Y.}~\bibnamefont {Lee}}, \bibinfo {author} {\bibfnamefont {J.}~\bibnamefont {Yun}}, \bibinfo {author} {\bibfnamefont {C.}~\bibnamefont {Lee}}, \bibinfo {author} {\bibfnamefont {M.}~\bibnamefont {Sirena}}, \bibinfo {author} {\bibfnamefont {J.}~\bibnamefont {Kim}},\ and\ \bibinfo {author} {\bibfnamefont {N.}~\bibnamefont {Haberkorn}},\ }\bibfield  {title} {\bibinfo {title} {{Penetration depth and critical fields in superconducting NbTi thin films grown by co-sputtering at room temperature}},\ }\href {https://doi.org/10.1088/1402-4896/ad4690} {\bibfield  {journal} {\bibinfo  {journal} {Phys. Scr.}\ }\textbf {\bibinfo {volume} {99}},\ \bibinfo {pages} {065963} (\bibinfo {year} {2024})}\BibitemShut {NoStop}%
\bibitem [{\citenamefont {McFadden}\ \emph {et~al.}(2022)\citenamefont {McFadden}, \citenamefont {Szunyogh}, \citenamefont {Bravo-Frank}, \citenamefont {Chatzichristos}, \citenamefont {Dehn}, \citenamefont {Fujimoto}, \citenamefont {Jancsó}, \citenamefont {Johannsen}, \citenamefont {Kálomista}, \citenamefont {Karner}, \citenamefont {Kiefl}, \citenamefont {Larsen}, \citenamefont {Lassen}, \citenamefont {Levy}, \citenamefont {Li}, \citenamefont {McKenzie}, \citenamefont {McPhee}, \citenamefont {Morris}, \citenamefont {Pearson}, \citenamefont {Sauer}, \citenamefont {Sigel}, \citenamefont {Thulstrup}, \citenamefont {MacFarlane}, \citenamefont {Hemmingsen},\ and\ \citenamefont {Stachura}}]{2022-McFadden-ACIE-61-e202207137}%
  \BibitemOpen
  \bibfield  {author} {\bibinfo {author} {\bibfnamefont {R.~M.~L.}\ \bibnamefont {McFadden}}, \bibinfo {author} {\bibfnamefont {D.}~\bibnamefont {Szunyogh}}, \bibinfo {author} {\bibfnamefont {N.}~\bibnamefont {Bravo-Frank}}, \bibinfo {author} {\bibfnamefont {A.}~\bibnamefont {Chatzichristos}}, \bibinfo {author} {\bibfnamefont {M.~H.}\ \bibnamefont {Dehn}}, \bibinfo {author} {\bibfnamefont {D.}~\bibnamefont {Fujimoto}}, \bibinfo {author} {\bibfnamefont {A.}~\bibnamefont {Jancsó}}, \bibinfo {author} {\bibfnamefont {S.}~\bibnamefont {Johannsen}}, \bibinfo {author} {\bibfnamefont {I.}~\bibnamefont {Kálomista}}, \bibinfo {author} {\bibfnamefont {V.~L.}\ \bibnamefont {Karner}}, \bibinfo {author} {\bibfnamefont {R.~F.}\ \bibnamefont {Kiefl}}, \bibinfo {author} {\bibfnamefont {F.~H.}\ \bibnamefont {Larsen}}, \bibinfo {author} {\bibfnamefont {J.}~\bibnamefont {Lassen}}, \bibinfo {author} {\bibfnamefont {C.~D.~P.}\ \bibnamefont {Levy}}, \bibinfo {author} {\bibfnamefont {R.}~\bibnamefont {Li}}, \bibinfo {author} {\bibfnamefont {I.}~\bibnamefont {McKenzie}}, \bibinfo {author} {\bibfnamefont {H.}~\bibnamefont {McPhee}}, \bibinfo {author} {\bibfnamefont {G.~D.}\ \bibnamefont {Morris}}, \bibinfo {author} {\bibfnamefont {M.~R.}\ \bibnamefont {Pearson}}, \bibinfo {author} {\bibfnamefont {S.~P.~A.}\ \bibnamefont {Sauer}}, \bibinfo {author} {\bibfnamefont {R.~K.~O.}\ \bibnamefont {Sigel}}, \bibinfo {author} {\bibfnamefont {P.~W.}\ \bibnamefont {Thulstrup}}, \bibinfo {author} {\bibfnamefont {W.~A.}\ \bibnamefont {MacFarlane}}, \bibinfo {author} {\bibfnamefont {L.}~\bibnamefont {Hemmingsen}},\ and\ \bibinfo {author} {\bibfnamefont {M.}~\bibnamefont {Stachura}},\ }\bibfield  {title} {\bibinfo {title} {{Magnesium(II)-ATP Complexes in 1-Ethyl-3-Methylimidazolium Acetate Solutions Characterized by $^{31}$Mg $\beta$-Radiation-Detected NMR Spectroscopy}},\ }\href {https://doi.org/10.1002/anie.202207137} {\bibfield  {journal} {\bibinfo  {journal} {Angew. Chem. Int. Ed.}\ }\textbf {\bibinfo {volume} {61}},\ \bibinfo {pages} {e202207137} (\bibinfo {year} {2022})}\BibitemShut {NoStop}%
\bibitem [{Note5()}]{Note5}%
  \BibitemOpen
  \bibinfo {note} {Note that the form of \protect \Cref {eq:corrected-shift} correctly accounts for the general case when both the sample and reference have different magnetic susceptibilities and demagnetization factors.}\BibitemShut {Stop}%
\bibitem [{Note6()}]{Note6}%
  \BibitemOpen
  \bibinfo {note} {The ``raw'' (i.e., uncorrected) \ch {^{8}Li} \gls {nmr} shifts were measured to range from \qtyrange [retain-explicit-plus=true]{-10}{+10}{\ppm }.}\BibitemShut {Stop}%
\bibitem [{\citenamefont {Parolin}\ \emph {et~al.}(2019)\citenamefont {Parolin}, \citenamefont {Salman}, \citenamefont {Chow}, \citenamefont {Song}, \citenamefont {Valiani}, \citenamefont {Saadaoui}, \citenamefont {O'Halloran}, \citenamefont {Hossain}, \citenamefont {Keeler}, \citenamefont {Kiefl}, \citenamefont {Kreitzman}, \citenamefont {Levy}, \citenamefont {Miller}, \citenamefont {Morris}, \citenamefont {Pearson}, \citenamefont {Smadella}, \citenamefont {Wang}, \citenamefont {Xu},\ and\ \citenamefont {MacFarlane}}]{2019-Parolin-PRB-100-209904}%
  \BibitemOpen
  \bibfield  {author} {\bibinfo {author} {\bibfnamefont {T.~J.}\ \bibnamefont {Parolin}}, \bibinfo {author} {\bibfnamefont {Z.}~\bibnamefont {Salman}}, \bibinfo {author} {\bibfnamefont {K.~H.}\ \bibnamefont {Chow}}, \bibinfo {author} {\bibfnamefont {Q.}~\bibnamefont {Song}}, \bibinfo {author} {\bibfnamefont {J.}~\bibnamefont {Valiani}}, \bibinfo {author} {\bibfnamefont {H.}~\bibnamefont {Saadaoui}}, \bibinfo {author} {\bibfnamefont {A.}~\bibnamefont {O'Halloran}}, \bibinfo {author} {\bibfnamefont {M.~D.}\ \bibnamefont {Hossain}}, \bibinfo {author} {\bibfnamefont {T.~A.}\ \bibnamefont {Keeler}}, \bibinfo {author} {\bibfnamefont {R.~F.}\ \bibnamefont {Kiefl}}, \bibinfo {author} {\bibfnamefont {S.~R.}\ \bibnamefont {Kreitzman}}, \bibinfo {author} {\bibfnamefont {C.~D.~P.}\ \bibnamefont {Levy}}, \bibinfo {author} {\bibfnamefont {R.~I.}\ \bibnamefont {Miller}}, \bibinfo {author} {\bibfnamefont {G.~D.}\ \bibnamefont {Morris}}, \bibinfo {author} {\bibfnamefont {M.~R.}\ \bibnamefont {Pearson}}, \bibinfo {author} {\bibfnamefont {M.}~\bibnamefont {Smadella}}, \bibinfo {author} {\bibfnamefont {D.}~\bibnamefont {Wang}}, \bibinfo {author} {\bibfnamefont {M.}~\bibnamefont {Xu}},\ and\ \bibinfo {author} {\bibfnamefont {W.~A.}\ \bibnamefont {MacFarlane}},\ }\bibfield  {title} {\bibinfo {title} {Erratum: {H}igh resolution $\ensuremath{\beta}$-{NMR} study of \ch{^{8}Li^{+}} implanted in gold {[Phys. Rev. B 77, 214107 (2008)]}},\ }\href {https://doi.org/10.1103/PhysRevB.100.209904} {\bibfield  {journal} {\bibinfo  {journal} {Phys. Rev. B}\ }\textbf {\bibinfo {volume} {100}},\ \bibinfo {pages} {209904} (\bibinfo {year} {2019})}\BibitemShut {NoStop}%
\bibitem [{\citenamefont {Morris}\ \emph {et~al.}(2004)\citenamefont {Morris}, \citenamefont {MacFarlane}, \citenamefont {Chow}, \citenamefont {Salman}, \citenamefont {Arseneau}, \citenamefont {Daviel}, \citenamefont {Hatakeyama}, \citenamefont {Kreitzman}, \citenamefont {Levy}, \citenamefont {Poutissou}, \citenamefont {Heffner}, \citenamefont {Elenewski}, \citenamefont {Greene},\ and\ \citenamefont {Kiefl}}]{2004-Morris-PRL-93-157601}%
  \BibitemOpen
  \bibfield  {author} {\bibinfo {author} {\bibfnamefont {G.~D.}\ \bibnamefont {Morris}}, \bibinfo {author} {\bibfnamefont {W.~A.}\ \bibnamefont {MacFarlane}}, \bibinfo {author} {\bibfnamefont {K.~H.}\ \bibnamefont {Chow}}, \bibinfo {author} {\bibfnamefont {Z.}~\bibnamefont {Salman}}, \bibinfo {author} {\bibfnamefont {D.~J.}\ \bibnamefont {Arseneau}}, \bibinfo {author} {\bibfnamefont {S.}~\bibnamefont {Daviel}}, \bibinfo {author} {\bibfnamefont {A.}~\bibnamefont {Hatakeyama}}, \bibinfo {author} {\bibfnamefont {S.~R.}\ \bibnamefont {Kreitzman}}, \bibinfo {author} {\bibfnamefont {C.~D.~P.}\ \bibnamefont {Levy}}, \bibinfo {author} {\bibfnamefont {R.}~\bibnamefont {Poutissou}}, \bibinfo {author} {\bibfnamefont {R.~H.}\ \bibnamefont {Heffner}}, \bibinfo {author} {\bibfnamefont {J.~E.}\ \bibnamefont {Elenewski}}, \bibinfo {author} {\bibfnamefont {L.~H.}\ \bibnamefont {Greene}},\ and\ \bibinfo {author} {\bibfnamefont {R.~F.}\ \bibnamefont {Kiefl}},\ }\bibfield  {title} {\bibinfo {title} {{Depth-Controlled $\ensuremath{\beta}$-NMR of $^{8}\mathrm{L}\mathrm{i}$ in a Thin Silver Film}},\ }\href {https://doi.org/10.1103/PhysRevLett.93.157601} {\bibfield  {journal} {\bibinfo  {journal} {Phys. Rev. Lett.}\ }\textbf {\bibinfo {volume} {93}},\ \bibinfo {pages} {157601} (\bibinfo {year} {2004})}\BibitemShut {NoStop}%
\bibitem [{\citenamefont {Parolin}\ \emph {et~al.}(2009)\citenamefont {Parolin}, \citenamefont {Shi}, \citenamefont {Salman}, \citenamefont {Chow}, \citenamefont {Dosanjh}, \citenamefont {Saadaoui}, \citenamefont {Song}, \citenamefont {Hossain}, \citenamefont {Kiefl}, \citenamefont {Levy}, \citenamefont {Pearson},\ and\ \citenamefont {MacFarlane}}]{2009-Parolin-PRB-80-174109}%
  \BibitemOpen
  \bibfield  {author} {\bibinfo {author} {\bibfnamefont {T.~J.}\ \bibnamefont {Parolin}}, \bibinfo {author} {\bibfnamefont {J.}~\bibnamefont {Shi}}, \bibinfo {author} {\bibfnamefont {Z.}~\bibnamefont {Salman}}, \bibinfo {author} {\bibfnamefont {K.~H.}\ \bibnamefont {Chow}}, \bibinfo {author} {\bibfnamefont {P.}~\bibnamefont {Dosanjh}}, \bibinfo {author} {\bibfnamefont {H.}~\bibnamefont {Saadaoui}}, \bibinfo {author} {\bibfnamefont {Q.}~\bibnamefont {Song}}, \bibinfo {author} {\bibfnamefont {M.~D.}\ \bibnamefont {Hossain}}, \bibinfo {author} {\bibfnamefont {R.~F.}\ \bibnamefont {Kiefl}}, \bibinfo {author} {\bibfnamefont {C.~D.~P.}\ \bibnamefont {Levy}}, \bibinfo {author} {\bibfnamefont {M.~R.}\ \bibnamefont {Pearson}},\ and\ \bibinfo {author} {\bibfnamefont {W.~A.}\ \bibnamefont {MacFarlane}},\ }\bibfield  {title} {\bibinfo {title} {{Nuclear magnetic resonance study of Li implanted in a thin film of niobium}},\ }\href {https://doi.org/10.1103/PhysRevB.80.174109} {\bibfield  {journal} {\bibinfo  {journal} {Phys. Rev. B}\ }\textbf {\bibinfo {volume} {80}},\ \bibinfo {pages} {174109} (\bibinfo {year} {2009})}\BibitemShut {NoStop}%
\bibitem [{\citenamefont {Wang}\ \emph {et~al.}(2006)\citenamefont {Wang}, \citenamefont {Hossain}, \citenamefont {Salman}, \citenamefont {Arseneau}, \citenamefont {Chow}, \citenamefont {Daviel}, \citenamefont {Keeler}, \citenamefont {Kiefl}, \citenamefont {Kreitzman}, \citenamefont {Levy}, \citenamefont {Morris}, \citenamefont {Miller}, \citenamefont {MacFarlane}, \citenamefont {Parolin},\ and\ \citenamefont {Saadaoui}}]{2006-Wang-PBC-374-239}%
  \BibitemOpen
  \bibfield  {author} {\bibinfo {author} {\bibfnamefont {D.}~\bibnamefont {Wang}}, \bibinfo {author} {\bibfnamefont {M.}~\bibnamefont {Hossain}}, \bibinfo {author} {\bibfnamefont {Z.}~\bibnamefont {Salman}}, \bibinfo {author} {\bibfnamefont {D.}~\bibnamefont {Arseneau}}, \bibinfo {author} {\bibfnamefont {K.}~\bibnamefont {Chow}}, \bibinfo {author} {\bibfnamefont {S.}~\bibnamefont {Daviel}}, \bibinfo {author} {\bibfnamefont {T.}~\bibnamefont {Keeler}}, \bibinfo {author} {\bibfnamefont {R.}~\bibnamefont {Kiefl}}, \bibinfo {author} {\bibfnamefont {S.}~\bibnamefont {Kreitzman}}, \bibinfo {author} {\bibfnamefont {C.}~\bibnamefont {Levy}}, \bibinfo {author} {\bibfnamefont {G.}~\bibnamefont {Morris}}, \bibinfo {author} {\bibfnamefont {R.}~\bibnamefont {Miller}}, \bibinfo {author} {\bibfnamefont {W.}~\bibnamefont {MacFarlane}}, \bibinfo {author} {\bibfnamefont {T.}~\bibnamefont {Parolin}},\ and\ \bibinfo {author} {\bibfnamefont {H.}~\bibnamefont {Saadaoui}},\ }\bibfield  {title} {\bibinfo {title} {{$\beta$-detected NMR of $^8$Li in the normal state of 2H-NbSe$_2$}},\ }\href {https://doi.org/10.1016/j.physb.2005.11.064} {\bibfield  {journal} {\bibinfo  {journal} {Physica B}\ }\textbf {\bibinfo {volume} {374-375}},\ \bibinfo {pages} {239} (\bibinfo {year} {2006})}\BibitemShut {NoStop}%
\bibitem [{\citenamefont {McFadden}\ \emph {et~al.}(2019)\citenamefont {McFadden}, \citenamefont {Chatzichristos}, \citenamefont {Chow}, \citenamefont {Cortie}, \citenamefont {Dehn}, \citenamefont {Fujimoto}, \citenamefont {Hossain}, \citenamefont {Ji}, \citenamefont {Karner}, \citenamefont {Kiefl}, \citenamefont {Levy}, \citenamefont {Li}, \citenamefont {McKenzie}, \citenamefont {Morris}, \citenamefont {Ofer}, \citenamefont {Pearson}, \citenamefont {Stachura}, \citenamefont {Cava},\ and\ \citenamefont {MacFarlane}}]{2019-McFadden-PRB-99-125201}%
  \BibitemOpen
  \bibfield  {author} {\bibinfo {author} {\bibfnamefont {R.~M.~L.}\ \bibnamefont {McFadden}}, \bibinfo {author} {\bibfnamefont {A.}~\bibnamefont {Chatzichristos}}, \bibinfo {author} {\bibfnamefont {K.~H.}\ \bibnamefont {Chow}}, \bibinfo {author} {\bibfnamefont {D.~L.}\ \bibnamefont {Cortie}}, \bibinfo {author} {\bibfnamefont {M.~H.}\ \bibnamefont {Dehn}}, \bibinfo {author} {\bibfnamefont {D.}~\bibnamefont {Fujimoto}}, \bibinfo {author} {\bibfnamefont {M.~D.}\ \bibnamefont {Hossain}}, \bibinfo {author} {\bibfnamefont {H.}~\bibnamefont {Ji}}, \bibinfo {author} {\bibfnamefont {V.~L.}\ \bibnamefont {Karner}}, \bibinfo {author} {\bibfnamefont {R.~F.}\ \bibnamefont {Kiefl}}, \bibinfo {author} {\bibfnamefont {C.~D.~P.}\ \bibnamefont {Levy}}, \bibinfo {author} {\bibfnamefont {R.}~\bibnamefont {Li}}, \bibinfo {author} {\bibfnamefont {I.}~\bibnamefont {McKenzie}}, \bibinfo {author} {\bibfnamefont {G.~D.}\ \bibnamefont {Morris}}, \bibinfo {author} {\bibfnamefont {O.}~\bibnamefont {Ofer}}, \bibinfo {author} {\bibfnamefont {M.~R.}\ \bibnamefont {Pearson}}, \bibinfo {author} {\bibfnamefont {M.}~\bibnamefont {Stachura}}, \bibinfo {author} {\bibfnamefont {R.~J.}\ \bibnamefont {Cava}},\ and\ \bibinfo {author} {\bibfnamefont {W.~A.}\ \bibnamefont {MacFarlane}},\ }\bibfield  {title} {\bibinfo {title} {{Ionic and electronic properties of the topological insulator ${\mathrm{Bi}}_{2}{\mathrm{Te}}_{2}\mathrm{Se}$ investigated via $\ensuremath{\beta}$-detected nuclear magnetic relaxation and resonance of $^{8}\mathrm{Li}$}},\ }\href {https://doi.org/10.1103/PhysRevB.99.125201} {\bibfield  {journal} {\bibinfo  {journal} {Phys. Rev. B}\ }\textbf {\bibinfo {volume} {99}},\ \bibinfo {pages} {125201} (\bibinfo {year} {2019})}\BibitemShut {NoStop}%
\bibitem [{\citenamefont {McFadden}\ \emph {et~al.}(2020)\citenamefont {McFadden}, \citenamefont {Chatzichristos}, \citenamefont {Cortie}, \citenamefont {Fujimoto}, \citenamefont {Hor}, \citenamefont {Ji}, \citenamefont {Karner}, \citenamefont {Kiefl}, \citenamefont {Levy}, \citenamefont {Li}, \citenamefont {McKenzie}, \citenamefont {Morris}, \citenamefont {Pearson}, \citenamefont {Stachura}, \citenamefont {Cava},\ and\ \citenamefont {MacFarlane}}]{2020-McFadden-PRB-102-235206}%
  \BibitemOpen
  \bibfield  {author} {\bibinfo {author} {\bibfnamefont {R.~M.~L.}\ \bibnamefont {McFadden}}, \bibinfo {author} {\bibfnamefont {A.}~\bibnamefont {Chatzichristos}}, \bibinfo {author} {\bibfnamefont {D.~L.}\ \bibnamefont {Cortie}}, \bibinfo {author} {\bibfnamefont {D.}~\bibnamefont {Fujimoto}}, \bibinfo {author} {\bibfnamefont {Y.~S.}\ \bibnamefont {Hor}}, \bibinfo {author} {\bibfnamefont {H.}~\bibnamefont {Ji}}, \bibinfo {author} {\bibfnamefont {V.~L.}\ \bibnamefont {Karner}}, \bibinfo {author} {\bibfnamefont {R.~F.}\ \bibnamefont {Kiefl}}, \bibinfo {author} {\bibfnamefont {C.~D.~P.}\ \bibnamefont {Levy}}, \bibinfo {author} {\bibfnamefont {R.}~\bibnamefont {Li}}, \bibinfo {author} {\bibfnamefont {I.}~\bibnamefont {McKenzie}}, \bibinfo {author} {\bibfnamefont {G.~D.}\ \bibnamefont {Morris}}, \bibinfo {author} {\bibfnamefont {M.~R.}\ \bibnamefont {Pearson}}, \bibinfo {author} {\bibfnamefont {M.}~\bibnamefont {Stachura}}, \bibinfo {author} {\bibfnamefont {R.~J.}\ \bibnamefont {Cava}},\ and\ \bibinfo {author} {\bibfnamefont {W.~A.}\ \bibnamefont {MacFarlane}},\ }\bibfield  {title} {\bibinfo {title} {{Local electronic and magnetic properties of the doped topological insulators ${\mathrm{Bi}}_{2}{\mathrm{Se}}_{3}:\mathrm{Ca}$ and ${\mathrm{Bi}}_{2}{\mathrm{Te}}_{3}:\mathrm{Mn}$ investigated using ion-implanted ${}^{8}\mathrm{Li}$ $\ensuremath{\beta}\text{\ensuremath{-}}\mathrm{NMR}$}},\ }\href {https://doi.org/10.1103/PhysRevB.102.235206} {\bibfield  {journal} {\bibinfo  {journal} {Phys. Rev. B}\ }\textbf {\bibinfo {volume} {102}},\ \bibinfo {pages} {235206} (\bibinfo {year} {2020})}\BibitemShut {NoStop}%
\bibitem [{\citenamefont {Stöckmann}\ and\ \citenamefont {Heitjans}(1984)}]{1984-Stockmann-JNCS-66-501}%
  \BibitemOpen
  \bibfield  {author} {\bibinfo {author} {\bibfnamefont {H.-J.}\ \bibnamefont {Stöckmann}}\ and\ \bibinfo {author} {\bibfnamefont {P.}~\bibnamefont {Heitjans}},\ }\bibfield  {title} {\bibinfo {title} {Low-temperature nuclear spin-lattice relaxation in glasses --- homogeneous and inhomogeneous averaging},\ }\href {https://doi.org/10.1016/0022-3093(84)90373-9} {\bibfield  {journal} {\bibinfo  {journal} {J. Non-Cryst. Solids}\ }\textbf {\bibinfo {volume} {66}},\ \bibinfo {pages} {501} (\bibinfo {year} {1984})}\BibitemShut {NoStop}%
\bibitem [{Note7()}]{Note7}%
  \BibitemOpen
  \bibinfo {note} {This suggests a significant fraction of \ch {^8Li} relax much more slowly than the fitted $1/T_1$, likely due to weak coupling to the electronic system in low-density or poorly metallic regions, beyond simple site-to-site Korringa variation.}\BibitemShut {Stop}%
\bibitem [{\citenamefont {Sugiyama}\ \emph {et~al.}(2017)\citenamefont {Sugiyama}, \citenamefont {Umegaki}, \citenamefont {Uyama}, \citenamefont {McFadden}, \citenamefont {Shiraki}, \citenamefont {Hitosugi}, \citenamefont {Salman}, \citenamefont {Saadaoui}, \citenamefont {Morris}, \citenamefont {MacFarlane},\ and\ \citenamefont {Kiefl}}]{2017-Sugiyama-PRB-96-094402}%
  \BibitemOpen
  \bibfield  {author} {\bibinfo {author} {\bibfnamefont {J.}~\bibnamefont {Sugiyama}}, \bibinfo {author} {\bibfnamefont {I.}~\bibnamefont {Umegaki}}, \bibinfo {author} {\bibfnamefont {T.}~\bibnamefont {Uyama}}, \bibinfo {author} {\bibfnamefont {R.~M.~L.}\ \bibnamefont {McFadden}}, \bibinfo {author} {\bibfnamefont {S.}~\bibnamefont {Shiraki}}, \bibinfo {author} {\bibfnamefont {T.}~\bibnamefont {Hitosugi}}, \bibinfo {author} {\bibfnamefont {Z.}~\bibnamefont {Salman}}, \bibinfo {author} {\bibfnamefont {H.}~\bibnamefont {Saadaoui}}, \bibinfo {author} {\bibfnamefont {G.~D.}\ \bibnamefont {Morris}}, \bibinfo {author} {\bibfnamefont {W.~A.}\ \bibnamefont {MacFarlane}},\ and\ \bibinfo {author} {\bibfnamefont {R.~F.}\ \bibnamefont {Kiefl}},\ }\bibfield  {title} {\bibinfo {title} {{Lithium diffusion in spinel ${\mathrm{Li}}_{4}{\mathrm{Ti}}_{5}{\mathrm{O}}_{12}$ and ${\mathrm{LiTi}}_{2}{\mathrm{O}}_{4}$ films detected with $^{8}\mathrm{Li}\phantom{\rule{4pt}{0ex}}\ensuremath{\beta}$-NMR}},\ }\href {https://doi.org/10.1103/PhysRevB.96.094402} {\bibfield  {journal} {\bibinfo  {journal} {Phys. Rev. B}\ }\textbf {\bibinfo {volume} {96}},\ \bibinfo {pages} {094402} (\bibinfo {year} {2017})}\BibitemShut {NoStop}%
\bibitem [{\citenamefont {Cortie}\ \emph {et~al.}(2016)\citenamefont {Cortie}, \citenamefont {Buck}, \citenamefont {Dehn}, \citenamefont {Karner}, \citenamefont {Kiefl}, \citenamefont {Levy}, \citenamefont {McFadden}, \citenamefont {Morris}, \citenamefont {McKenzie}, \citenamefont {Pearson}, \citenamefont {Wang},\ and\ \citenamefont {MacFarlane}}]{2016-Cortie-PRL-116-106103}%
  \BibitemOpen
  \bibfield  {author} {\bibinfo {author} {\bibfnamefont {D.~L.}\ \bibnamefont {Cortie}}, \bibinfo {author} {\bibfnamefont {T.}~\bibnamefont {Buck}}, \bibinfo {author} {\bibfnamefont {M.~H.}\ \bibnamefont {Dehn}}, \bibinfo {author} {\bibfnamefont {V.~L.}\ \bibnamefont {Karner}}, \bibinfo {author} {\bibfnamefont {R.~F.}\ \bibnamefont {Kiefl}}, \bibinfo {author} {\bibfnamefont {C.~D.~P.}\ \bibnamefont {Levy}}, \bibinfo {author} {\bibfnamefont {R.~M.~L.}\ \bibnamefont {McFadden}}, \bibinfo {author} {\bibfnamefont {G.~D.}\ \bibnamefont {Morris}}, \bibinfo {author} {\bibfnamefont {I.}~\bibnamefont {McKenzie}}, \bibinfo {author} {\bibfnamefont {M.~R.}\ \bibnamefont {Pearson}}, \bibinfo {author} {\bibfnamefont {X.~L.}\ \bibnamefont {Wang}},\ and\ \bibinfo {author} {\bibfnamefont {W.~A.}\ \bibnamefont {MacFarlane}},\ }\bibfield  {title} {\bibinfo {title} {{$\ensuremath{\beta}$-NMR Investigation of the Depth-Dependent Magnetic Properties of an Antiferromagnetic Surface}},\ }\href {https://doi.org/10.1103/PhysRevLett.116.106103} {\bibfield  {journal} {\bibinfo  {journal} {Phys. Rev. Lett.}\ }\textbf {\bibinfo {volume} {116}},\ \bibinfo {pages} {106103} (\bibinfo {year} {2016})}\BibitemShut {NoStop}%
\bibitem [{Note8()}]{Note8}%
  \BibitemOpen
  \bibinfo {note} {For example, sharing $A_0$ instead of $\beta $ yields a very similar fit, with a virtually identical $T$-dependence to $1/T_1$.}\BibitemShut {Stop}%
\bibitem [{Note9()}]{Note9}%
  \BibitemOpen
  \bibinfo {note} {We note that the small $\beta $ value suggests that a biexponental relaxation model would also work; however, we find that it's ``extra'' degrees-of-freedom lead to overparameterization when applied to the present data.}\BibitemShut {Stop}%
\bibitem [{\citenamefont {Hossain}\ \emph {et~al.}(2009)\citenamefont {Hossain}, \citenamefont {Salman}, \citenamefont {Wang}, \citenamefont {Chow}, \citenamefont {Kreitzman}, \citenamefont {Keeler}, \citenamefont {Levy}, \citenamefont {MacFarlane}, \citenamefont {Miller}, \citenamefont {Morris}, \citenamefont {Parolin}, \citenamefont {Pearson}, \citenamefont {Saadaoui},\ and\ \citenamefont {Kiefl}}]{2009-Hossain-PRB-79-144518}%
  \BibitemOpen
  \bibfield  {author} {\bibinfo {author} {\bibfnamefont {M.~D.}\ \bibnamefont {Hossain}}, \bibinfo {author} {\bibfnamefont {Z.}~\bibnamefont {Salman}}, \bibinfo {author} {\bibfnamefont {D.}~\bibnamefont {Wang}}, \bibinfo {author} {\bibfnamefont {K.~H.}\ \bibnamefont {Chow}}, \bibinfo {author} {\bibfnamefont {S.}~\bibnamefont {Kreitzman}}, \bibinfo {author} {\bibfnamefont {T.~A.}\ \bibnamefont {Keeler}}, \bibinfo {author} {\bibfnamefont {C.~D.~P.}\ \bibnamefont {Levy}}, \bibinfo {author} {\bibfnamefont {W.~A.}\ \bibnamefont {MacFarlane}}, \bibinfo {author} {\bibfnamefont {R.~I.}\ \bibnamefont {Miller}}, \bibinfo {author} {\bibfnamefont {G.~D.}\ \bibnamefont {Morris}}, \bibinfo {author} {\bibfnamefont {T.~J.}\ \bibnamefont {Parolin}}, \bibinfo {author} {\bibfnamefont {M.}~\bibnamefont {Pearson}}, \bibinfo {author} {\bibfnamefont {H.}~\bibnamefont {Saadaoui}},\ and\ \bibinfo {author} {\bibfnamefont {R.~F.}\ \bibnamefont {Kiefl}},\ }\bibfield  {title} {\bibinfo {title} {Low-field cross spin relaxation of \ch{^{8}Li} in superconducting \ch{NbSe2}},\ }\href {https://doi.org/10.1103/PhysRevB.79.144518} {\bibfield  {journal} {\bibinfo  {journal} {Phys. Rev. B}\ }\textbf {\bibinfo {volume} {79}},\ \bibinfo {pages} {144518} (\bibinfo {year} {2009})}\BibitemShut {NoStop}%
\bibitem [{\citenamefont {Bloembergen}\ \emph {et~al.}(1948)\citenamefont {Bloembergen}, \citenamefont {Purcell},\ and\ \citenamefont {Pound}}]{1948-Bloembergen-PR-73-679}%
  \BibitemOpen
  \bibfield  {author} {\bibinfo {author} {\bibfnamefont {N.}~\bibnamefont {Bloembergen}}, \bibinfo {author} {\bibfnamefont {E.~M.}\ \bibnamefont {Purcell}},\ and\ \bibinfo {author} {\bibfnamefont {R.~V.}\ \bibnamefont {Pound}},\ }\bibfield  {title} {\bibinfo {title} {{Relaxation Effects in Nuclear Magnetic Resonance Absorption}},\ }\href {https://doi.org/10.1103/PhysRev.73.679} {\bibfield  {journal} {\bibinfo  {journal} {Phys. Rev.}\ }\textbf {\bibinfo {volume} {73}},\ \bibinfo {pages} {679} (\bibinfo {year} {1948})}\BibitemShut {NoStop}%
\bibitem [{\citenamefont {Kiefl}\ \emph {et~al.}(1993)\citenamefont {Kiefl}, \citenamefont {MacFarlane}, \citenamefont {Chow}, \citenamefont {Dunsiger}, \citenamefont {Duty}, \citenamefont {Johnston}, \citenamefont {Schneider}, \citenamefont {Sonier}, \citenamefont {Brard}, \citenamefont {Strongin}, \citenamefont {Fischer},\ and\ \citenamefont {Smith}}]{1993-Kiefl-PRL-70-3987}%
  \BibitemOpen
  \bibfield  {author} {\bibinfo {author} {\bibfnamefont {R.~F.}\ \bibnamefont {Kiefl}}, \bibinfo {author} {\bibfnamefont {W.~A.}\ \bibnamefont {MacFarlane}}, \bibinfo {author} {\bibfnamefont {K.~H.}\ \bibnamefont {Chow}}, \bibinfo {author} {\bibfnamefont {S.}~\bibnamefont {Dunsiger}}, \bibinfo {author} {\bibfnamefont {T.~L.}\ \bibnamefont {Duty}}, \bibinfo {author} {\bibfnamefont {T.~M.~S.}\ \bibnamefont {Johnston}}, \bibinfo {author} {\bibfnamefont {J.~W.}\ \bibnamefont {Schneider}}, \bibinfo {author} {\bibfnamefont {J.}~\bibnamefont {Sonier}}, \bibinfo {author} {\bibfnamefont {L.}~\bibnamefont {Brard}}, \bibinfo {author} {\bibfnamefont {R.~M.}\ \bibnamefont {Strongin}}, \bibinfo {author} {\bibfnamefont {J.~E.}\ \bibnamefont {Fischer}},\ and\ \bibinfo {author} {\bibfnamefont {A.~B.}\ \bibnamefont {Smith}},\ }\bibfield  {title} {\bibinfo {title} {{Coherence peak and superconducting energy gap in ${\mathrm{Rb}}_{3}$${\mathrm{C}}_{60}$ observed by muon spin relaxation}},\ }\href {https://doi.org/10.1103/PhysRevLett.70.3987} {\bibfield  {journal} {\bibinfo  {journal} {Phys. Rev. Lett.}\ }\textbf {\bibinfo {volume} {70}},\ \bibinfo {pages} {3987} (\bibinfo {year} {1993})}\BibitemShut {NoStop}%
\bibitem [{\citenamefont {Kotegawa}\ \emph {et~al.}(2001)\citenamefont {Kotegawa}, \citenamefont {Ishida}, \citenamefont {Kitaoka}, \citenamefont {Muranaka},\ and\ \citenamefont {Akimitsu}}]{2001-Kotegawa-PRL-87-127001}%
  \BibitemOpen
  \bibfield  {author} {\bibinfo {author} {\bibfnamefont {H.}~\bibnamefont {Kotegawa}}, \bibinfo {author} {\bibfnamefont {K.}~\bibnamefont {Ishida}}, \bibinfo {author} {\bibfnamefont {Y.}~\bibnamefont {Kitaoka}}, \bibinfo {author} {\bibfnamefont {T.}~\bibnamefont {Muranaka}},\ and\ \bibinfo {author} {\bibfnamefont {J.}~\bibnamefont {Akimitsu}},\ }\bibfield  {title} {\bibinfo {title} {{Evidence for Strong-Coupling $\mathit{s}$-Wave Superconductivity in ${\mathrm{MgB}}_{2}$: $^{11}$B NMR Study}},\ }\href {https://doi.org/10.1103/PhysRevLett.87.127001} {\bibfield  {journal} {\bibinfo  {journal} {Phys. Rev. Lett.}\ }\textbf {\bibinfo {volume} {87}},\ \bibinfo {pages} {127001} (\bibinfo {year} {2001})}\BibitemShut {NoStop}%
\bibitem [{\citenamefont {Curro}\ \emph {et~al.}(2005)\citenamefont {Curro}, \citenamefont {Caldwell}, \citenamefont {Bauer}, \citenamefont {Morales}, \citenamefont {Graf}, \citenamefont {Bang}, \citenamefont {Balatsky}, \citenamefont {Thompson},\ and\ \citenamefont {Sarrao}}]{2005-Curro-Nature-434-622}%
  \BibitemOpen
  \bibfield  {author} {\bibinfo {author} {\bibfnamefont {N.~J.}\ \bibnamefont {Curro}}, \bibinfo {author} {\bibfnamefont {T.}~\bibnamefont {Caldwell}}, \bibinfo {author} {\bibfnamefont {E.~D.}\ \bibnamefont {Bauer}}, \bibinfo {author} {\bibfnamefont {L.~A.}\ \bibnamefont {Morales}}, \bibinfo {author} {\bibfnamefont {M.~J.}\ \bibnamefont {Graf}}, \bibinfo {author} {\bibfnamefont {Y.}~\bibnamefont {Bang}}, \bibinfo {author} {\bibfnamefont {A.~V.}\ \bibnamefont {Balatsky}}, \bibinfo {author} {\bibfnamefont {J.~D.}\ \bibnamefont {Thompson}},\ and\ \bibinfo {author} {\bibfnamefont {J.~L.}\ \bibnamefont {Sarrao}},\ }\bibfield  {title} {\bibinfo {title} {{Unconventional superconductivity in \ch{PuCoGa5}}},\ }\href {https://doi.org/10.1038/nature03428} {\bibfield  {journal} {\bibinfo  {journal} {Nature}\ }\textbf {\bibinfo {volume} {434}},\ \bibinfo {pages} {622–625} (\bibinfo {year} {2005})}\BibitemShut {NoStop}%
\bibitem [{\citenamefont {Sheahen}(1966)}]{1966-Sheahen-PR-149-368}%
  \BibitemOpen
  \bibfield  {author} {\bibinfo {author} {\bibfnamefont {T.~P.}\ \bibnamefont {Sheahen}},\ }\bibfield  {title} {\bibinfo {title} {{Rules for the Energy Gap and Critical Field of Superconductors}},\ }\href {https://doi.org/10.1103/PhysRev.149.368} {\bibfield  {journal} {\bibinfo  {journal} {Phys. Rev.}\ }\textbf {\bibinfo {volume} {149}},\ \bibinfo {pages} {368} (\bibinfo {year} {1966})}\BibitemShut {NoStop}%
\bibitem [{\citenamefont {Beckmann}(1988)}]{1988-Beckmann-PR-171-85}%
  \BibitemOpen
  \bibfield  {author} {\bibinfo {author} {\bibfnamefont {P.~A.}\ \bibnamefont {Beckmann}},\ }\bibfield  {title} {\bibinfo {title} {Spectral densities and nuclear spin relaxation in solids},\ }\href {https://doi.org/10.1016/0370-1573(88)90073-7} {\bibfield  {journal} {\bibinfo  {journal} {Phys. Rep.}\ }\textbf {\bibinfo {volume} {171}},\ \bibinfo {pages} {85} (\bibinfo {year} {1988})}\BibitemShut {NoStop}%
\bibitem [{\citenamefont {Barends}\ \emph {et~al.}(2009)\citenamefont {Barends}, \citenamefont {Hortensius}, \citenamefont {Zijlstra}, \citenamefont {Baselmans}, \citenamefont {Yates}, \citenamefont {Gao},\ and\ \citenamefont {Klapwijk}}]{2009-Barends-IEEETAS-19-936}%
  \BibitemOpen
  \bibfield  {author} {\bibinfo {author} {\bibfnamefont {R.}~\bibnamefont {Barends}}, \bibinfo {author} {\bibfnamefont {H.~L.}\ \bibnamefont {Hortensius}}, \bibinfo {author} {\bibfnamefont {T.}~\bibnamefont {Zijlstra}}, \bibinfo {author} {\bibfnamefont {J.~J.~A.}\ \bibnamefont {Baselmans}}, \bibinfo {author} {\bibfnamefont {S.~J.~C.}\ \bibnamefont {Yates}}, \bibinfo {author} {\bibfnamefont {J.~R.}\ \bibnamefont {Gao}},\ and\ \bibinfo {author} {\bibfnamefont {T.~M.}\ \bibnamefont {Klapwijk}},\ }\bibfield  {title} {\bibinfo {title} {{Noise in NbTiN, Al, and Ta Superconducting Resonators on Silicon and Sapphire Substrates}},\ }\href {https://doi.org/10.1109/TASC.2009.2018086} {\bibfield  {journal} {\bibinfo  {journal} {IEEE Trans. Appl. Supercond.}\ }\textbf {\bibinfo {volume} {19}},\ \bibinfo {pages} {936} (\bibinfo {year} {2009})}\BibitemShut {NoStop}%
\bibitem [{\citenamefont {Barends}\ \emph {et~al.}(2008)\citenamefont {Barends}, \citenamefont {Hortensius}, \citenamefont {Zijlstra}, \citenamefont {Baselmans}, \citenamefont {Yates}, \citenamefont {Gao},\ and\ \citenamefont {Klapwijk}}]{2008-Barends-APL-92-223502}%
  \BibitemOpen
  \bibfield  {author} {\bibinfo {author} {\bibfnamefont {R.}~\bibnamefont {Barends}}, \bibinfo {author} {\bibfnamefont {H.~L.}\ \bibnamefont {Hortensius}}, \bibinfo {author} {\bibfnamefont {T.}~\bibnamefont {Zijlstra}}, \bibinfo {author} {\bibfnamefont {J.~J.~A.}\ \bibnamefont {Baselmans}}, \bibinfo {author} {\bibfnamefont {S.~J.~C.}\ \bibnamefont {Yates}}, \bibinfo {author} {\bibfnamefont {J.~R.}\ \bibnamefont {Gao}},\ and\ \bibinfo {author} {\bibfnamefont {T.~M.}\ \bibnamefont {Klapwijk}},\ }\bibfield  {title} {\bibinfo {title} {{Contribution of dielectrics to frequency and noise of NbTiN superconducting resonators}},\ }\href {https://doi.org/10.1063/1.2937837} {\bibfield  {journal} {\bibinfo  {journal} {Appl. Phys. Lett.}\ }\textbf {\bibinfo {volume} {92}},\ \bibinfo {pages} {223502} (\bibinfo {year} {2008})}\BibitemShut {NoStop}%
\bibitem [{\citenamefont {Tinkham}(1996)}]{1996-Tinkham-Book-2-McGraw}%
  \BibitemOpen
  \bibfield  {author} {\bibinfo {author} {\bibfnamefont {M.}~\bibnamefont {Tinkham}},\ }\href@noop {} {\emph {\bibinfo {title} {Introduction to Superconductivity}}},\ \bibinfo {edition} {2nd}\ ed.,\ International Series in Pure and Applied Physics\ (\bibinfo  {publisher} {McGraw-Hill},\ \bibinfo {address} {New York},\ \bibinfo {year} {1996})\BibitemShut {NoStop}%
\bibitem [{\citenamefont {Junginger}(2012)}]{2012-Tobi-thesis}%
  \BibitemOpen
  \bibfield  {author} {\bibinfo {author} {\bibfnamefont {T.}~\bibnamefont {Junginger}},\ }\emph {\bibinfo {title} {{Investigations of the surface resistance of superconducting materials}}},\ \href {https://doi.org/10.11588/heidok.00013728} {Ph.D. thesis},\ \bibinfo  {school} {University of Heidelberg} (\bibinfo {year} {2012})\BibitemShut {NoStop}%
\bibitem [{\citenamefont {Slichter}(1990)}]{1990-Slichter-Book-3-Springer}%
  \BibitemOpen
  \bibfield  {author} {\bibinfo {author} {\bibfnamefont {C.~P.}\ \bibnamefont {Slichter}},\ }\href {https://doi.org/10.1007/978-3-662-09441-9} {\emph {\bibinfo {title} {Principles of Magnetic Resonance}}}\ (\bibinfo  {publisher} {Springer Berlin Heidelberg},\ \bibinfo {year} {1990})\BibitemShut {NoStop}%
\bibitem [{\citenamefont {G{\"o}tze}\ and\ \citenamefont {Ketterle}(1983)}]{1983-Gotze-ZPB-54-49}%
  \BibitemOpen
  \bibfield  {author} {\bibinfo {author} {\bibfnamefont {W.}~\bibnamefont {G{\"o}tze}}\ and\ \bibinfo {author} {\bibfnamefont {W.}~\bibnamefont {Ketterle}},\ }\bibfield  {title} {\bibinfo {title} {Nuclear spin relaxation in disordered conductors},\ }\href {https://doi.org/10.1007/BF01507949} {\bibfield  {journal} {\bibinfo  {journal} {Z. Phys. B: Condens. Matter}\ }\textbf {\bibinfo {volume} {54}},\ \bibinfo {pages} {49} (\bibinfo {year} {1983})}\BibitemShut {NoStop}%
\bibitem [{\citenamefont {Shastry}\ and\ \citenamefont {Abrahams}(1994)}]{1994-Shastry-PRL-72-1933}%
  \BibitemOpen
  \bibfield  {author} {\bibinfo {author} {\bibfnamefont {B.~S.}\ \bibnamefont {Shastry}}\ and\ \bibinfo {author} {\bibfnamefont {E.}~\bibnamefont {Abrahams}},\ }\bibfield  {title} {\bibinfo {title} {{What does the Korringa ratio measure?}},\ }\href {https://doi.org/10.1103/PhysRevLett.72.1933} {\bibfield  {journal} {\bibinfo  {journal} {Phys. Rev. Lett.}\ }\textbf {\bibinfo {volume} {72}},\ \bibinfo {pages} {1933} (\bibinfo {year} {1994})}\BibitemShut {NoStop}%
\bibitem [{Note10()}]{Note10}%
  \BibitemOpen
  \bibinfo {note} {If different values were chosen, such as $K^\protect \mathrm {c} (\qty {20}{\kelvin }) = \qty {22.5 \pm 1.4}{\ppm } $ or $K^\protect \mathrm {c} (\qty {120}{\kelvin }) = \qty {20.3 \pm 1.3}{\ppm }$, the corresponding Korringa ratios would be $\protect \mathcal {K} = \num {3.4 \pm 0.5}$ and $\protect \mathcal {K} = \num {2.7 \pm 0.4}$, respectively.}\BibitemShut {Stop}%
\bibitem [{\citenamefont {Nishihara}\ \emph {et~al.}(2004)\citenamefont {Nishihara}, \citenamefont {Furutani}, \citenamefont {Yokota}, \citenamefont {Ohyanagi},\ and\ \citenamefont {Kumashiro}}]{2004-Nishihara-JAC-383-308}%
  \BibitemOpen
  \bibfield  {author} {\bibinfo {author} {\bibfnamefont {H.}~\bibnamefont {Nishihara}}, \bibinfo {author} {\bibfnamefont {Y.}~\bibnamefont {Furutani}}, \bibinfo {author} {\bibfnamefont {S.}~\bibnamefont {Yokota}}, \bibinfo {author} {\bibfnamefont {M.}~\bibnamefont {Ohyanagi}},\ and\ \bibinfo {author} {\bibfnamefont {Y.}~\bibnamefont {Kumashiro}},\ }\bibfield  {title} {\bibinfo {title} {Nuclear spin-lattice relaxation of \ch{^{93}Nb} in a superconducting \ch{NbN} synthesized by {SHS}},\ }\href {https://doi.org/10.1016/j.jallcom.2004.04.032} {\bibfield  {journal} {\bibinfo  {journal} {J. Alloys Compd.}\ }\textbf {\bibinfo {volume} {383}},\ \bibinfo {pages} {308} (\bibinfo {year} {2004})}\BibitemShut {NoStop}%
\bibitem [{\citenamefont {Lascialfari}\ \emph {et~al.}(2009)\citenamefont {Lascialfari}, \citenamefont {Rigamonti}, \citenamefont {Bernardi}, \citenamefont {Corti}, \citenamefont {Gauzzi},\ and\ \citenamefont {Villegier}}]{2009-Lascialfari-PRB-80-104505}%
  \BibitemOpen
  \bibfield  {author} {\bibinfo {author} {\bibfnamefont {A.}~\bibnamefont {Lascialfari}}, \bibinfo {author} {\bibfnamefont {A.}~\bibnamefont {Rigamonti}}, \bibinfo {author} {\bibfnamefont {E.}~\bibnamefont {Bernardi}}, \bibinfo {author} {\bibfnamefont {M.}~\bibnamefont {Corti}}, \bibinfo {author} {\bibfnamefont {A.}~\bibnamefont {Gauzzi}},\ and\ \bibinfo {author} {\bibfnamefont {J.~C.}\ \bibnamefont {Villegier}},\ }\bibfield  {title} {\bibinfo {title} {Superconducting properties of a textured \ch{NbN} film from \ch{^{93}Nb} {NMR} relaxation and magnetization measurements},\ }\href {https://doi.org/10.1103/PhysRevB.80.104505} {\bibfield  {journal} {\bibinfo  {journal} {Phys. Rev. B}\ }\textbf {\bibinfo {volume} {80}},\ \bibinfo {pages} {104505} (\bibinfo {year} {2009})}\BibitemShut {NoStop}%
\bibitem [{\citenamefont {Bahlouli}(1992)}]{1992-Bahlouli-PLA-164-206}%
  \BibitemOpen
  \bibfield  {author} {\bibinfo {author} {\bibfnamefont {H.}~\bibnamefont {Bahlouli}},\ }\bibfield  {title} {\bibinfo {title} {Nuclear spin relaxation rate in disordered superconductors},\ }\href {https://doi.org/10.1016/0375-9601(92)90704-P} {\bibfield  {journal} {\bibinfo  {journal} {Phys. Lett. A}\ }\textbf {\bibinfo {volume} {164}},\ \bibinfo {pages} {206} (\bibinfo {year} {1992})}\BibitemShut {NoStop}%
\bibitem [{\citenamefont {Devereaux}(1993)}]{1993-Devereaux-ZPBCM-90-65}%
  \BibitemOpen
  \bibfield  {author} {\bibinfo {author} {\bibfnamefont {T.~P.}\ \bibnamefont {Devereaux}},\ }\bibfield  {title} {\bibinfo {title} {Nuclear spin relaxation in strongly disordered superconductors},\ }\href {https://doi.org/10.1007/BF01321033} {\bibfield  {journal} {\bibinfo  {journal} {Z. Phys. B: Condens. Matter}\ }\textbf {\bibinfo {volume} {90}},\ \bibinfo {pages} {65} (\bibinfo {year} {1993})}\BibitemShut {NoStop}%
\bibitem [{Note11()}]{Note11}%
  \BibitemOpen
  \bibinfo {note} {This value reflects a spatial average over many vortex unit cells within the $\qty {\sim 3}{\milli \meter }$ \gls {bnmr} beam spot. Given the short~\gls {gl} coherence length $\xi _\protect \mathrm {GL} = \qty {4.3 \pm 0.5}{\nm }$, the vortex cores occupy a negligible volume fraction, so the spatial variation of $\Delta $ in the mixed state can be neglected.}\BibitemShut {Stop}%
\bibitem [{\citenamefont {Mattis}\ and\ \citenamefont {Bardeen}(1958)}]{1958-Mattis-PR-111-412}%
  \BibitemOpen
  \bibfield  {author} {\bibinfo {author} {\bibfnamefont {D.~C.}\ \bibnamefont {Mattis}}\ and\ \bibinfo {author} {\bibfnamefont {J.}~\bibnamefont {Bardeen}},\ }\bibfield  {title} {\bibinfo {title} {Theory of the anomalous skin effect in normal and superconducting metals},\ }\href {https://doi.org/10.1103/PhysRev.111.412} {\bibfield  {journal} {\bibinfo  {journal} {Phys. Rev.}\ }\textbf {\bibinfo {volume} {111}},\ \bibinfo {pages} {412} (\bibinfo {year} {1958})}\BibitemShut {NoStop}%
\bibitem [{\citenamefont {Chockalingam}\ \emph {et~al.}(2009)\citenamefont {Chockalingam}, \citenamefont {Chand}, \citenamefont {Kamlapure}, \citenamefont {Jesudasan}, \citenamefont {Mishra}, \citenamefont {Tripathi},\ and\ \citenamefont {Raychaudhuri}}]{2009-Chockalingam-PRB-79-094509}%
  \BibitemOpen
  \bibfield  {author} {\bibinfo {author} {\bibfnamefont {S.~P.}\ \bibnamefont {Chockalingam}}, \bibinfo {author} {\bibfnamefont {M.}~\bibnamefont {Chand}}, \bibinfo {author} {\bibfnamefont {A.}~\bibnamefont {Kamlapure}}, \bibinfo {author} {\bibfnamefont {J.}~\bibnamefont {Jesudasan}}, \bibinfo {author} {\bibfnamefont {A.}~\bibnamefont {Mishra}}, \bibinfo {author} {\bibfnamefont {V.}~\bibnamefont {Tripathi}},\ and\ \bibinfo {author} {\bibfnamefont {P.}~\bibnamefont {Raychaudhuri}},\ }\bibfield  {title} {\bibinfo {title} {{Tunneling studies in a homogeneously disordered $s$-wave superconductor: NbN}},\ }\href {https://doi.org/10.1103/PhysRevB.79.094509} {\bibfield  {journal} {\bibinfo  {journal} {Phys. Rev. B}\ }\textbf {\bibinfo {volume} {79}},\ \bibinfo {pages} {094509} (\bibinfo {year} {2009})}\BibitemShut {NoStop}%
\bibitem [{\citenamefont {Sac\'ep\'e}\ \emph {et~al.}(2008)\citenamefont {Sac\'ep\'e}, \citenamefont {Chapelier}, \citenamefont {Baturina}, \citenamefont {Vinokur}, \citenamefont {Baklanov},\ and\ \citenamefont {Sanquer}}]{2008-Sacepe-PRL-101-157006}%
  \BibitemOpen
  \bibfield  {author} {\bibinfo {author} {\bibfnamefont {B.}~\bibnamefont {Sac\'ep\'e}}, \bibinfo {author} {\bibfnamefont {C.}~\bibnamefont {Chapelier}}, \bibinfo {author} {\bibfnamefont {T.~I.}\ \bibnamefont {Baturina}}, \bibinfo {author} {\bibfnamefont {V.~M.}\ \bibnamefont {Vinokur}}, \bibinfo {author} {\bibfnamefont {M.~R.}\ \bibnamefont {Baklanov}},\ and\ \bibinfo {author} {\bibfnamefont {M.}~\bibnamefont {Sanquer}},\ }\bibfield  {title} {\bibinfo {title} {{Disorder-Induced Inhomogeneities of the Superconducting State Close to the Superconductor-Insulator Transition}},\ }\href {https://doi.org/10.1103/PhysRevLett.101.157006} {\bibfield  {journal} {\bibinfo  {journal} {Phys. Rev. Lett.}\ }\textbf {\bibinfo {volume} {101}},\ \bibinfo {pages} {157006} (\bibinfo {year} {2008})}\BibitemShut {NoStop}%
\bibitem [{Note12()}]{Note12}%
  \BibitemOpen
  \bibinfo {note} {Particularly in ellipsoidal samples, which prevent field penetration from both sides of the superconducting layers and ensure a uniform magnetic response, making them a relevant proxy for~\gls {srf} cavities.}\BibitemShut {Stop}%
\bibitem [{\citenamefont {Thoeng}\ \emph {et~al.}(2024)\citenamefont {Thoeng}, \citenamefont {Asaduzzaman}, \citenamefont {Kolb}, \citenamefont {McFadden}, \citenamefont {Morris}, \citenamefont {Ticknor}, \citenamefont {Dunsiger}, \citenamefont {Karner}, \citenamefont {Fujimoto}, \citenamefont {Junginger}, \citenamefont {Kiefl}, \citenamefont {MacFarlane}, \citenamefont {Li}, \citenamefont {Saminathan},\ and\ \citenamefont {Laxdal}}]{2024-Thoeng-SR-14-21487}%
  \BibitemOpen
  \bibfield  {author} {\bibinfo {author} {\bibfnamefont {E.}~\bibnamefont {Thoeng}}, \bibinfo {author} {\bibfnamefont {M.}~\bibnamefont {Asaduzzaman}}, \bibinfo {author} {\bibfnamefont {P.}~\bibnamefont {Kolb}}, \bibinfo {author} {\bibfnamefont {R.~M.~L.}\ \bibnamefont {McFadden}}, \bibinfo {author} {\bibfnamefont {G.~D.}\ \bibnamefont {Morris}}, \bibinfo {author} {\bibfnamefont {J.~O.}\ \bibnamefont {Ticknor}}, \bibinfo {author} {\bibfnamefont {S.~R.}\ \bibnamefont {Dunsiger}}, \bibinfo {author} {\bibfnamefont {V.~L.}\ \bibnamefont {Karner}}, \bibinfo {author} {\bibfnamefont {D.}~\bibnamefont {Fujimoto}}, \bibinfo {author} {\bibfnamefont {T.}~\bibnamefont {Junginger}}, \bibinfo {author} {\bibfnamefont {R.~F.}\ \bibnamefont {Kiefl}}, \bibinfo {author} {\bibfnamefont {W.~A.}\ \bibnamefont {MacFarlane}}, \bibinfo {author} {\bibfnamefont {R.}~\bibnamefont {Li}}, \bibinfo {author} {\bibfnamefont {S.}~\bibnamefont {Saminathan}},\ and\ \bibinfo {author} {\bibfnamefont {R.~E.}\ \bibnamefont {Laxdal}},\ }\bibfield  {title} {\bibinfo {title} {Depth-resolved characterization of {Meissner} screening breakdown in surface treated niobium},\ }\href {https://doi.org/10.1038/s41598-024-71724-5} {\bibfield  {journal} {\bibinfo  {journal} {Sci. Rep.}\ }\textbf {\bibinfo {volume} {14}},\ \bibinfo {pages} {21487} (\bibinfo {year} {2024})}\BibitemShut {NoStop}%
\bibitem [{\citenamefont {Thoeng}\ \emph {et~al.}(2023)\citenamefont {Thoeng}, \citenamefont {McFadden}, \citenamefont {Saminathan}, \citenamefont {Morris}, \citenamefont {Kolb}, \citenamefont {Matheson}, \citenamefont {Asaduzzaman}, \citenamefont {Baartman}, \citenamefont {Dunsiger}, \citenamefont {Fujimoto}, \citenamefont {Junginger}, \citenamefont {Karner}, \citenamefont {Kiy}, \citenamefont {Li}, \citenamefont {Stachura}, \citenamefont {Ticknor}, \citenamefont {Kiefl}, \citenamefont {MacFarlane},\ and\ \citenamefont {Laxdal}}]{2023-Thoeng-RSI-94-023305}%
  \BibitemOpen
  \bibfield  {author} {\bibinfo {author} {\bibfnamefont {E.}~\bibnamefont {Thoeng}}, \bibinfo {author} {\bibfnamefont {R.~M.~L.}\ \bibnamefont {McFadden}}, \bibinfo {author} {\bibfnamefont {S.}~\bibnamefont {Saminathan}}, \bibinfo {author} {\bibfnamefont {G.~D.}\ \bibnamefont {Morris}}, \bibinfo {author} {\bibfnamefont {P.}~\bibnamefont {Kolb}}, \bibinfo {author} {\bibfnamefont {B.}~\bibnamefont {Matheson}}, \bibinfo {author} {\bibfnamefont {M.}~\bibnamefont {Asaduzzaman}}, \bibinfo {author} {\bibfnamefont {R.}~\bibnamefont {Baartman}}, \bibinfo {author} {\bibfnamefont {S.~R.}\ \bibnamefont {Dunsiger}}, \bibinfo {author} {\bibfnamefont {D.}~\bibnamefont {Fujimoto}}, \bibinfo {author} {\bibfnamefont {T.}~\bibnamefont {Junginger}}, \bibinfo {author} {\bibfnamefont {V.~L.}\ \bibnamefont {Karner}}, \bibinfo {author} {\bibfnamefont {S.}~\bibnamefont {Kiy}}, \bibinfo {author} {\bibfnamefont {R.}~\bibnamefont {Li}}, \bibinfo {author} {\bibfnamefont {M.}~\bibnamefont {Stachura}}, \bibinfo {author} {\bibfnamefont {J.~O.}\ \bibnamefont {Ticknor}}, \bibinfo {author} {\bibfnamefont {R.~F.}\ \bibnamefont {Kiefl}}, \bibinfo {author} {\bibfnamefont {W.~A.}\ \bibnamefont {MacFarlane}},\ and\ \bibinfo {author} {\bibfnamefont {R.~E.}\ \bibnamefont {Laxdal}},\ }\bibfield  {title} {\bibinfo {title} {A new high parallel-field spectrometer at {TRIUMF}'s {$\beta$-NMR} facility},\ }\href {https://doi.org/10.1063/5.0137368} {\bibfield  {journal} {\bibinfo  {journal} {Rev. Sci. Instrum.}\ }\textbf {\bibinfo {volume} {94}},\ \bibinfo {pages} {023305} (\bibinfo {year} {2023})}\BibitemShut {NoStop}%
\bibitem [{\citenamefont {MacFarlane}\ \emph {et~al.}(2014{\natexlab{b}})\citenamefont {MacFarlane}, \citenamefont {Tschense}, \citenamefont {Buck}, \citenamefont {Chow}, \citenamefont {Cortie}, \citenamefont {Hariwal}, \citenamefont {Kiefl}, \citenamefont {Koumoulis}, \citenamefont {Levy}, \citenamefont {McKenzie}, \citenamefont {McGee}, \citenamefont {Morris}, \citenamefont {Pearson}, \citenamefont {Song}, \citenamefont {Wang}, \citenamefont {Hor},\ and\ \citenamefont {Cava}}]{2014-MacFarlane-PRB-90-214422}%
  \BibitemOpen
  \bibfield  {author} {\bibinfo {author} {\bibfnamefont {W.~A.}\ \bibnamefont {MacFarlane}}, \bibinfo {author} {\bibfnamefont {C.~B.~L.}\ \bibnamefont {Tschense}}, \bibinfo {author} {\bibfnamefont {T.}~\bibnamefont {Buck}}, \bibinfo {author} {\bibfnamefont {K.~H.}\ \bibnamefont {Chow}}, \bibinfo {author} {\bibfnamefont {D.~L.}\ \bibnamefont {Cortie}}, \bibinfo {author} {\bibfnamefont {A.~N.}\ \bibnamefont {Hariwal}}, \bibinfo {author} {\bibfnamefont {R.~F.}\ \bibnamefont {Kiefl}}, \bibinfo {author} {\bibfnamefont {D.}~\bibnamefont {Koumoulis}}, \bibinfo {author} {\bibfnamefont {C.~D.~P.}\ \bibnamefont {Levy}}, \bibinfo {author} {\bibfnamefont {I.}~\bibnamefont {McKenzie}}, \bibinfo {author} {\bibfnamefont {F.~H.}\ \bibnamefont {McGee}}, \bibinfo {author} {\bibfnamefont {G.~D.}\ \bibnamefont {Morris}}, \bibinfo {author} {\bibfnamefont {M.~R.}\ \bibnamefont {Pearson}}, \bibinfo {author} {\bibfnamefont {Q.}~\bibnamefont {Song}}, \bibinfo {author} {\bibfnamefont {D.}~\bibnamefont {Wang}}, \bibinfo {author} {\bibfnamefont {Y.~S.}\ \bibnamefont {Hor}},\ and\ \bibinfo {author} {\bibfnamefont {R.~J.}\ \bibnamefont {Cava}},\ }\bibfield  {title} {\bibinfo {title} {{$\beta$}-detected {NMR} of \ch{^{8}Li^{+}} in \ch{Bi}, \ch{Sb}, and the topological insulator \ch{Bi_{0.9}Sb_{0.1}}},\ }\href {https://doi.org/10.1103/PhysRevB.90.214422} {\bibfield  {journal} {\bibinfo  {journal} {Phys. Rev. B}\ }\textbf {\bibinfo {volume} {90}},\ \bibinfo {pages} {214422} (\bibinfo {year} {2014}{\natexlab{b}})}\BibitemShut {NoStop}%
\bibitem [{\citenamefont {Shannon}(1976)}]{1976-Shannon-AC-A32-751}%
  \BibitemOpen
  \bibfield  {author} {\bibinfo {author} {\bibfnamefont {R.~D.}\ \bibnamefont {Shannon}},\ }\bibfield  {title} {\bibinfo {title} {{Revised effective ionic radii and systematic studies of interatomic distances in halides and chalcogenides}},\ }\href {https://doi.org/10.1107/S0567739476001551} {\bibfield  {journal} {\bibinfo  {journal} {Acta Cryst.}\ }\textbf {\bibinfo {volume} {A32}},\ \bibinfo {pages} {751} (\bibinfo {year} {1976})}\BibitemShut {NoStop}%
\bibitem [{\citenamefont {Chow}\ \emph {et~al.}(2012)\citenamefont {Chow}, \citenamefont {Mansour}, \citenamefont {Fan}, \citenamefont {Kiefl}, \citenamefont {Morris}, \citenamefont {Salman}, \citenamefont {Dunlop}, \citenamefont {MacFarlane}, \citenamefont {Saadaoui}, \citenamefont {Mosendz}, \citenamefont {Kardasz}, \citenamefont {Heinrich}, \citenamefont {Jung}, \citenamefont {Levy}, \citenamefont {Pearson}, \citenamefont {Parolin}, \citenamefont {Wang}, \citenamefont {Hossain}, \citenamefont {Song},\ and\ \citenamefont {Smadella}}]{2012-Chow-PRB-85-092103}%
  \BibitemOpen
  \bibfield  {author} {\bibinfo {author} {\bibfnamefont {K.~H.}\ \bibnamefont {Chow}}, \bibinfo {author} {\bibfnamefont {A.~I.}\ \bibnamefont {Mansour}}, \bibinfo {author} {\bibfnamefont {I.}~\bibnamefont {Fan}}, \bibinfo {author} {\bibfnamefont {R.~F.}\ \bibnamefont {Kiefl}}, \bibinfo {author} {\bibfnamefont {G.~D.}\ \bibnamefont {Morris}}, \bibinfo {author} {\bibfnamefont {Z.}~\bibnamefont {Salman}}, \bibinfo {author} {\bibfnamefont {T.}~\bibnamefont {Dunlop}}, \bibinfo {author} {\bibfnamefont {W.~A.}\ \bibnamefont {MacFarlane}}, \bibinfo {author} {\bibfnamefont {H.}~\bibnamefont {Saadaoui}}, \bibinfo {author} {\bibfnamefont {O.}~\bibnamefont {Mosendz}}, \bibinfo {author} {\bibfnamefont {B.}~\bibnamefont {Kardasz}}, \bibinfo {author} {\bibfnamefont {B.}~\bibnamefont {Heinrich}}, \bibinfo {author} {\bibfnamefont {J.}~\bibnamefont {Jung}}, \bibinfo {author} {\bibfnamefont {C.~D.~P.}\ \bibnamefont {Levy}}, \bibinfo {author} {\bibfnamefont {M.~R.}\ \bibnamefont {Pearson}}, \bibinfo {author} {\bibfnamefont {T.~J.}\ \bibnamefont {Parolin}}, \bibinfo {author} {\bibfnamefont {D.}~\bibnamefont {Wang}}, \bibinfo {author} {\bibfnamefont {M.~D.}\ \bibnamefont {Hossain}}, \bibinfo {author} {\bibfnamefont {Q.}~\bibnamefont {Song}},\ and\ \bibinfo {author} {\bibfnamefont {M.}~\bibnamefont {Smadella}},\ }\bibfield  {title} {\bibinfo {title} {Detection and decoherence of level-crossing resonances of \ch{^{8}Li} in \ch{Cu}},\ }\href {https://doi.org/10.1103/PhysRevB.85.092103} {\bibfield  {journal} {\bibinfo  {journal} {Phys. Rev. B}\ }\textbf {\bibinfo {volume} {85}},\ \bibinfo {pages} {092103} (\bibinfo {year} {2012})}\BibitemShut {NoStop}%
\bibitem [{\citenamefont {Xu}\ \emph {et~al.}(2008)\citenamefont {Xu}, \citenamefont {Hossain}, \citenamefont {Saadaoui}, \citenamefont {Parolin}, \citenamefont {Chow}, \citenamefont {Keeler}, \citenamefont {Kiefl}, \citenamefont {Morris}, \citenamefont {Salman}, \citenamefont {Song}, \citenamefont {Wang},\ and\ \citenamefont {MacFarlane}}]{2008-Xu-JMR-191-47}%
  \BibitemOpen
  \bibfield  {author} {\bibinfo {author} {\bibfnamefont {M.}~\bibnamefont {Xu}}, \bibinfo {author} {\bibfnamefont {M.}~\bibnamefont {Hossain}}, \bibinfo {author} {\bibfnamefont {H.}~\bibnamefont {Saadaoui}}, \bibinfo {author} {\bibfnamefont {T.}~\bibnamefont {Parolin}}, \bibinfo {author} {\bibfnamefont {K.}~\bibnamefont {Chow}}, \bibinfo {author} {\bibfnamefont {T.}~\bibnamefont {Keeler}}, \bibinfo {author} {\bibfnamefont {R.}~\bibnamefont {Kiefl}}, \bibinfo {author} {\bibfnamefont {G.}~\bibnamefont {Morris}}, \bibinfo {author} {\bibfnamefont {Z.}~\bibnamefont {Salman}}, \bibinfo {author} {\bibfnamefont {Q.}~\bibnamefont {Song}}, \bibinfo {author} {\bibfnamefont {D.}~\bibnamefont {Wang}},\ and\ \bibinfo {author} {\bibfnamefont {W.}~\bibnamefont {MacFarlane}},\ }\bibfield  {title} {\bibinfo {title} {{Proximal magnetometry in thin films using $\beta$NMR}},\ }\href {https://doi.org/10.1016/j.jmr.2007.11.022} {\bibfield  {journal} {\bibinfo  {journal} {J. Magn. Reson.}\ }\textbf {\bibinfo {volume} {191}},\ \bibinfo {pages} {47} (\bibinfo {year} {2008})}\BibitemShut {NoStop}%
\bibitem [{\citenamefont {Prozorov}\ and\ \citenamefont {Kogan}(2018)}]{2018-Prozorov-PRA-10-014030}%
  \BibitemOpen
  \bibfield  {author} {\bibinfo {author} {\bibfnamefont {R.}~\bibnamefont {Prozorov}}\ and\ \bibinfo {author} {\bibfnamefont {V.~G.}\ \bibnamefont {Kogan}},\ }\bibfield  {title} {\bibinfo {title} {{Effective Demagnetizing Factors of Diamagnetic Samples of Various Shapes}},\ }\href {https://doi.org/10.1103/PhysRevApplied.10.014030} {\bibfield  {journal} {\bibinfo  {journal} {Phys. Rev. Appl.}\ }\textbf {\bibinfo {volume} {10}},\ \bibinfo {pages} {014030} (\bibinfo {year} {2018})}\BibitemShut {NoStop}%
\bibitem [{\citenamefont {Rietschel}\ \emph {et~al.}(1980)\citenamefont {Rietschel}, \citenamefont {Winter},\ and\ \citenamefont {Reichardt}}]{1980-Rietschel-PRB-22-4284}%
  \BibitemOpen
  \bibfield  {author} {\bibinfo {author} {\bibfnamefont {H.}~\bibnamefont {Rietschel}}, \bibinfo {author} {\bibfnamefont {H.}~\bibnamefont {Winter}},\ and\ \bibinfo {author} {\bibfnamefont {W.}~\bibnamefont {Reichardt}},\ }\bibfield  {title} {\bibinfo {title} {{Strong depression of superconductivity in VN by spin fluctuations}},\ }\href {https://doi.org/10.1103/PhysRevB.22.4284} {\bibfield  {journal} {\bibinfo  {journal} {Phys. Rev. B}\ }\textbf {\bibinfo {volume} {22}},\ \bibinfo {pages} {4284} (\bibinfo {year} {1980})}\BibitemShut {NoStop}%
\bibitem [{\citenamefont {Haynes}(2016)}]{2016-CRC-handbook}%
  \BibitemOpen
  \bibfield  {author} {\bibinfo {author} {\bibfnamefont {W.~M.}\ \bibnamefont {Haynes}},\ }\href {https://doi.org/10.1201/9781315380476} {\emph {\bibinfo {title} {{CRC Handbook of Chemistry and Physics}}}}\ (\bibinfo  {publisher} {CRC Press},\ \bibinfo {year} {2016})\BibitemShut {NoStop}%
\end{thebibliography}%


\begin{thebibliography}{4}%
\makeatletter
\providecommand \@ifxundefined [1]{%
 \@ifx{#1\undefined}
}%
\providecommand \@ifnum [1]{%
 \ifnum #1\expandafter \@firstoftwo
 \else \expandafter \@secondoftwo
 \fi
}%
\providecommand \@ifx [1]{%
 \ifx #1\expandafter \@firstoftwo
 \else \expandafter \@secondoftwo
 \fi
}%
\providecommand \natexlab [1]{#1}%
\providecommand \enquote  [1]{``#1''}%
\providecommand \bibnamefont  [1]{#1}%
\providecommand \bibfnamefont [1]{#1}%
\providecommand \citenamefont [1]{#1}%
\providecommand \href@noop [0]{\@secondoftwo}%
\providecommand \href [0]{\begingroup \@sanitize@url \@href}%
\providecommand \@href[1]{\@@startlink{#1}\@@href}%
\providecommand \@@href[1]{\endgroup#1\@@endlink}%
\providecommand \@sanitize@url [0]{\catcode `\\12\catcode `\$12\catcode `\&12\catcode `\#12\catcode `\^12\catcode `\_12\catcode `\%12\relax}%
\providecommand \@@startlink[1]{}%
\providecommand \@@endlink[0]{}%
\providecommand \url  [0]{\begingroup\@sanitize@url \@url }%
\providecommand \@url [1]{\endgroup\@href {#1}{\urlprefix }}%
\providecommand \urlprefix  [0]{URL }%
\providecommand \Eprint [0]{\href }%
\providecommand \doibase [0]{https://doi.org/}%
\providecommand \selectlanguage [0]{\@gobble}%
\providecommand \bibinfo  [0]{\@secondoftwo}%
\providecommand \bibfield  [0]{\@secondoftwo}%
\providecommand \translation [1]{[#1]}%
\providecommand \BibitemOpen [0]{}%
\providecommand \bibitemStop [0]{}%
\providecommand \bibitemNoStop [0]{.\EOS\space}%
\providecommand \EOS [0]{\spacefactor3000\relax}%
\providecommand \BibitemShut  [1]{\csname bibitem#1\endcsname}%
\let\auto@bib@innerbib\@empty
\bibitem [{\citenamefont {Miikkulainen}\ \emph {et~al.}(2013)\citenamefont {Miikkulainen}, \citenamefont {Leskelä}, \citenamefont {Ritala},\ and\ \citenamefont {Puurunen}}]{2013-Miikkulainen-JAP-113-021301}%
  \BibitemOpen
  \bibfield  {author} {\bibinfo {author} {\bibfnamefont {V.}~\bibnamefont {Miikkulainen}}, \bibinfo {author} {\bibfnamefont {M.}~\bibnamefont {Leskelä}}, \bibinfo {author} {\bibfnamefont {M.}~\bibnamefont {Ritala}},\ and\ \bibinfo {author} {\bibfnamefont {R.~L.}\ \bibnamefont {Puurunen}},\ }\bibfield  {title} {\bibinfo {title} {{Crystallinity of inorganic films grown by atomic layer deposition: Overview and general trends}},\ }\href {https://doi.org/10.1063/1.4757907} {\bibfield  {journal} {\bibinfo  {journal} {J. Appl. Phys.}\ }\textbf {\bibinfo {volume} {113}},\ \bibinfo {pages} {021301} (\bibinfo {year} {2013})}\BibitemShut {NoStop}%
\bibitem [{\citenamefont {Groll}\ \emph {et~al.}(2015)\citenamefont {Groll}, \citenamefont {Pellin}, \citenamefont {Zasadzinksi},\ and\ \citenamefont {Proslier}}]{2015-Groll-RSI-86-095111}%
  \BibitemOpen
  \bibfield  {author} {\bibinfo {author} {\bibfnamefont {N.}~\bibnamefont {Groll}}, \bibinfo {author} {\bibfnamefont {M.~J.}\ \bibnamefont {Pellin}}, \bibinfo {author} {\bibfnamefont {J.~F.}\ \bibnamefont {Zasadzinksi}},\ and\ \bibinfo {author} {\bibfnamefont {T.}~\bibnamefont {Proslier}},\ }\bibfield  {title} {\bibinfo {title} {Point contact tunneling spectroscopy apparatus for large scale mapping of surface superconducting properties},\ }\href {https://doi.org/10.1063/1.4931066} {\bibfield  {journal} {\bibinfo  {journal} {Rev. Sci. Instrum.}\ }\textbf {\bibinfo {volume} {86}},\ \bibinfo {pages} {095111} (\bibinfo {year} {2015})}\BibitemShut {NoStop}%
\bibitem [{\citenamefont {Kalboussi}\ \emph {et~al.}(2025)\citenamefont {Kalboussi}, \citenamefont {Curci}, \citenamefont {Miserque}, \citenamefont {Troadec}, \citenamefont {Brun}, \citenamefont {Walls}, \citenamefont {Jullien}, \citenamefont {Eozenou}, \citenamefont {Baudrier}, \citenamefont {Maurice}, \citenamefont {Bertrand}, \citenamefont {Sahuquet},\ and\ \citenamefont {Proslier}}]{2025-Kalboussi-PRA-23-044023}%
  \BibitemOpen
  \bibfield  {author} {\bibinfo {author} {\bibfnamefont {Y.}~\bibnamefont {Kalboussi}}, \bibinfo {author} {\bibfnamefont {I.}~\bibnamefont {Curci}}, \bibinfo {author} {\bibfnamefont {F.}~\bibnamefont {Miserque}}, \bibinfo {author} {\bibfnamefont {D.}~\bibnamefont {Troadec}}, \bibinfo {author} {\bibfnamefont {N.}~\bibnamefont {Brun}}, \bibinfo {author} {\bibfnamefont {M.}~\bibnamefont {Walls}}, \bibinfo {author} {\bibfnamefont {G.}~\bibnamefont {Jullien}}, \bibinfo {author} {\bibfnamefont {F.}~\bibnamefont {Eozenou}}, \bibinfo {author} {\bibfnamefont {M.}~\bibnamefont {Baudrier}}, \bibinfo {author} {\bibfnamefont {L.}~\bibnamefont {Maurice}}, \bibinfo {author} {\bibfnamefont {Q.}~\bibnamefont {Bertrand}}, \bibinfo {author} {\bibfnamefont {P.}~\bibnamefont {Sahuquet}},\ and\ \bibinfo {author} {\bibfnamefont {T.}~\bibnamefont {Proslier}},\ }\bibfield  {title} {\bibinfo {title} {{Crystallinity in niobium oxides: A pathway to mitigate two-level-system defects in niobium three-dimensional resonators for quantum applications}},\ }\href {https://doi.org/10.1103/PhysRevApplied.23.044023} {\bibfield  {journal} {\bibinfo  {journal} {Phys. Rev. Appl.}\ }\textbf {\bibinfo {volume} {23}},\ \bibinfo {pages} {044023} (\bibinfo {year} {2025})}\BibitemShut {NoStop}%
\bibitem [{\citenamefont {Dynes}\ \emph {et~al.}(1978)\citenamefont {Dynes}, \citenamefont {Narayanamurti},\ and\ \citenamefont {Garno}}]{1978-Dynes-PRL-41-1509}%
  \BibitemOpen
  \bibfield  {author} {\bibinfo {author} {\bibfnamefont {R.~C.}\ \bibnamefont {Dynes}}, \bibinfo {author} {\bibfnamefont {V.}~\bibnamefont {Narayanamurti}},\ and\ \bibinfo {author} {\bibfnamefont {J.~P.}\ \bibnamefont {Garno}},\ }\bibfield  {title} {\bibinfo {title} {{Direct Measurement of Quasiparticle-Lifetime Broadening in a Strong-Coupled Superconductor}},\ }\href {https://doi.org/10.1103/PhysRevLett.41.1509} {\bibfield  {journal} {\bibinfo  {journal} {Phys. Rev. Lett.}\ }\textbf {\bibinfo {volume} {41}},\ \bibinfo {pages} {1509} (\bibinfo {year} {1978})}\BibitemShut {NoStop}%
\end{thebibliography}%

\end{document}


\title{
	Supplementary material for \\``Superconducting properties of thin film \ch{Nb_{1-x}Ti_xN} studied via the NMR of implanted \ch{^8Li}''
}

\author{Md~Asaduzzaman}
\email[E-mail: ]{asadm@uvic.ca}
\affiliation{Department of Physics and Astronomy, University of Victoria, 3800 Finnerty Road, Victoria, BC V8P~5C2, Canada}
\affiliation{TRIUMF, 4004 Wesbrook Mall, Vancouver, BC V6T~2A3, Canada}

\author{Ryan~M.~L.~McFadden}
\affiliation{Department of Physics and Astronomy, University of Victoria, 3800 Finnerty Road, Victoria, BC V8P~5C2, Canada}
\affiliation{TRIUMF, 4004 Wesbrook Mall, Vancouver, BC V6T~2A3, Canada}

\author{Edward~Thoeng}
\affiliation{Department of Physics and Astronomy, University of British Columbia, 6224 Agricultural Road, Vancouver, British Columbia V6T 1Z1, Canada}
\affiliation{TRIUMF, 4004 Wesbrook Mall, Vancouver, BC V6T~2A3, Canada}

\author{Yasmine~Kalboussi}
\affiliation{Institut des lois fondamentales de l’univers, Commissariat de l’énergie atomique-centre de saclay, Paris-Saclay University, 91191 Gif-sur-Yvette, France}

\author{Ivana~Curci}
\affiliation{Institut des lois fondamentales de l’univers, Commissariat de l’énergie atomique-centre de saclay, Paris-Saclay University, 91191 Gif-sur-Yvette, France}

\author{Thomas~Proslier}
\affiliation{Institut des lois fondamentales de l’univers, Commissariat de l’énergie atomique-centre de saclay, Paris-Saclay University, 91191 Gif-sur-Yvette, France}

\author{Sarah~R.~Dunsiger}
\affiliation{TRIUMF, 4004 Wesbrook Mall, Vancouver, BC V6T~2A3, Canada}
\affiliation{Department of Physics, Simon Fraser University, 8888 University Drive, Burnaby, BC V5A~1S6, Canada}

\author{W.~Andrew~MacFarlane}
\affiliation{TRIUMF, 4004 Wesbrook Mall, Vancouver, BC V6T~2A3, Canada}
\affiliation{Department of Chemistry, University of British Columbia, 2036 Main Mall, Vancouver, BC V6T~1Z1, Canada}
\affiliation{Stewart Blusson Quantum Matter Institute, University of British Columbia, Vancouver, BC V6T~1Z4, Canada}

\author{Gerald~D.~Morris}
\affiliation{TRIUMF, 4004 Wesbrook Mall, Vancouver, BC V6T~2A3, Canada}

\author{Ruohong~Li}
\affiliation{TRIUMF, 4004 Wesbrook Mall, Vancouver, BC V6T~2A3, Canada}

\author{John~O.~Ticknor}
\affiliation{Department of Chemistry, University of British Columbia, 2036 Main Mall, Vancouver, BC V6T 1Z1, Canada}
\affiliation{Stewart Blusson Quantum Matter Institute, University of British Columbia, Vancouver, BC V6T~1Z4, Canada}

\author{Robert~E.~Laxdal}
\affiliation{Department of Physics and Astronomy, University of Victoria, 3800 Finnerty Road, Victoria, BC V8P~5C2, Canada}
\affiliation{TRIUMF, 4004 Wesbrook Mall, Vancouver, BC V6T~2A3, Canada}
	
\author{Tobias~Junginger}
\email[E-mail: ]{junginger@uvic.ca}
\affiliation{Department of Physics and Astronomy, University of Victoria, 3800 Finnerty Road, Victoria, BC V8P~5C2, Canada}
\affiliation{TRIUMF, 4004 Wesbrook Mall, Vancouver, BC V6T~2A3, Canada}

\date{\today}

\maketitle
\section{Structural and Thin Film Characterization
\label{sec:xrr-gixrd}
}

\begin{figure}
	\centering
	\includegraphics*[width=0.5\columnwidth]{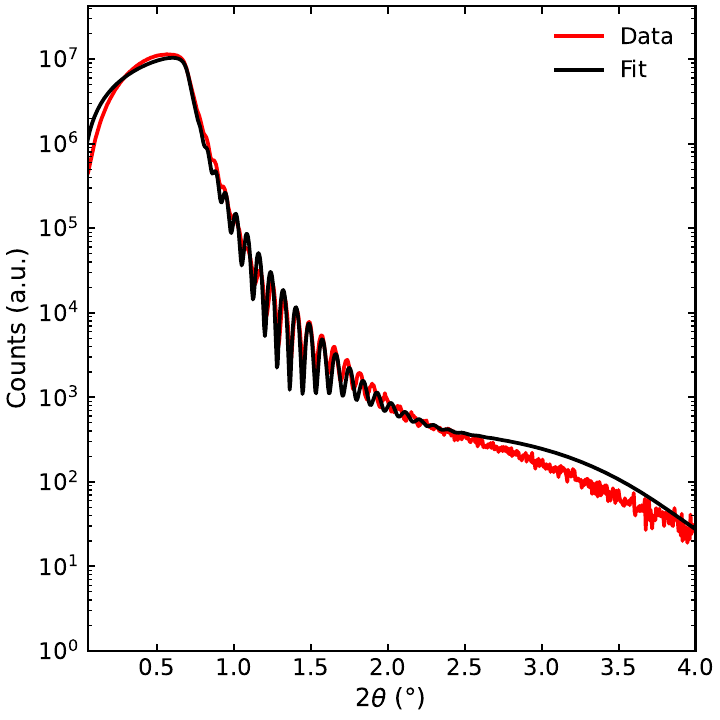}
	\caption{
	\label{fig:xrr} 
	\Gls{xrr} measurement of the \ch{Nb_{0.75}Ti_{0.25}N}(\qty{91}{\nm})/\ch{AlN}(\qty{4}{\nm}) bilayer deposited on an \ch{Al2O3} substrate. Experimental data (red curve) and the corresponding fit model (black curve) are shown.
	}
\end{figure}

\begin{figure}
	\centering
	\includegraphics*[width=0.5\columnwidth]{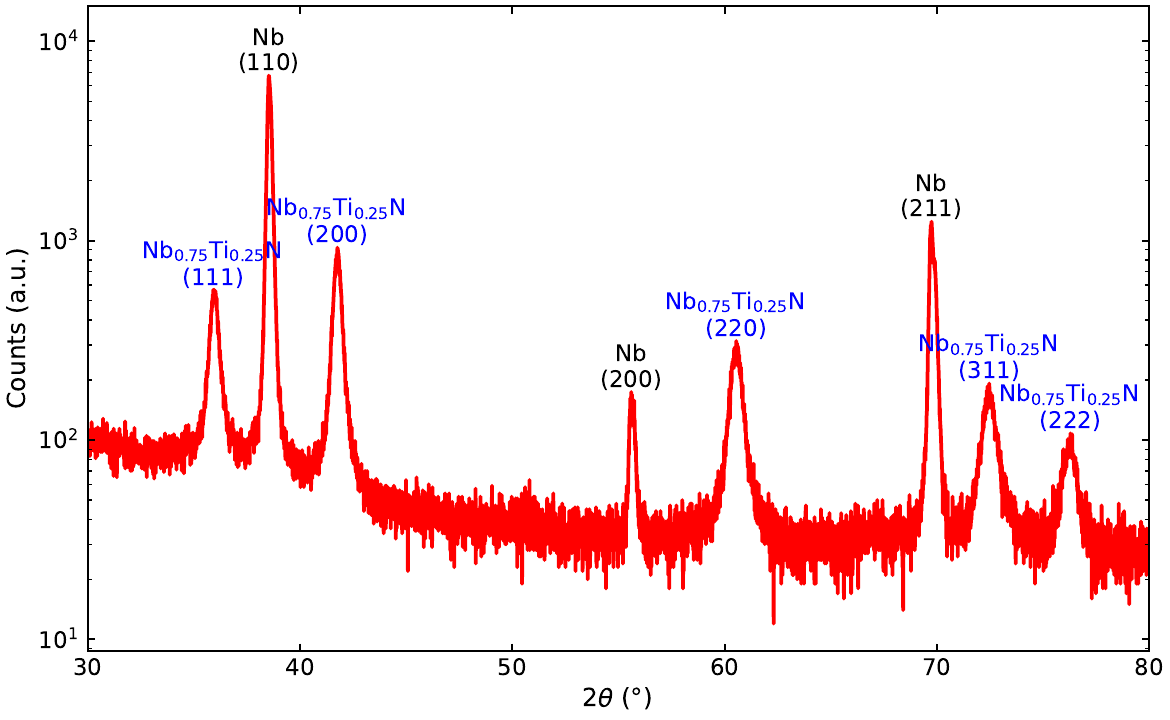}
	\caption{
	\label{fig:xrd} 
	\Gls{gixrd} pattern of the \ch{Nb_{0.75}Ti_{0.25}N}/\ch{AlN}/\ch{Nb} multilayer structure. 
	From the measured diffraction peak positions, the lattice parameter was determined to be $a = \qty{4.313}{\angstrom}$.
	The diffraction peaks corresponding to the \ch{Nb_{0.75}Ti_{0.25}N} film (shown in blue) indexed according to the cubic $B1$ \ch{NaCl}-type structure. The prominent reflections appear at positions consistent with the (111), (200), (220), (311), and (222) planes of \ch{Nb_{0.75}Ti_{0.25}N}. Peaks corresponding to the \ch{Nb} substrate (shown in black) indexed as (110), (200), and (211) reflections, confirming \gls{fcc} structure of the space group $Fm\bar{3}m$ as expected.	
	The logarithmic intensity scale highlights both strong and weak reflections across the range of $2\theta$ values from \qty{30}{\degree} to \qty{80}{\degree}.
	}
\end{figure}

The \ch{Nb_{1-x}Ti_{x}N}/\ch{AlN} bilayer was deposited on a \ch{Nb} substrate using thermal~\gls{ald} in a custom-built reactor at CEA Saclay. ~\gls{ald} is a chemical-phase film deposition technique based on sequential, self-limiting gas-surface reactions, enabling atomic-scale thickness control. During this process, two or more chemical precursors are introduced to the surface separately, one at a time, following a cyclic sequence~\cite{2013-Miikkulainen-JAP-113-021301}. 
The thickness and density of the deposited \ch{Nb_{0.75}Ti_{0.25}N}(\qty{91}{\nm})/\ch{AlN}(\qty{4}{\nm}) bilayer were determined using~\gls{xrr} on a witness \ch{Al2O3} sample, measured with a Rigaku Smartlab diffractometer using  Cu K-alpha radiation. ~\Cref{fig:xrr} represents fitting the~\gls{xrr} data 
yielded a \ch{Nb_{0.75}Ti_{0.25}N} thickness of~\qty{91}{\nm} with a density of ~\qty{6.5}{\gram\per\centi\meter\cubed} and an \ch{AlN} thickness of ~\qty{4}{\nm} with a density of ~\qty{3.26}{\gram\per\centi\meter\cubed}. ~\gls{gixrd} confirmed a cubic $B1$ structure for the \ch{Nb_{0.75}Ti_{0.25}N} film,
with a measured lattice parameter of $a =\qty{4.313}{\angstrom}$, as shown in~\Cref{fig:xrd}.  
Additionally, the \ch{Nb_{0.75}Ti_{0.25}N} film resistivity was measured to be \qty{124}{\micro\ohm\centi\metre} on the same witness sample using a standard four-point probe method.

\begin{figure}
	\centering
	\includegraphics*[width=0.5\columnwidth]{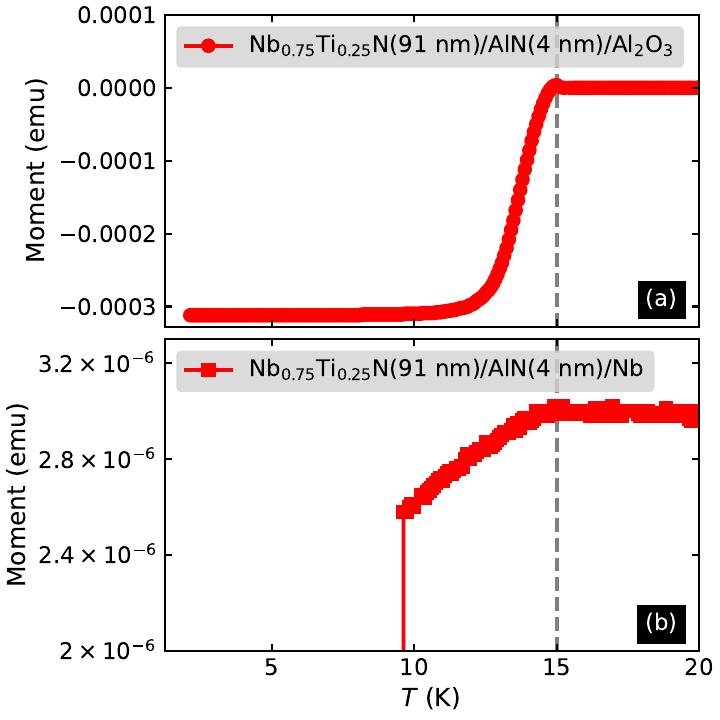}
	\caption{
	\label{fig:tc-sapphire-Nb} 
	Superconducting transition temperature $T_\mathrm{c}$ measurements using VSM for (a) the \ch{Nb_{0.75}Ti_{0.25}N}(\qty{91}{\nm})/\ch{AlN}(\qty{4}{\nm})/\ch{Al2O3} sample, showing a single transition at~\qty{15}{\kelvin}, and (b) the \ch{Nb_{0.75}Ti_{0.25}N}(\qty{91}{\nm})/\ch{AlN}(\qty{4}{\nm})/\ch{Nb} sample, displaying two transitions at~\qty{15}{\kelvin} and~\qty{9.3}{\kelvin}. The dashed lines indicate the corresponding $T_\mathrm{c}$ values in each plot.
	}
\end{figure}

\section{Superconducting Properties Characterization
\label{sec:vsm-pct}
}

The superconducting critical temperature $T_\mathrm{c}$ was measured using \gls{vsm}, which detects the ``bulk'' magnetization as a function of temperature, identifying $T_\mathrm{c}$ by the sharp drop in magnetization as the sample enters the Meissner state. For the \ch{Nb_{0.75}Ti_{0.25}N}(\qty{91}{\nano\meter})/\ch{AlN}(\qty{4}{\nano\meter})/\ch{Al2O3} sample, a clear superconducting transition was observed at approximately~\qty{15}{\kelvin} (see~\Cref{fig:tc-sapphire-Nb}(a)). In contrast, the \ch{Nb_{0.75}Ti_{0.25}N}(\qty{91}{\nano\meter})/\ch{AlN}(\qty{4}{\nano\meter})/\ch{Nb} multilayer exhibited two distinct transitions: a first at \qty{15}{\kelvin}, attributed to the \ch{Nb_{0.75}Ti_{0.25}N} layer, and a second at \qty{9.3}{\kelvin}, corresponding to the \ch{Nb} substrate, as shown in~\Cref{fig:tc-sapphire-Nb}(b).

The surface superconducting properties were characterized on similarly prepared \ch{Nb_{0.75}Ti_{0.25}N}(\qty{50}{\nm})/\ch{AlN}(\qty{6}{\nm})/\ch{Nb} multilayer structure using~\gls{pct}, following the methodology described in Ref.~\cite{2015-Groll-RSI-86-095111}. This technique is a powerful and sensitive probe of the~\gls{dos} in superconducting materials and has recently been applied in~\gls{srf} contexts to reveal metallic inclusions and crystalline inhomogeneities in niobium surfaces~\cite{2025-Kalboussi-PRA-23-044023}.
In our setup, junctions were formed by approaching the sample surface with an \ch{Al} tip, creating a superconductor-insulator-normal (SIN) junction where the insulator is the oxide layer on the sample surface.
The ~\gls{pct} system used in this study is capable of measuring junction resistances spanning from a quasi-ohmic regime (a few hundred \unit{\ohm}) to a tunneling regime (up to ~\qty{1}{\giga\ohm}), while enabling lateral mapping of superconducting properties over areas ranging from tens of~\unit{\micro\meter\squared} to several~\unit{\milli\meter\squared}.
In the tunneling regime, the current $I_\mathrm{ns}$ flowing between a normal (n) metal electrode (in this case, an \ch{Al} tip) and a superconducting (s) sample (e.g., a \ch{Nb_{0.75}Ti_{0.25}N} film) through an insulator (i.e., the oxide layer on the surface which is thin and formed during deposition) is measured as a function of the applied bias voltage $V$. The resulting~\gls{pct} spectra provide information about the superconducting DOS. To quantitatively interpret these measurements, the data were analyzed using the standard expression for the differential tunneling conductance, expressed as:
\begin{equation}
    \frac{dI_\mathrm{ns}(V)}{dV} \propto \int_{-\infty}^{\infty} N_\mathrm{s}(E) \left( - \frac{\partial f (E + eV)}{\partial (eV)} \right) \, \mathrm{d}E,
    \label{eq:current-derivative}
\end{equation}
where 
$N_\mathrm{s}(E)$ is the superconducting ~\gls{dos}, and $f(E)$ is the Fermi-Dirac distribution function.
Here the superconducting ~\gls{dos}, $N_\mathrm{s}(E)$, was modeled using the Dynes formula~\cite{1978-Dynes-PRL-41-1509}:
\begin{equation}
    N_\mathrm{s}(E) = N_\mathrm{n} \, \mathrm{Re} \left\{ \frac{E + i\Gamma_\mathrm{D}}{\sqrt{(E - i\Gamma_\mathrm{D})^2 - \Delta^2}} \right\},
	\label{eq:dos-dynes}
\end{equation}
where $N_\mathrm{n}$ is the normal-state DOS at the Fermi level, $\Delta$ is the superconducting energy gap, and $\Gamma_\mathrm{D}$ is the phenomenological quasiparticle lifetime broadening parameter.

\begin{figure}
	\centering
	\includegraphics*[width=0.5\columnwidth]{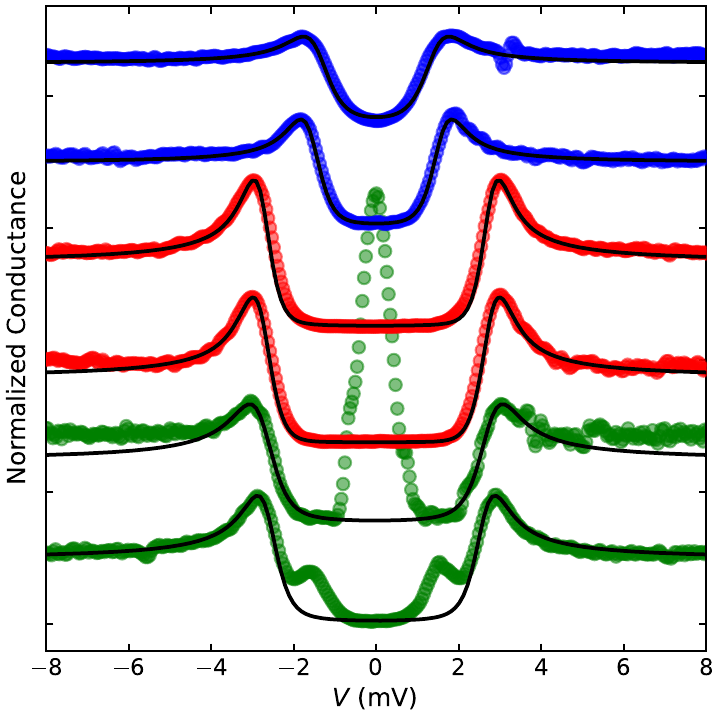}
	\caption{
	\label{fig:pct-spectra} 
	Representative tunneling conductance spectra measured on the \ch{Nb_{0.75}Ti_{0.25}N}(\qty{46}{\nm})/\ch{AlN}(\qty{8}{\nm})/\ch{Nb} multilayer sample at~\qty{1.8}{\kelvin} within a~\qtyproduct{100 x 100}{\micro\meter} area. Closed circles are measured data points, while solid lines represent fit to the data using~\Cref{eq:current-derivative,eq:dos-dynes}. The data illustrate the range of junction behaviors observed: two with smaller superconducting gaps (top), two characteristic of ideal \ch{Nb_{0.75}Ti_{0.25}N} spectra (middle), and two exhibiting enhanced subgap conductance (bottom).  
	Solid lines represent fits to the Dynes formula. The extracted parameters $(\Delta,\,\Gamma_\mathrm{D})$ in~\si{\milli\electronvolt}, from top to bottom, are: (1.44, 0.25), (1.56, 0.10), (2.74, 0.02), (2.75, 0.06), (2.74, 0.19), and (2.61, 0.143).
	}
\end{figure}

A total of approximately 100 junctions were measured within a~\qtyproduct{100 x 100}{\micro\meter} area at \qty{1.8}{\kelvin}. Representative tunneling conductance spectra alongside Dynes fits are presented in~\Cref{fig:pct-spectra}. The data reveal a range of junction behaviors: some with smaller superconducting gaps (two top curves), some consistent with ideal \ch{Nb_{0.75}Ti_{0.25}N} spectra (middle two curves), and others showing subgap conductance features (bottom two curves), which may arise from magnetic impurities or local stoichiometric fluctuations within the \ch{Nb_{0.75}Ti_{0.25}N} film.

\begin{figure}
	\centering
	\includegraphics*[width=0.5\columnwidth]{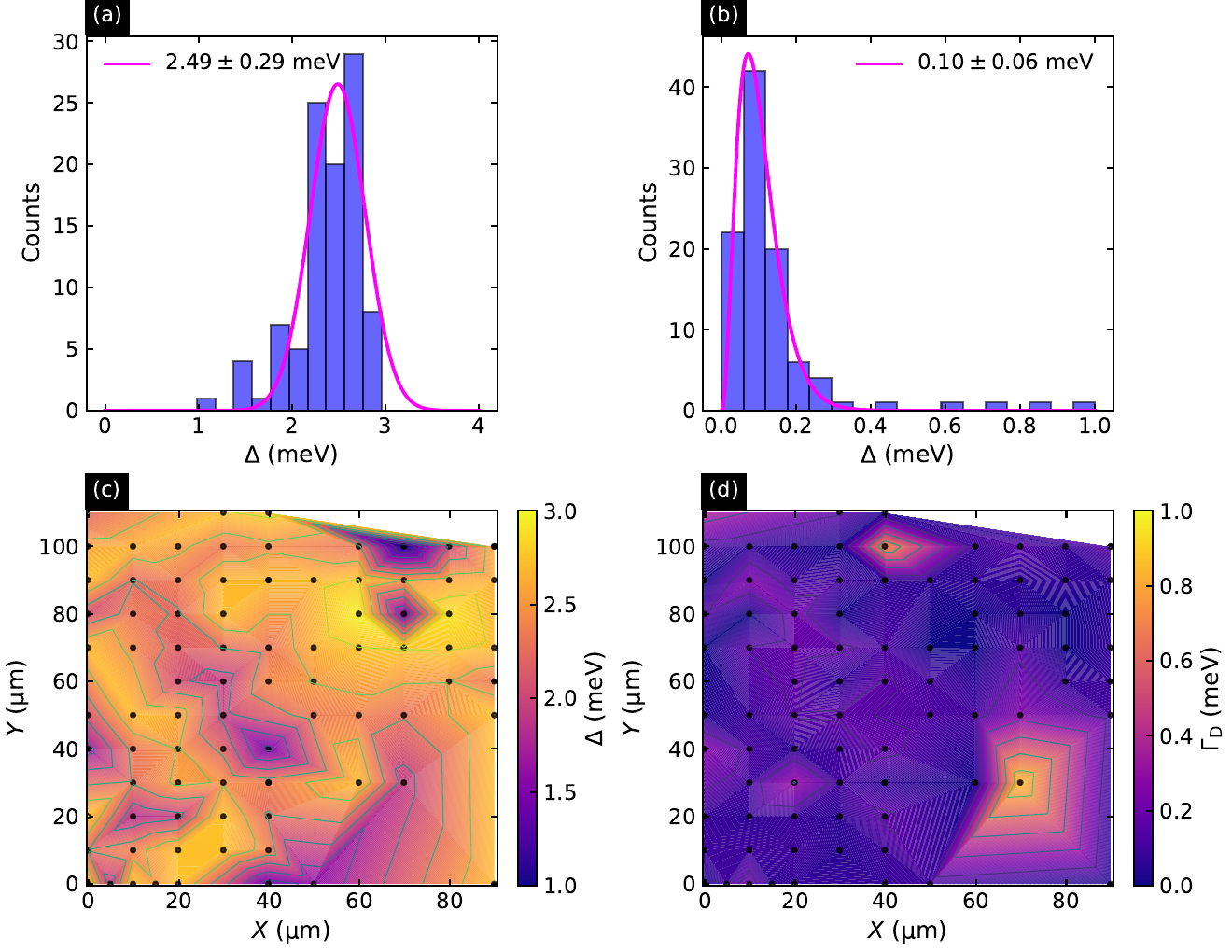}
	\caption{
	\label{fig:pct-delta-gamma} 
	Statistics of the superconducting gap $\Delta$ and Dynes broadening parameter $\Gamma_\mathrm{D}$ were extracted from fits to tunneling conductance spectra measured at \qty{1.8}{\kelvin}.
	(a) Histogram of the $\Delta$, with a Gaussian fit (solid line) giving a mean value of $\Delta = \qty{2.49 \pm 0.29}{\milli\electronvolt}$. (b) Histogram of $\Gamma_\mathrm{D}$, fitted with a Gamma distribution (solid line), yielding a mean value of $\qty{0.10 \pm 0.06}{\milli\electronvolt}$. (c) Spatial map of $\Delta$ and (d) spatial map of $\Gamma_\mathrm{D}$ across the sample surface. Black dots indicate the positions of individual tunnel junctions. The color bars reflect the local values of $\Delta$ and $\Gamma_\mathrm{D}$, revealing the spatial distribution. 
	}
\end{figure}

The statistical distributions of $\Delta$ and $\Gamma_\mathrm{D}$, extracted from Dynes fits, along with their spatial maps, are shown in~\Cref{fig:pct-delta-gamma}. 
The histogram of $\Delta$ values was fitted with a Gaussian function (solid line in~\Cref{fig:pct-delta-gamma}(a)), yielding a distribution centered at \qty{2.49 \pm 0.29}{\milli\electronvolt}. Similarly, the distribution of $\Gamma_\mathrm{D}$ values was fitted with a Gamma distribution (solid line in~\Cref{fig:pct-delta-gamma}(b)), resulting in a mean value of \qty{0.10 \pm 0.06}{\milli\electronvolt}. 
The corresponding spatial maps of $\Delta$ and $\Gamma_\mathrm{D}$ (shown in~\Cref{fig:pct-delta-gamma}(c, d)) reveal some some local variations of superconducting properties.

\begin{figure}
	\centering
	\includegraphics*[width=0.5\columnwidth]{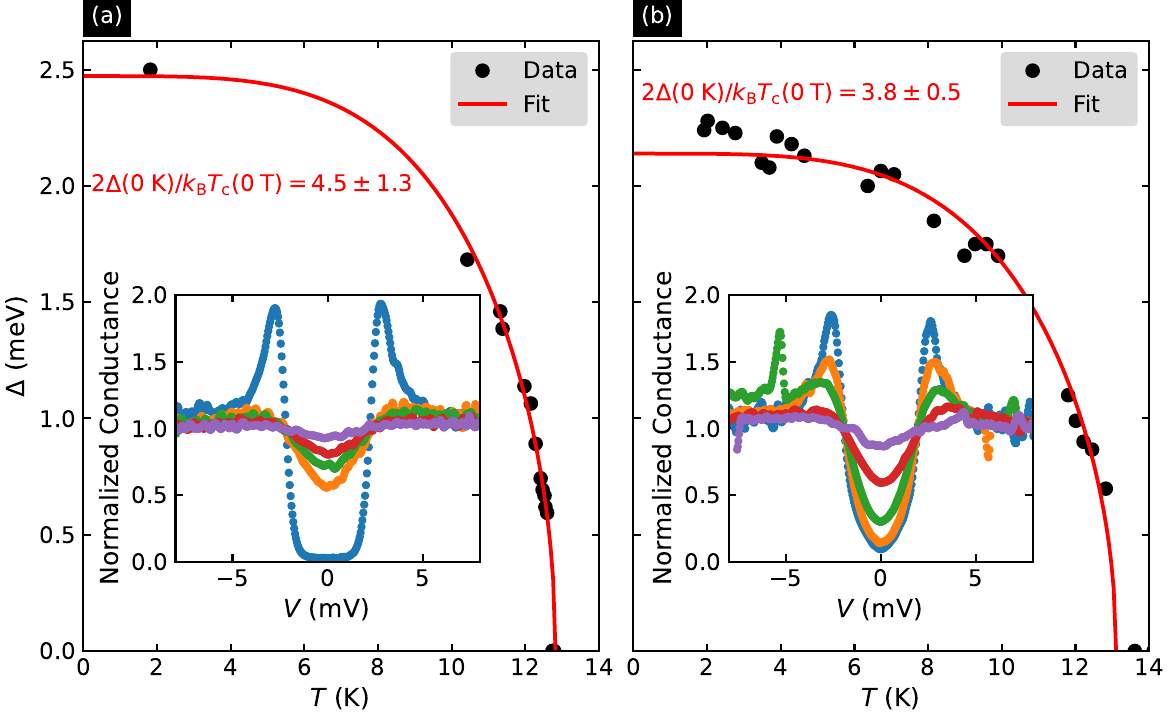}
	\caption{
	\label{fig:pct-two-loc} 
	$T$ dependence of the $\Delta$ measured by~\gls{pct} spectroscopy on the \ch{Nb_{0.75}Ti_{0.25}N}/\ch{AlN}/\ch{Nb} multilayer sample at two different locations. Black circles represent the extracted $\Delta(T)$ values obtained by fitting the differential conductance spectra at each temperature using~\Cref{eq:current-derivative,eq:dos-dynes}. The solid red lines correspond to fits using Equation (17) of the main text, yielding $2\Delta(\qty{0}{\kelvin})/k_\mathrm{B}T_\mathrm{c}(\qty{0}{\tesla})=\num{4.5 \pm 1.3}$ in (a) and \num{3.8 \pm 0.5} in (b), indicating slight spatial variations. The insets show representative normalized tunneling conductance spectra at select temperatures, highlighting the superconducting gap and coherence peaks.
	}
\end{figure}

Additionally, temperature-dependent measurements of the superconducting energy gap $\Delta(T)$ were performed at two distinct locations on the sample, as displayed in~\Cref{fig:pct-two-loc}. In both cases, the temperature evolution of the gap closely follows the expected~\gls{bcs} behavior, yielding $2\Delta(\qty{0}{\kelvin})/k_\mathrm{B} T_\mathrm{c}(\qty{0}{\tesla}) = \num{4.5 \pm 1.3}$ in~\Cref{fig:pct-two-loc}(a) and $2\Delta(\qty{0}{\kelvin})/k_\mathrm{B} T_\mathrm{c}(\qty{0}{\tesla}) = \num{3.8 \pm 0.5}$ in~\Cref{fig:pct-two-loc}(b) confirming conventional $s$-wave superconductivity in the \ch{Nb_{0.75}Ti_{0.25}N} layer.

\bibliography{references.bib}